\documentclass[11pt]{article}

\usepackage{graphicx}
\usepackage{epstopdf}
\begin{document}
\bibliographystyle{unsrt}
%
%%%%%%%%%%%%%%%%%%%%%%%%%%%%%%%%%%%%%%%%%%%%%%%%%%%%%%%%%%%%%%%%%%%%%
%%%%%%%%%%%%%%%%%%%%%%%%%%%%% my commands %%%%%%%%%%%%%%%%%%%%%%%%%%%
%%%%%%%%%%%%%%%%%%%%%%%%%%%%%%%%%%%%%%%%%%%%%%%%%%%%%%%%%%%%%%%%%%%%%
%
% indent first line of new sections
%
% \let\@afterindentfalse\@afterindenttrue
% \@afterindenttrue
%
% grouping commands
\newcommand{\brac}[1]{\left[ #1 \right]}
\newcommand{\cbrace}[1]{\left\{ #1 \right\}}
\newcommand{\select}[1]{\left\{\begin{array}{ll} #1 \end{array} \right.}
\newcommand{\kase}[1]{\begin{array}{ll} #1 \end{array}}
\newcommand{\fn}[1]{\left( #1 \right)}
\newcommand{\ave}[1]{\left\langle #1\right\rangle}
\newcommand{\abs}[1]{\left| #1 \right|}
\newcommand{\norm}[1]{\left\| #1 \right\|}
\newcommand{\Frac}[2]{\frac{\textstyle #1}{\textstyle #2}}
% partial derivative helpers
\newcommand{\dx}{\Delta x}
\newcommand{\dy}{\Delta y}
\newcommand{\dz}{\Delta z}
\newcommand{\dt}{\Delta t}
\newcommand{\D}[2]{\Delta_{#1}#2}
\newcommand{\dxdy}[2]{\frac{\partial #1}{\partial #2}}
\newcommand{\dxdt}[1]{\frac{\partial #1}{\partial t}}
\newcommand{\dxdz}[1]{\frac{\partial #1}{\partial z}}
\newcommand{\ddt}[1]{\frac{\partial}{\partial t} \fn{#1}}
\newcommand{\ddz}[1]{\frac{\partial}{\partial z} \fn{#1}}
\newcommand{\ddx}[1]{\frac{\partial}{\partial x} \fn{#1}}
\newcommand{\dxdydz}[3]{\frac{\partial^2 #1}{\partial #2 \partial #3}}
\newcommand{\dxdys}[2]{\frac{\partial^2 #1}{\partial #2^2}}
\newcommand{\dxdyt}[2]{\frac{\partial^3 #1}{\partial #2^3}}
\newcommand{\dxdyf}[2]{\frac{\partial^4 #1}{\partial #2^4}}
\newcommand{\Dt}[1]{\frac{D#1}{D t}}
\newcommand{\ddxi}[1]{\frac{\partial}{\partial \xi} \fn{#1}}
\newcommand{\ddn}[1]{\frac{\partial}{\partial \eta} \fn{#1}}
\newcommand{\ddtb}[1]{\frac{\partial}{\partial t} \brac{#1}}
\newcommand{\ddzb}[1]{\frac{\partial}{\partial z} \brac{#1}}
\newcommand{\ddxb}[1]{\frac{\partial}{\partial x} \brac{#1}}
\newcommand{\ddxib}[1]{\frac{\partial}{\partial \xi} \brac{#1}}
\newcommand{\ddnb}[1]{\frac{\partial}{\partial \eta} \brac{#1}}
\newcommand{\Grad}[1]{\nabla #1}
\newcommand{\Div}[1]{\nabla\cdot #1}
\newcommand{\Curl}[1]{\nabla\times #1}
% special symbols etc...
\newcommand{\half}{\frac{1}{2}}
\newcommand{\Half}{\Frac{1}{2}}
\newcommand{\sm}[1]{\tilde{s}_{#1}}
\newcommand{\mdot}{\dot{m}}
\newcommand{\ten}[1]{\times 10^{#1}\,}
% other stuff
\def\boldsymbol#1{\mbox{\boldmath$#1$}}
\newcommand{\bls}[1]{\baselineskip #1pt}
\newcommand{\PH}[1]{\begin{center}{\huge \bf #1}\end{center}}
\newcommand{\bc}{\begin{center}}
\newcommand{\ec}{\end{center}}
\newcommand{\bea}{\begin{eqnarray}}
\newcommand{\eea}{\end{eqnarray}}
\newcommand{\beas}{\begin{eqnarray*}}
\newcommand{\eeas}{\end{eqnarray*}}
\newcommand{\be}{\begin{equation}}
\newcommand{\ee}{\end{equation}}
\newcommand{\bseq}{\begin{subequations}}
\newcommand{\eseq}{\end{subequations}}
\newcommand{\bdm}{\begin{displaymath}}
\newcommand{\edm}{\end{displaymath}}
\newcommand{\ben}{\begin{enumerate}}
\newcommand{\een}{\end{enumerate}}
\newcommand{\bit}{\begin{itemize}}
\newcommand{\eit}{\end{itemize}}
\newcommand{\bds}{\begin{description}}
\newcommand{\eds}{\end{description}}
\newcommand{\bvb}{\begin{verbatim}}
\newcommand{\evb}{\end{verbatim}}
\newcommand{\BF}{\begin{figure}}
\newcommand{\EF}{\end{figure}}
\newcommand{\BTB}{\begin{table}}
\newcommand{\ETB}{\end{table}}
\renewcommand{\arraystretch}{1.5}
\renewcommand{\textfraction}{0.0}
\hyphenation{God-u-nov}
\hyphenation{anti-diff-u-sive}
\def\proof{\par{\it Proof}. \ignorespaces}
\def\endproof{{\ \vbox{\hrule\hbox{%
   \vrule height1.3ex\hskip0.8ex\vrule}\hrule
  }}\par}
\newtheorem{theorem}{Theorem}
\newtheorem{remark}{Remark}
\newtheorem{lemma}{Lemma}
\newtheorem{definition}{Definition}
\newtheorem{corollary}{Corollary}
\newtheorem{algorithm}{Algorithm}
\newcommand{\BT}{\begin{theorem}}
\newcommand{\ET}{\end{theorem}}
\newcommand{\BA}{\begin{algorithm}\vspace{24pt}}
\newcommand{\EA}{\end{algorithm}}
\newcommand{\BL}{\begin{lemma}}
\newcommand{\EL}{\end{lemma}}
\newcommand{\BR}{\begin{remark}}
\newcommand{\ER}{\end{remark}}
\newcommand{\BC}{\begin{corollary}}
\newcommand{\EC}{\end{corollary}}
\newcommand{\BD}{\begin{definition}}
\newcommand{\ED}{\end{definition}}
%
%%%%%%%%%%%%%%%%%%%%%%%%%%%%%%%%%%%%%%%%%%%%%%%%%%%%%%%%%%%%%%%%%%%%%
%%%%%%%%%%%%%%%%%%%%%%%%%%%%%% document %%%%%%%%%%%%%%%%%%%%%%%%%%%%%
%%%%%%%%%%%%%%%%%%%%%%%%%%%%%%%%%%%%%%%%%%%%%%%%%%%%%%%%%%%%%%%%%%%%%
%
\begin{center}
{\Large\bf A Generalized Flux-Corrected Transport Algorithm I: A 
Finite-Difference Formulation}\\ 
\vspace{0,5in}
{William J. Rider and Dennis R. Liles\\ 
Reactor Design and Analysis Group\\
Los Alamos National Laboratory\\ 
Los Alamos, NM 87545\\} 
\end{center}
Classification Index Numbers: 65M05, 76N10\\
Keywords: flux-corrected transport, conservation laws, high-resolution\\

\begin{abstract}
This paper presents a generalized flux-corrected transport (FCT) 
algorithm, which is shown to be total variation diminishing under some 
conditions.  The new algorithm has improved properties from the 
standpoint of use and analysis.  Results show that the new FCT algorithm 
performs better than the older FCT algorithms and is comparable with 
other modern methods.  This reformulation will also allow the FCT to be 
used effectively with exact or approximate Riemann solvers and as an 
implicit algorithm.
\end{abstract}

\section{Introduction}\label{sect: intro}
Godunov~\cite{godunov59} showed that the monotonic solution of first-
order hyperbolic conservation laws is at most first-order accurate for 
linear differencing schemes.  The first algorithm to successfully 
address this difficulty was the flux-corrected transport (FCT) algorithm 
of Boris, Book, and Hain!\cite{boris73,boris75,boris76a,boris76b} .  
This algorithm performed quite well on 
linear advection problems and paved the way for future developments in 
the field.  It essentially consisted of computing a solution with a 
nondiffusive transport method followed by a stabilizing diffusive step.  
This monotone solution is then used to aid in the construction of an 
antidiffusive step in which the solution from the first part of the 
algorithm is locally sampled and corrections are ``patched'' to it.  
This is accomplished with a flux limiter that only applies the flux 
corrections in the smooth part of the flow.   As a result, the solution 
will be of a high-order in smooth parts of the convected profile, but 
first-order near discontinuities and steep gradients.  Extension of the 
FCT algorithm to systems of conservation laws, however, has proved less 
successful.

Further developments on this topic were achieved by van 
Leer~\cite{vanleer79} in his higher order extensions of Godunov's method 
often referred to as MUSCL.  The prescription of slope-limiting used by 
van Leer has great similarity to the flux-limiting used in the original 
FCT.  The difficulties associated with FCT with systems equations is not 
shared by MUSCL because an exact or approximate solution to the local 
Riemann problem is used to construct the convective fluxes.  While this 
approach adds complexity and cost to the solution procedure, the 
corresponding quality of the solution is greatly improved.  Recently, 
researchers have extended the ideas of van Leer to arbitrarily 
high-order spatially or temporally and christened these methods as 
uniformly~\cite{harten87b} or essentially~\cite{harten87a} 
nonoscillatory (UNO or ENO) schemes.  

The effort to put the new modern algorithms on firmer theoretical 
footing resulted in the concept of total variation diminishing (TVD) 
methods~\cite{harten84}, which have a number of desirable properties.  
To be total variation diminishing, a scheme must satisfy the following 
inequalities,
\bdm
TV\fn{u^{n+1}}\leq TV\fn{u^n}\;,
\edm
where
\bdm
TV\fn{u}=\sum_{j=-\infty}^\infty \abs{u_{j+1} - u_j}\;.
\edm
While these methods include classic monotone schemes, they can also be 
extended to include methods that are second-order in the $L_1$ norm.  By 
construction, these methods are still first-order at points of extrema 
(in the $L_\infty$ norm).  A second property of TVD schemes, which is 
both useful and satisfying, is that they can be extended to include 
implicit temporal differencing~\cite{yee85}.  This generality is quite 
desirable as it allows a more general use of TVD algorithms for a wide 
range of problems.  It should be noted that MUSCL schemes have also been 
extended to include implicit temporal 
differencing~\cite{vanleer85,yee90}.

Zalesak~\cite{zalesak79} redefined the FCT in such a way as to make it 
more general.  A standard low-order solution, similar to that obtained 
by donor-cell differencing, is used to define a monotonic solution.  
This solution is then used to limit an antidiffusive flux, which is 
defined as the difference between a high-order and low-order flux.  As 
with the earlier versions of the FCT, the limiter is designed to give no 
antidiffusive flux when an extrema or a discontinuity is reached.  This 
prescription of the FCT can allow the user to specify a wide range of 
low-order fluxes as well as a large variety of high-order fluxes.  These 
have included central differencing of second or higher order, 
Lax-Wendroff, and spectral fluxes~\cite{mcdonald89}.  Recently, several 
researchers~\cite{steinle89} have introduced an implicit FCT algorithm; 
however, this algorithm is limited to small multiples of the 
Courant-Friedrichs-Lewy (CFL) number.  This is because the low-order 
solution is produced by multiple sub-cycles with an explicit donor-cell 
(or other monotonic) solution and an implicit high-order solution.  The 
high-order solution is only stable for small multiples of the  CFL 
number, thus limiting the applicability of this algorithm.  The FCT has 
also been extended for use with a finite-element solution method with 
great success~\cite{lohner87}.

The performance of the explicit FCT algorithm is the subject of this 
paper.  Several investigators~\cite{ikeda79}~\cite{woodward84} have 
noted for the older FCT algorithm that a lower CFL limit is required for 
stability.  The FCT algorithm also suffers from being overcompressive 
(as will be shown in Section~\ref{sect: results}). This was shown in a 
test of the FCT on a shock tube problem~\cite{zalesak81}, where even at 
a CFL number of 0.1, the solution was of relatively poor quality.  This 
probably is due to the handling of the pressure-related terms in the 
momentum and energy equations.  This work aims to address these 
problems, first through making several improvements to the FCT and then 
by showing the extension of this modified FCT to systems of equations.  
In accomplishing this, we will make extensive use of approximate Riemann 
solvers of the type introduced by Roe~\cite{roe81}.

We have several objectives to be addressed in this paper: the 
performance of FCT on systems of equations, which needs to be improved 
in terms of solution quality, and efficiency (the necessity of using 
small CFL numbers), more direct ties to other modern algorithms  (such 
as TVD algorithms), and analysis of the potential TVD properties of FCT 
methods.  The first objective will follow the last two objectives in 
treatment.  The first two objectives are complementary in nature and 
should follow from one another.  In producing new FCT algorithms, we 
will seek one step methods, not requiring the diffusive first step used 
in older FCT methods.

This paper is organized into four sections.  The following section 
provides an overview of the numerical solution of hyperbolic 
conservation laws.  Later in that section, the FCT method according to 
Zalesak is introduced.  This method is analyzed and suggestions for 
improvements are made including the extension of FCT to systems of 
equations.  This takes two forms: one method is denoted as the ``new 
FCT'' method, and the second is denoted as the ``modified-flux FCT'' 
method.  The ``new FCT'' method is similar to symmetric TVD methods, and 
the ``modified-flux FCT'' is similar to Harten's modified-flux TVD 
method.  In the third section, results are presented for the methods 
discussed in the paper.  These results are for a scalar wave equation, 
Burgers' equation and a shock tube problem for the Euler equations.  
Finally, some closing remarks will be made.

\section{Method Development}\label{sect: theory}
Before describing the changes we will make in the FCT algorithm, we will 
make several introductory points.  Consider
\be
\Frac{\partial u}{\partial t} + \Frac{\partial f\fn{u}}{\partial x} = 
0\;,
\label{eq: scalar}
\ee
which is a first-order hyperbolic transport equation for $u$ where $f$ 
is
the flux of $u$.  Equation~(\ref{eq: scalar}) can be written as
\be
\Frac{\partial u}{\partial t} + a\Frac{\partial u}{\partial x} = 0\;,
\label{eq: scalar char}
\ee
where
\bdm
a = \Frac{\partial f}{\partial u}\;.
\edm
The characteristic speed, $a$, is particularly useful in defining 
finite-difference solutions to Eq~\ (\ref{eq: scalar}).  A system of 
conservation laws can be similarly defined; however, the construction of 
effective finite-difference solution for systems of equations requires 
more care. Consider
\be
\Frac{\partial {\bf U}}{\partial t} + \Frac{\partial {\bf F}\fn{\bf 
U}}{\partial x} = 0\;,
\label{eq: hscl}
\ee
which is a set of hyperbolic conservation laws where ${\bf U}$ is a 
column vector $\fn{u^1,u^2,\ldots,u^m}^T$ of conserved quantities and 
${\bf F}\fn{\bf U}$ is a column vector $\fn{f^1,f^2,\ldots,f^m}^T$of 
fluxes of ${\bf U}$.  Equation (\ref{eq: hscl}) can be written as
\be
\Frac{\partial {\bf U}}{\partial t} + A\Frac{\partial {\bf U}}{\partial 
x} = 0\;,
\label{eq: hscl char}
\ee
where
\bdm
A = \Frac{\partial {\bf F}\fn{\bf U}}{\partial {\bf 
U}}=\brac{\begin{array}{ccc} \partial f^1/\partial u^1         & \ldots      
& \partial f^1/\partial u^m\\ 
                             \vdots         & \ddots  & \vdots\\ 
                             \partial f^m/\partial u^1 & \ldots & 
\partial f^m/\partial u^m \end{array}}\;.
\edm
The matrix $A$ is the flux Jacobian for the system defined by Eq.\ 
(\ref{eq: hscl}), which is quite useful in the construction of 
finite-difference solutions of this system of equations, as will be 
shown in Section~\ref{sect: syst} of this paper.

In general, equations of the type considered above can develop 
discontinuous solutions even when the initial data is smooth.  Because 
of this, the solutions are not unique.  To rectify this, the admissible 
solutions must satisfy an entropy condition (for details on this 
see~\cite{lax72}~\cite{smoller82}).  It is the formation of 
discontinuities in the solution that causes the difficulties for 
finite-difference solutions of Eq.\ (\ref{eq: scalar}).  At these 
discontinuities, the function ceases to be smooth and the usual 
assumptions made in constructing finite-difference approximations 
collapse.  As a result, more physical information needs to be 
incorporated into the solution procedure.

The FCT (and most other finite-difference methods) was constructed to 
solve Eq.\ (\ref{eq: scalar}).  This approach will be followed 
initially, but will eventually be abandoned to some extent when systems 
of equations are considered.  First, the basics of Zalesak's FCT will be 
reviewed, followed by several basic suggestions for improvements in this 
algorithm.  These improvements will be discussed briefly with 
comparisons being drawn between these new FCT methods and the symmetric 
TVD schemes~\cite{yee87a}.  Finally, the FCT will be extended to systems 
of hyperbolic conservation laws using an approximate Riemann solver.

For the remainder of the presentation, the following nomenclature will 
be used: $\Delta_{j+\half}u = u_{j+1} - u_j$.  A conservative 
finite-difference solution to Eq.\ (\ref{eq: scalar}) using a simple 
forward Euler time discretization is
\be
u_j^{n+1} = u_j^n - \lambda \fn{\hat{f}_{j+\half} - \hat{f}_{j-
\half}}\;,
\label{eq: cfd}
\ee
where $\lambda = \Delta t / \Delta x$.  The temporal spacing is $\Delta 
t$ and $\Delta x$ is the spatial mesh spacing that will be assumed 
constant for the remainder of the paper (varying mesh widths will result 
is somewhat more complex expressions).  The superscript $n$ refers to 
time, $n+1$ refers to the time $t+\Delta t$, and the subscript $j$ 
refers to space with $j$ being a cell center and $j\pm\half$ being the 
cell edges.  The construction of the numerical fluxes 
$\hat{f}_{j\pm\half}$ will be the subject of this section.  The cell 
edge flux is defined as
\be
\hat{f}_{j+\half} = \Half \fn{f_j + f_{j+1}} + \phi_{j+\half}\;,
\ee
where $\phi$ is a numerical dissipation term.  For a system of equations 
the flux is written
\be
\hat{\bf F}_{j+\half} = \Half \fn{{\bf F}_j + {\bf F}_{j+1}} + 
\Phi_{j+\half}\;,
\ee
where ${\bf F}$ and $\Phi$ are vectors, but are defined similarly to the 
single equation case.  For instance, the first-order donor-cell flux is
\be
\hat{f}_{j+\half}^{DC} = \Half \fn{f_j + f_{j+1} -
\abs{a_{j+\half}}\Delta_{j+\half}u}\;,
\label{eq: dc}
\ee
thus
\bdm
\phi_{j+\half}^{DC} = - \Half\abs{a_{j+\half}}\Delta_{j+\half}u\;.
\edm
For the remainder of the paper, the prescription of the numerical 
dissipation term, $\phi$ or $\Phi$, will be used to define algorithms.  
One slight modification of the above methodology is used for nonlinear 
equations and systems; as suggested by Yee~\cite{yee87a} an entropy fix 
is implemented for the donor-cell differencing, which modifies the use 
of the absolute value in donor-cell differencing  of a characteristic 
speed, by
\be
\psi\fn{z} = \select{\abs{z} & \mbox{if $\abs{z} \geq \epsilon$}\\ 
\fn{z^2 + \epsilon^2}/2\epsilon & \mbox{if $\abs{z} < \epsilon$}}\;,
\ee
if one is dealing with a linear equation set $\epsilon =0$.  The 
parameter $\epsilon$ is determined by the following 
equation~\cite{harten83c},
\bdm
\epsilon = \max\brac{0,a_{j+\half} - a_j,a_{j+1} - a_{j+\half}}\;.
\edm
Thus, the numerical diffusion term in the donor-cell flux becomes
\bdm
\phi_{j+\half}^{DC} = - \Half\psi\fn{a_{j+\half}}\Delta_{j+\half}u\;.
\edm
This description of the donor-cell method reduces to Roe's 
method~\cite{roe81} for scalar equations.  The term donor-cell methods 
and Roe's methods should be viewed as equivalent in this sense.

\subsection{Zalesak's FCT Algorithm}
Zalesak's FCT has been classified as a hybrid method that is applied in 
two steps.  By being hybrid, the algorithm is based on the blending of 
high- and low-order difference schemes together.  Step one is 
accomplished with a first-order monotonic solution such as donor-cell 
plus some additional diffusion (the entropy fix discussed in the 
previous section adds such dissipation).  This could be accomplished 
with other first-order algorithms such as Godunov's~\cite{godunov59} or 
Engquist and Osher's~\cite{engquist81}.  These fluxes are used to 
produce a transported diffused solution $\tilde{u}$ as follows:
\be
\tilde{u}_j = u_j^n - \lambda\fn{\hat{f}_{j+\half}^{DC} - \hat{f}_{j-
\half}^{DC}}\;.
\label{eq: utilde}
\ee
A high-order flux, $f^H$, is defined in some way and then the low-order 
flux is subtracted from the high-order flux to define the antidiffusive 
flux as
\bdm
\hat{f}_{j+\half}^{AD} = \hat{f}_{j+\half}^H - \hat{f}_{j+\half}^L\;,
\edm
where in this paper, we have defined $\hat{f}^L = \hat{f}^{DC}$.
The antidiffusive flux is then limited with respect to the local 
gradients of the conserved variable computed with the transported and 
diffused solution.  Zalesak defined his limiter as a prelude to a truly 
multidimensional limiter, but also defined an equivalent limiter as
\be
\hat{f}_{j+\half}^C = S_{j+\half}\max\cbrace{0,
\min\brac{S_{j+\half}\lambda^{-1}\Delta_{j-\half}\tilde{u},
\abs{\hat{f}_{j+\half}^{AD}},S_{j+\half}\lambda^{-
1}\Delta_\frac{3}{2}\tilde{u}}}\;,
\label{eq: old limiter}
\ee
where $S_{j+\half} = \hat{f}_{j+\half}^{AD} / 
\abs{\hat{f}_{j+\half}^{AD}}$ is the sign of the conserved variable's 
gradient spatially.  This limiter is identical to the limiter defined by 
Boris and Book~\cite{boris73}, but with a different definition of 
$\hat{f}^{AD}$. The final cell-edge numerical diffusion is defined by
\be
\phi_{j+\half}^{FCT} = \hat{f}_{j+\half}^C + \phi_{j+\half}^{DC}\;.
\ee
The FCT generally carries a stability limit on its time step of
\bdm
\lambda \abs{a} \leq 1\;.
\edm

Before going further, several critical comments need to be made 
concerning this algorithm.  Despite the striking generality, which is 
driven by the prescription of the antidiffusive fluxes, the algorithm 
has some deficiencies.  By its formulation as a two-step method it has 
some disadvantages in terms of analytical analysis and efficiency of 
implementation.  By the use of the inverse grid ratio $\lambda^{-1}$ in 
the flux limiter, the algorithm is effectively limited to explicit time 
discretization (as will be shown in the following section).  The use of 
a diffused solution in the limiter is important in stabilizing the 
solution, which could yield oscillatory solutions without this step.  
Under closer examination, the use of a diffused solution acts as an 
upwind weighted artificial diffusion term.  This sort of definition 
could lead to a fairly complex one-step FCT algorithm, which has, at 
first glance similarity to UNO-type schemes.  The diffusive terms in the 
FCT algorithm's limiter are upwind weighted rather than centered as with 
UNO based algorithms.  This can be seen by direct substitution of the 
diffusive first step into the second step of the algorithm.  
Additionally, numerical experiments with a scalar advection equation 
show that the total variation for the FCT solution can increase with 
time for a CFL number less than one.  

The use of higher order antidiffusive fluxes with this prescription of 
the FCT also raises some questions about the actual order of the 
approximation.  The antidiffusive flux is of the higher order, but the 
local gradients in the limiter are only accurate to second-order.  This 
suggests that the solution may actually be of only second-order 
spatially (in the $L_1$ norm).  This also holds for temporal order as 
the local gradient terms are only first-order in space, thus an 
antidiffusive flux based on a Lax-Wendroff flux may actually yield a 
first-order accurate temporal approximation.  Thus the form of the local 
gradients used in the limiter may also need to be modified to accomplish 
the goal of true higher order accuracy.  

This seems to contradict results reported in~\cite{oran87,zalesak87}, 
but these results are for a scalar wave equation where the solution is 
translated by the flow field.  With nonlinear equations or systems this 
may cause difficulties because of the formation of shocks and 
discontinuities.  This is explored further in Section~\ref{sect: burg}.

\subsection{A New FCT Algorithm}
The first and simplest change is to rewrite the flux limiter as
\be
\hat{f}_{j+\half}^C=S_{j+\half}\max\left\{\rule{0ex}{2.5ex}0,\min\right[
S_{j+\half} \tilde{\sigma}_{j-\half}\Delta_{j-\half}\tilde{u},
\abs{\hat{f}_{j+\half}^{AD}}, 
S_{j+\half}
\tilde{\sigma}_{j+\frac{3}{2}}\Delta_{j+\frac{3}{2}}\tilde{u} 
\left]\rule{0ex}{2.5ex}\right\} \;,
\label{eq: new fct limiter}
\ee
where
\be
\tilde{\sigma}_{j+\half} = \psi\fn{\tilde{a}_{j+\half}}\;,
\label{eq: sig2}
\ee
or
\be
\tilde{\sigma}_{j+\half} = \psi\fn{\tilde{a}_{j+\half}} - \lambda 
\tilde{a}_{j+\half}^2\;,
\label{eq: sig22}
\ee
and $S_{j+\half}$ has the same definition as before.  The second choice 
for $\tilde{\sigma}_{j+\half}$ gives second-order accuracy in both time 
and space if $\hat{f}_{j+\half}^{AD}$ is of similar or higher 
accuracy~\cite{harten84}.  This relatively small change has a 
significant impact on the FCT algorithm as will be shown later both 
analytically and computationally.   This form is also a great deal 
closer to the definition of limiters used in TVD algorithms.  However, 
this still leaves a two-step method which poses some problems from the 
standpoint of efficiency and extension to systems of conservation laws.

The similarities of this modification of the FCT with symmetric TVD 
schemes~\cite{yee87a} are quite strong.  The necessary changes to 
convert this scheme into one equivalent to the one described by Yee are 
simple.  This consists of dividing the local gradient terms in the 
limiter by two and removing the first step of the FCT.  Yee writes the 
numerical flux for the symmetric TVD method as
\be
\hat{f}_{j+\half} = \Half \brac{a_{j+\half} \fn{u_j + u_{j+1}} - 
\psi\fn{a_{j+\half}}\Delta_{j+\half}u + Q_{j+\half}}\;.
\label{eq: sym tvd}
\ee
An example of the $Q_{j+\half}$ function would be
\bea
Q_{j+\half}=S_{j+\half}\max\left\{\rule{0ex}{2.5ex}0,\min\right[& S_{j+\half} 
&\psi\fn{a_{j+\frac{3}{2}}}\Delta_{j+\frac{3}{2}}u,
\psi\fn{a_{j+\half}}\Delta_{j+\half}u, \nonumber\\  
& S_{j+\half} & \psi\fn{a_{j-\half}}\Delta_{j-\half}u 
\left]\rule{0ex}{2.5ex}\right\} \;,
\label{eq: Q 1}
\eea
which strikes a strong resemblance with Eq.\ (\ref{eq: new fct limiter}) for 
an antidiffusive flux defined with a second-order central difference.  For 
ease of analysis, this method is rewritten in the following form:
\be
\hat{f}_{j+\half} = \Half \brac{a_{j+\half} \fn{u_j + u_{j+1}} - 
\psi\fn{a_{j+\half}} \fn{1- Q_{j+\half}}\Delta_{j+\half}u}\;,
\label{eq: sym tvd2}
\ee
where 
\bdm
Q_{j+\half}=\mbox{minmod}\fn{1,r^+_{j+\half}, r^-_{j+\half}}\;,
\edm
with $r^+_{j+\half} = \Delta_{j+\frac{3}{2}}u / \Delta_{j+\half}u$ and $r^-
_{j+\half} = \Delta_{j-\half}u / \Delta_{j+\half}u$.  The minmod limiter used 
with symmetric TVD schemes is defined by Yee, but has the same effect as Eq.\ 
(\ref{eq: Q 1}).

\BT
The new FCT algorithm without the monotone (TVD) first step is TVD under 
the following conditions: for $\tilde{\sigma} = \psi\fn{a}$ 
\be
\lambda \abs{a} < \Frac{1}{2\fn{1-\theta}}\;,
\ee
and for $\tilde{\sigma} = \psi\fn{a} -\lambda a^2$
\be
\lambda \abs{a} < \Frac{1}{1-\theta}\;.
\ee
Here $\theta$ is an implicitness parameter (see Yee~\cite{yee87a}) with 
$\theta = 1$ the equation is fully implicit, and with $\theta=0$ the 
equation is fully explicit.  Without the monotone first step, the fully 
explicit old FCT algorithm is not TVD for any CFL number, but the 
implicit FCT algorithm is TVD under the condition
\be
\lambda \abs{a} < \Frac{\theta}{1-\theta}\;.
\ee
\ET

\begin{proof}
The FCT cell-edge flux can be written in the same way as the flux for a 
symmetric TVD scheme by defining
\be
\hat{f}^C_{j+\half} = \Half\abs{a_{j+\half}} Q_{j+\half} 
\Delta_{j+\half}u\;,
\ee
if $Q_{j+\half}$ is based on Eq. (~\ref{eq: sig2})
\bdm
Q_{j+\half} = \mbox{minmod}\fn{1,2\tilde{r}^+,2\tilde{r}^-}\;,
\edm
and if $Q_{j+\half}$ is based on Eq. (~\ref{eq: sig22})
\bdm
Q_{j+\half} = \fn{1-\lambda \abs{a_{j+\half}}} 
\mbox{minmod}\fn{1,2\tilde{r}^+,2\tilde{r}^-}\;,
\edm
and
\bdm
\tilde{r}^+ = 
\Frac{\Delta_{j+\frac{3}{2}}\tilde{u}}{\Delta_{j+\half}u}\;,
\edm
\bdm
\tilde{r}^- = \Frac{\Delta_{j-\half}\tilde{u}}{\Delta_{j+\half}u}\;.
\edm
In~\cite{yee87a} the inequalities that need to be satisfied in order for 
a flux of the form given in Eq.\ (\ref{eq: sym tvd}) to be TVD are
\be
Q_{j+\half} < 2\;,
\ee
and
\be
\Frac{Q_{j+\half}}{r_{j+\half}^\pm} < \Frac{2}{\lambda\fn{1-\theta} 
\abs{a_{j+\half}}} - 2\;,
\label{eq: q2}
\ee
\be
\lambda \abs{a}<\Frac{1}{1-\theta}\;.
\ee
The FCT limiter given in Eq.\ (\ref{eq: new fct limiter}) satisfies the 
first and last of these relations, but satisfaction of the other 
relation (\ref{eq: q2}) in a rigorous manner has proved to be more 
difficult.  To establish some bounds on the properties of the FCT 
solutions, the first step of the FCT will be ignored for the time being.  
Given this, the worst cases for the limiter are $Q = 2r^{\pm}$ or 
$2\fn{1-\lambda \abs{a}}r^{\pm}$.  Comparing the first of these cases 
with Eq.\ (\ref{eq: q2}) gives
\bdm
2 < \Frac{2}{\lambda\fn{1-\theta} \abs{a}} - 2\;,
\edm
or
\bdm
\lambda \abs{a} < \Frac{1}{2\fn{1-\theta}}\;.
\edm
For the second of the two cases,
\bdm
2\fn{1-\lambda \abs{a}} < \Frac{2}{\lambda\abs{a}\fn{1-\theta}} - 2\;,
\edm
or
\bdm
\lambda \abs{a} < \Frac{1}{1-\theta}\;.
\edm

\BR
Thus, even without the first step, the new FCT algorithm is TVD under 
some conditions (given above).  Thus the implementations of this method 
should not include the diffusive first step because it is not necessary.  
It is also unconditionally stable for fully implicit temporal 
discretization.  The first step adds more dissipation into the 
algorithm, which should result in higher CFL limits for the first case.  
Numerical experiments confirm this and show that the new FCT is TVD for 
all CFL numbers less than one.  
\ER

Zalesak's FCT can be subjected to a similar test after a reformulation 
of its limiter.  Given the same definition as before for 
$f^C_{j+\half}$, 
\be
Q_{j+\half} = \mbox{minmod}\fn{1,\Frac{2\tilde{r}^+}{\lambda 
\abs{a}},\Frac{2\tilde{r}^-}{\lambda \abs{a}}}\;,
\ee
where $\tilde{r}^{\pm}$ are defined as before.  Using Eq.\ (\ref{eq: 
q2}), and again neglecting the first step, one can show that
\be
\lambda \abs{a} < \Frac{\theta}{1-\theta}\;.
\ee
\end{proof}

\BR
Thus, for a fully explicit approximation without the first step, 
Zalesak's FCT is never TVD.  However, as the degree of implicitness 
increases, the algorithm becomes TVD for some CFL numbers and eventually 
becomes unconditionally TVD at $\theta = 1$.  If one looks at the form 
of the limiter as the CFL number increases, the effective antidiffusive 
flux reduces in an inversely proportional fashion.  Therefore, at large 
CFL numbers, Zalesak's FCT is largely ineffective as a high-order 
implicit algorithm.  Numerical experiments have shown that with the 
first step, Zalesak's FCT produces results that diminish in total 
variation up to a CFL number of about 0.95.  
\ER

The new FCT method given above is an analog to the symmetric TVD method.  
Another form of common TVD method is Harten's modified flux method.  The 
next section describes a FCT method developed along those lines.

\subsection{A Modified-Flux FCT Algorithm}
To attain these goals, the FCT will be recast in the form of Harten's 
modified-flux TVD scheme~\cite{harten84}.  From this basis several FCT 
limiters can be shown to be TVD by the criteria given by~\cite{sweby84}, 
and the FCT can be written as a one-step method and extended to use as 
an implicit algorithm in the same way as TVD methods are~\cite{yee85}.  
This will be examined in a future paper.

The modified-flux TVD method is defined by computing cell-centered 
modified fluxes and making the overall flux upwind with respect to both 
the ``physical'' and modified fluxes.  Formally, the modified-flux 
formulation has a dissipation
term,
\be
\phi_{j+\half}^{MF} = \Half\brac{g_j + g_{j+1} - \psi\fn{a_{j+\half} + 
\gamma_{j+\half}}\Delta_{j+\half}u}\;,
\label{eq: mod flux}
\ee
where
\be
g_j = \mbox{minmod}\fn{\sigma_{j-\half}\Delta_{j-\half}u, 
\sigma_{j+\half}\Delta_{j+\half}u}\;,
\label{eq: def g}
\ee
and
\be
\gamma_{j+\half} = \select{\Frac{\Delta_{j+\half}g}{\Delta_{j+\half}u} & 
\mbox{if $\Delta_{j+\half}u \neq 0$}\\ 0 & \mbox{otherwise}}\;.
\ee
The minmod function of two arguments has the usual definition given 
in~\cite{yee87b}, which gives the same effect as the FCT limiter for 
three arguments.  A general form of the minmod function for two 
arguments is
\bea
\mbox{minmod}\fn{a,b;n} = \mbox{sign}\fn{a} \max\left[\right. 0, & \min 
& \fn{n\abs{a}, \mbox{sign}\fn{a}b},\nonumber \\ & \min & 
\fn{\abs{a},n\;\mbox{sign}\fn{a}b}\left]\right. \;,
\eea
which for $n=2$ gives the Superbee limiter developed by 
Roe~\cite{roe85a}.  The function $\sigma_{j+\half}$ can have several 
forms, including
\be
\sigma_{j+\half} = \Half\psi\fn{a_{j+\half}}\;,
\label{eq: sig upwind}
\ee
or
\be
\sigma_{j+\half} = \Half\brac{\psi\fn{a_{j+\half}} - \lambda 
a_{j+\half}^2}\;.
\label{eq: sig 2nd order}
\ee
For Eq.\ (~\ref{eq: sig upwind}), the stability limit depends on the 
form of the limiter, for instance the general minmod limiter yields a 
stability limit of 
\bdm
\lambda \abs{a} \leq \Frac{2}{\fn{2+n}\fn{1-\theta}}\;
\edm
for $n \leq 2$.  The use of Eq.\ (~\ref{eq: sig 2nd order}) gives a 
stability limit of
\bdm
\lambda \abs{a} \leq 1\;
\edm
for all values of $n \leq 2$.  The second definition has been 
recommended for explicit, time-accurate solutions~\cite{harten84,yee85}.

To formulate the FCT in a similar form, simply change the specification 
of the limiter.  The traditional limiter used with the FCT is 
effectively a cell-edged flux rather than a cell-centered flux as needed 
for the modified-flux formulation.  The definition of the antidiffusive 
flux must also be changed to a form more amenable to this formulation.  
This requires a more thoughtful statement of the antidiffusive flux, 
which can be easily incorporated with the type of formulation desired.  
For instance, the second-order central difference antidiffusive flux is
\be
\hat{f}_{j+\half}^{AD} = \Half \psi\fn{a_{j+\half}}\Delta_{j+\half}u\;,
\ee
or a Lax-Wendroff flux
\be
\hat{f}_{j+\half}^{AD} = \Half \brac{\psi\fn{a_{j+\half}} - \lambda
a_{j+\half}^2}\Delta_{j+\half}u\;,
\ee
or a fourth-order central difference
\be
\hat{f}_{j+\half}^{AD} = \Half\psi\fn{a_{j+\half}}\Delta_{j+\half}u + 
\Frac{1}{12} \fn{\Delta_{j-\half}f - \Delta_{j+\frac{3}{2}}f}\;,
\ee
which can be written
\bdm
\hat{f}_{j+\half}^{AD} = \Half\psi\fn{a_{j+\half}}\Delta_{j+\half}u + 
\Frac{1}{12} \fn{a_{j-\half}\Delta_{j-\half}u - 
a_{j+\frac{3}{2}}\Delta_{j+\frac{3}{2}}u}\;.
\edm
These forms can be incorporated with a new limiter that has the desired 
properties.  These properties are the centering of the flux about grid 
point $j$, and being TVD for second-order antidiffusive fluxes.  This 
limiter has the following form:
\bea
g_j = \mbox{minmod}\fn{\hat{f}_{j\pm\half}^{AD},\D{j\pm\half}{u};n} = & 
S_{j+\half} & \max\left[\rule{0ex}{2.5ex}\right. 0, \min \fn{\Half 
n\abs{\hat{f}_{j+\half}^{AD}}, n
S_{j+\half} \sigma_{j-\half}\Delta_{j-\half}u}, \nonumber \\ & \min & 
\fn{n\sigma_{j+\half}\abs{\Delta_{j+\half}u},\Half n S_{j+\half}
\hat{f}_{j-\half}^{AD}}\left. \rule{0ex}{2.5ex}\right ] \;,
\label{eq: fct tvd limiter}
\eea
where $\sigma_{j+\half}$ is defined by Eq.\ (\ref{eq: sig upwind}) or 
Eq.\ (\ref{eq: sig 2nd order}).  The function $S_{j+\half}$ is set to 
zero if the $\hat{f}_{j\pm\half}$ differ in sign.

\BR
It is important to note that this method does not require a diffusive 
first step to be successful.
\ER

Analysis of this limiter for the second-order central-difference-based 
antidiffusive flux follows that of Sweby~\cite{sweby84}.  For the values 
of $0\leq n \leq 2$ in Eq.\ (\ref{eq: fct tvd limiter}), the resulting 
limiter is in the TVD region of the curves shown in Fig.\ \ref{fig: 
limiters}.  For the value of $n=2$, the resulting limiter is identical 
to Roe's Superbee limiter~\cite{roe85a}.  Shown in this figure are the 
plots for $n=1$ and $n=1.5$; the plot for $n=2$ is identical to the 
upper boundary of the second-order TVD region.  The second-order TVD 
region is shown by the shaded region of the figure.  These limiters are 
second-order for all $n$ for $r \leq 1/2$ and also second-order for $r 
\geq 2/n$.  The only limiter of this class that is always second-order 
is the $n=2$ limiter.  In this figure the terms $r$ and $Q$ are defined
\bdm
r = \Frac{\D{j+\half}{u}}{\D{j-\half}{u}}\;,
\edm
and
\bdm
Q = \Frac{g_j}{\D{j-\half}{u}}\;.
\edm

\subsection{Extension of FCT to Systems of Equations}\label{sect: syst}
The extension of the previously described methods to systems of 
hyperbolic conservation laws is no simple matter.  We will consider the 
Euler equations where the vectors 
\bdm
{\bf U} = \brac{\begin{array}{c}u^1\\ u^2\\ u^3 \end{array}} = 
\brac{\begin{array}{c}\rho\\ m\\ E \end{array}}\;,
\edm
and
\bdm 
{\bf F}\fn{\bf U} = \brac{\begin{array}{c}f^1\\ f^2\\ f^3 \end{array}} = 
\brac{\begin{array}{c}m \\ m^2/\rho + p\\ m\fn{E+p}/\rho \end{array}}\;,
\edm
are defined for Eq.\ (\ref{eq: hscl}).  Here $m=\rho u$ where $u$ is the 
fluid velocity.  The pressure, $p$, and density, $\rho$, are related to 
the energy by an equation of state (for an ideal gas), 
\bdm
p = \rho \varepsilon \fn{\gamma - 1}\;,
\edm
where $\varepsilon = E/\rho - 1/2 u^2$ and $\gamma$ is the ratio of 
specific heats for the gas in question.  This will serve as an example 
of the implementation, but the use of the techniques discussed here is 
not limited to this equation set or equation of state.  The FCT 
currently is extended to systems in the simplest fashion.  Traditional 
implementations of the FCT take the pressure terms in $F$ as source 
terms and are handled with central differences.  This leads to a poor 
representation of the wave interactions and the results that follow are 
often less than satisfactory.

The use of exact and approximate Riemann solvers offers a way through 
which more of the physical nature of the solution can be integrated into 
the solution procedure.  To the authors' knowledge no attempt has been 
made to incorporate Riemann solvers with any of the previous FCT 
algorithms.  Using van Leer's Riemann 
solver~\cite{vanleer79,gottlieb88}, with Godunov's  first-order 
method~\cite{godunov59,sod78} as the low-order method with the first 
modification of the FCT limiter, was our first attempt to incorporate a 
Riemann solver with FCT.  While the results are better than the standard 
FCT implementation, they are worse than Godunov's method alone.  A 
second approach is detailed below using Roe's approximate Riemann 
solver.

Roe's approximate Riemann solver and its descendents use a decomposition 
of the characteristic field for the system of conservation laws.  Taking 
the form of the system of hyperbolic conservation laws given in Eq.\ 
(\ref{eq: hscl char}), the flux Jacobian, $A$, is decomposed into right 
and left eigenvectors and eigenvalues or characteristics as
\be
A = R\Lambda R^{-1}\;,
\ee
where $R$ is a matrix where the columns are the eigenvectors, ${\bf 
r}^k$, of the eigenvalues, $a^k$, which are the diagonal entries of 
$\Lambda$.  The matrix $R^{-1}$ is the inverse of $R$ whose rows will be 
denoted as ${\bf l}^k$, the left eigenvectors, where the index $k$ 
refers to the $k^{th}$ wave in the system.  

In Roe's formulation it is required that the matrix $A$ is averaged from 
its neighboring states so that
\bdm
\fn{{\bf F}_{j+1} - {\bf F}_j} = A_{j+\half} \fn{{\bf U}_{j+1} - {\bf 
U}_j}\;.
\edm
This averaging for the system in question requires that a parameter be 
defined by
\be
D_{j+\half}=\fn{\rho_{j+1}/\rho_j}^{1/2}\;,
\ee
which is in turn used to define the following cell edge values:
\be
u_{j+\half}=\Frac{D_{j+\half}u_{j+1} + u_j}{D_{j+\half} + 1}\;,
\ee
\be
H_{j+\half}=\Frac{D_{j+\half}H_{j+1} + H_j}{D_{j+\half} + 1}\;,
\ee
and
\be
c_{j+\half}=\brac{\fn{\gamma -1}\fn{H_{j+\half} - \Half 
u_{j+\half}^2}}^{1/2}\;,
\ee
where
\be
H=\Frac{\gamma p}{\fn{\gamma-1}\rho}+\Half u^2\;.
\ee
For the Euler equations, the eigenvalues of the flux Jacobian are
\be
\fn{a^1,a^2,a^3}=\fn{u,u+c,u-c}\;.
\ee
The right eigenvectors form a matrix 
\be
R = \fn{{\bf r}^1,{\bf r}^2,{\bf r}^3}= 
    \brac{\begin{array}{ccc} 1         & 1      & 1\\ 
                             u         & u + c  & u - c\\ 
                             \half u^2 & H + uc & H - uc \end{array}}\;,
\ee
and by using
\bdm
z_1 = \Half \fn{\gamma -1} \Frac{u^2}{c^2}\;,
\edm
\bdm
z_2 = \Frac{\gamma -1}{c^2}\;,
\edm
the left eigenvectors form a matrix
\be
R^{-1} = \brac{\begin{array}{c}{\bf l}^1\\ {\bf l}^2\\ {\bf l}^3 
\end{array}} = 
\brac{\begin{array}{ccc}  1-z_1&z_2 u & -z_2\\
\Half\fn{z_1-\Frac{u}{c}} & -\Half\fn{z_2 u-\Frac{1}{c}} & \Half z_2\\ 
\Half\fn{z_1+\Frac{u}{c}} & -\Half\fn{z_2 u+\Frac{1}{c}} & \Half z_2 
\end{array}}\;.
\ee
In the results presented in the next section, Roe's~\cite{roe81} 
averaging procedure was used.

The implementation of these Riemann solvers relies on the following 
transformations:
\be
\Delta_{j+\half}u^j = \sum_k r_{j+\half}^k \alpha_{j+\half}^k\;,
\ee
where
\be
\alpha_{j+\half}^k = \sum_j l_{j+\half}^k \Delta_{j+\half}u^j\;.
\ee
The numerical dissipation terms are then written as
\be
\Phi^{DC}_{j+\half} = \sum_k\Half r_{j+\half}^k \psi\fn{a_{j+\half}^k} 
\alpha_{j+\half}^k\;,
\ee
\be
\Phi^{FCT}_{j+\half} = \sum_k r_{j+\half}^k \fn{f_{j+\half}^{C\;k} + 
\Phi^{DC}_{j+\half}}\;,
\ee
and
\be
\Phi^{MF}_{j+\half} = \sum_k \Half r_{j+\half}^k \brac{g_j^k + g_{j+1}^k 
- \psi\fn{a_{j+\half}^k + \gamma_{j+\half}^k}\alpha_{j+\half}^k}\;,
\ee
where
\be
g_j^k = \mbox{minmod}\fn{\sigma_{j-\half}^k\alpha_{j-\half}^k, 
\sigma_{j+\half}^k\alpha_{j+\half}^k}\;,
\label{eq: def g2}
\ee
and
\be
\gamma_{j+\half}^k = 
\select{\Frac{\Delta_{j+\half}g^k}{\alpha_{j+\half}^k} & \mbox{if 
$\alpha_{j+\half} \neq 0$}\\ 0 & \mbox{otherwise}}\;.
\ee

Given these expressions for the numerical dissipation, the flux limiters 
used in the modified FCT (and for that matter classical FCT) Eqs. 
(\ref{eq: old limiter}),(\ref{eq: new fct limiter}), and (\ref{eq: fct 
tvd limiter}) are rewritten to take advantage of these forms.  When a 
monotone first step is required with the FCT, Roe's first-order 
method~\cite{roe81} plus the entropy correction is used for the low-
order method.  The antidiffusive fluxes for the $k^{th}$ wave are 
rewritten as
\be
\hat{f}_{j+\half}^{AD} = \Half 
\psi\fn{a_{j+\half}^k}\alpha^k_{j+\half}\;,
\ee
or a Lax-Wendroff flux
\be
\hat{f}_{j+\half}^{AD} = \Half \brac{\psi\fn{a_{j+\half}^k} - \lambda
\fn{a_{j+\half}^{k}}^2}\alpha^k_{j+\half}\;,
\ee
or a fourth-order central difference
\be
\hat{f}_{j+\half}^{AD} = \Half\psi\fn{a_{j+\half}^k}\alpha^k_{j+\half} + 
\Frac{1}{12} \fn{a_{j-\half}^k\alpha^k_{j-\half} -
a_{j+\frac{3}{2}}^k\alpha^k_{j+\frac{3}{2}}}\;.
\ee
For the classic FCT method, the flux limiter becomes
\be
\hat{f}_{j+\half}^{C}=S_{j+\half} \max\brac{0,\min\fn{
S_{j+\half}\lambda^{-1}\alpha_{j-\half}^k, 
\abs{\hat{f}_{j+\half}^{AD}},S_{j+\half}\lambda^{-
1}\alpha_{j+\frac{3}{2}}^k}}\;.
\ee
The new FCT limiter becomes
\be
\hat{f}_{j+\half}^{C\;k}=S_{j+\half} \max\brac{0,\min\fn{
S_{j+\half}\tilde{\sigma}_{j-\half}^k\tilde{\alpha}_{j-
\half}^k,\abs{\hat{f}_{j+\half}^{AD}},
S_{j+\half}\tilde{\sigma}_{j+\frac{3}{2}}^k\tilde{\alpha}_{j+\frac{3}{2}
}^k}}\;,
\label{eq: new fct syst}
\ee
where
\bdm
\tilde{\sigma}_{j+\half}^k = \psi\fn{\tilde{a}_{j+\half}^k}
\edm
or
\bdm
\tilde{\sigma}_{j+\half}^k = \psi\fn{\tilde{a}_{j+\half}^k} - \lambda 
\fn{a_{j+\half}^{k}}^2\;.
\edm
The modified-flux FCT method becomes
\bea
g_j = \mbox{minmod}\fn{\hat{f}_{j\pm\half}^{AD},\D{j\pm\half}{\alpha};n} 
= S_{j+\half}  & \max & \left[\rule{0ex}{2.5ex}\right. 0, \min \fn{\Half 
n\abs{\hat{f}_{j+\half}^{AD}}, n
S_{j+\half} \sigma_{j-\half}\alpha_{j-\half}^k}, \nonumber \\ & \min & 
\fn{n\sigma_{j+\half}\abs{\alpha_{j+\half}^k},\Half n S_{j+\half}
\hat{f}_{j-\half}^{AD}}\left. \rule{0ex}{2.5ex}\right ] \;,
\eea
where
\bdm
\sigma_{j+\half}^k = \Half{\psi\fn{a_{j+\half}^k}}
\edm
or
\bdm
\sigma_{j+\half}^k = \Half\brac{{\psi\fn{a_{j+\half}^k} - \lambda 
\fn{a_{j+\half}^{k}}^2}}\;.
\edm

In the next section, we will discuss the quality of solutions using 
these methods.

\section{Results}\label{sect: results}
To gauge the capability of the methods discussed in the previous 
sections, three test problems were solved with the FCT methods and 
several other high-resolution finite-difference methods.  The other 
methods used will not be described in detail here.   The first test 
problem will be to solve a scalar advection equation, Eq. (\ref{eq: 
scalar}), on a uniform grid.  Two problems will be considered: a square 
wave and a sine wave over a complete period.  Both waves have an 
amplitude of one.  The second problem will be the inviscid Burger's 
equation,
\bdm
\dxdt{u} + \ddx{\Half u^2} = 0\;,
\edm
with initial data of a sine wave on a periodic domain with an amplitude 
of one.  This solution will be compared with the exact solution and the 
corresponding error norms will be used to show convergence and order of 
approximation in these norms for the various methods.  Finally, the 
shock tube problem used by Sod~\cite{sod78} will be used as a vehicle 
for comparison of these methods for their use with systems of hyperbolic 
conservation laws.

\subsection{Scalar Advection Equation}
For the scalar advection of a square wave with a uniform velocity, the 
FCT performs quite well with very little numerical diffusion present in 
the solution.  These solutions are obtained for a CFL number held 
constant at $1/2$ after 80 time steps.  

\begin{figure}
\caption{
Solution of the scalar advection equation with Zalesak's FCT with 
the high-order flux defined by second-order central differencing.
}
\includegraphics[width=0.5\textwidth]{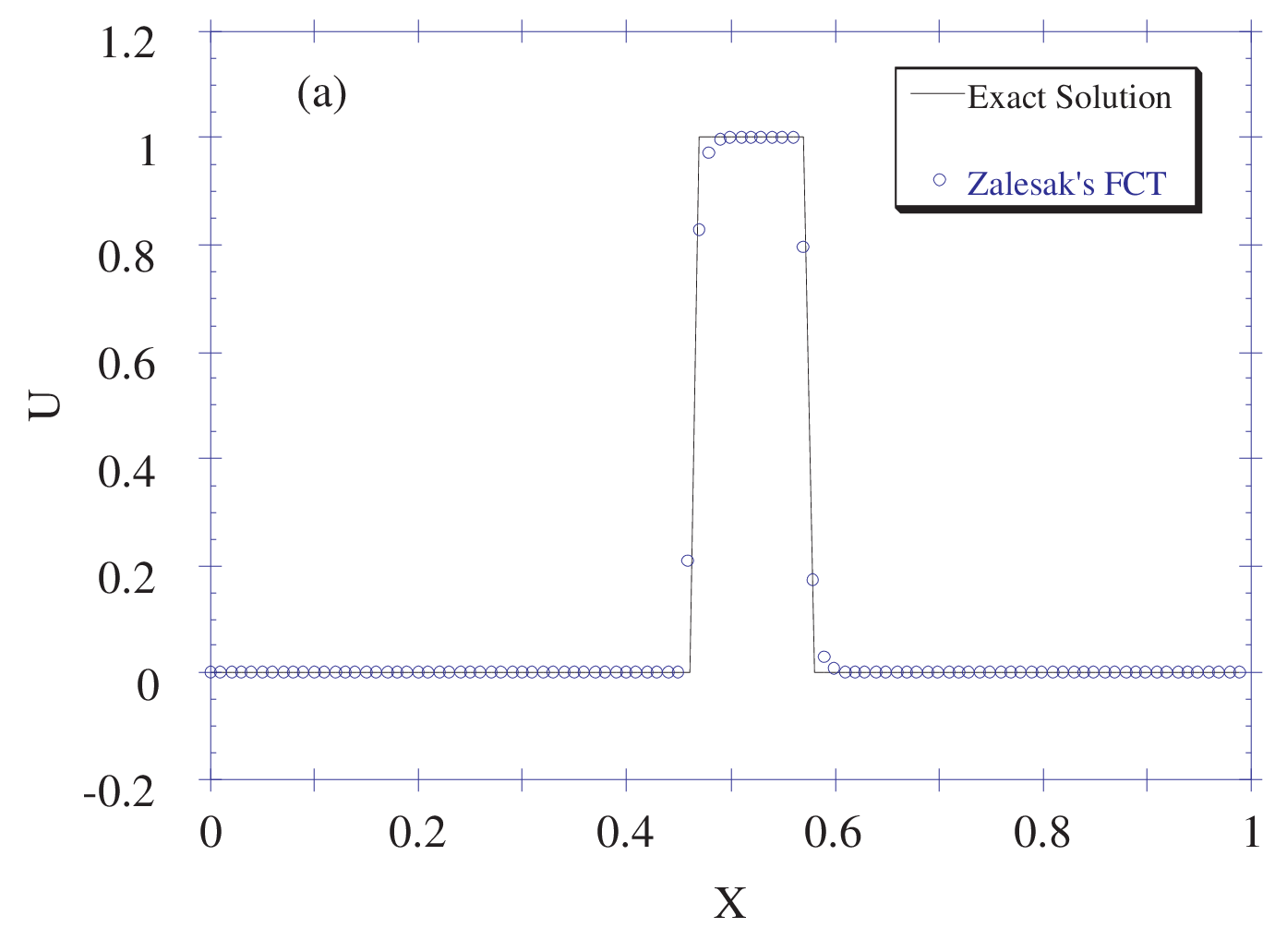}
\includegraphics[width=0.5\textwidth]{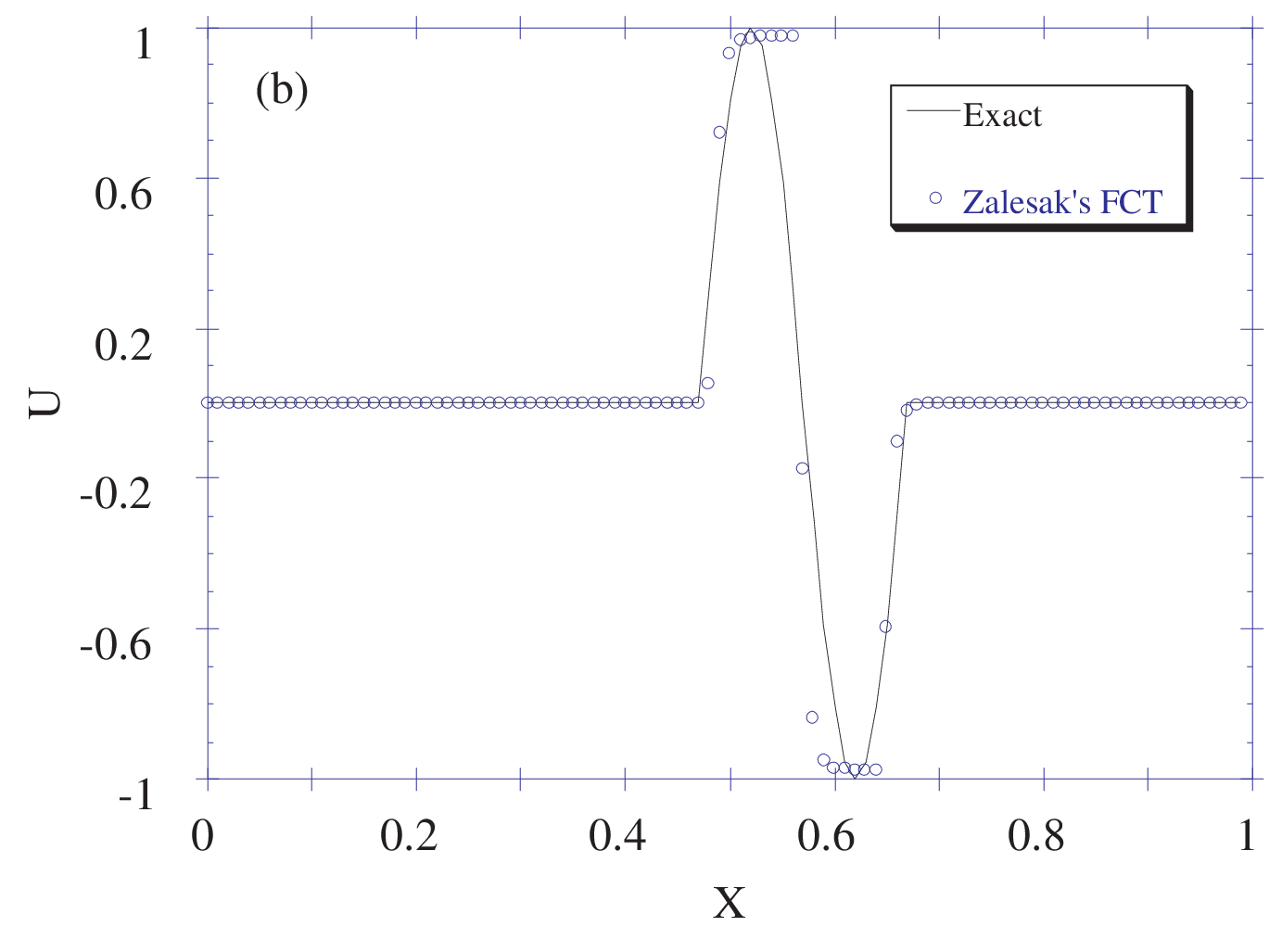}
\label{fig: zalesak fct scalar}
\end{figure}

\begin{figure}
\caption{
Solution of the scalar advection equation with the new FCT with 
the high-order flux defined by second-order central differencing.
}
\includegraphics[width=0.5\textwidth]{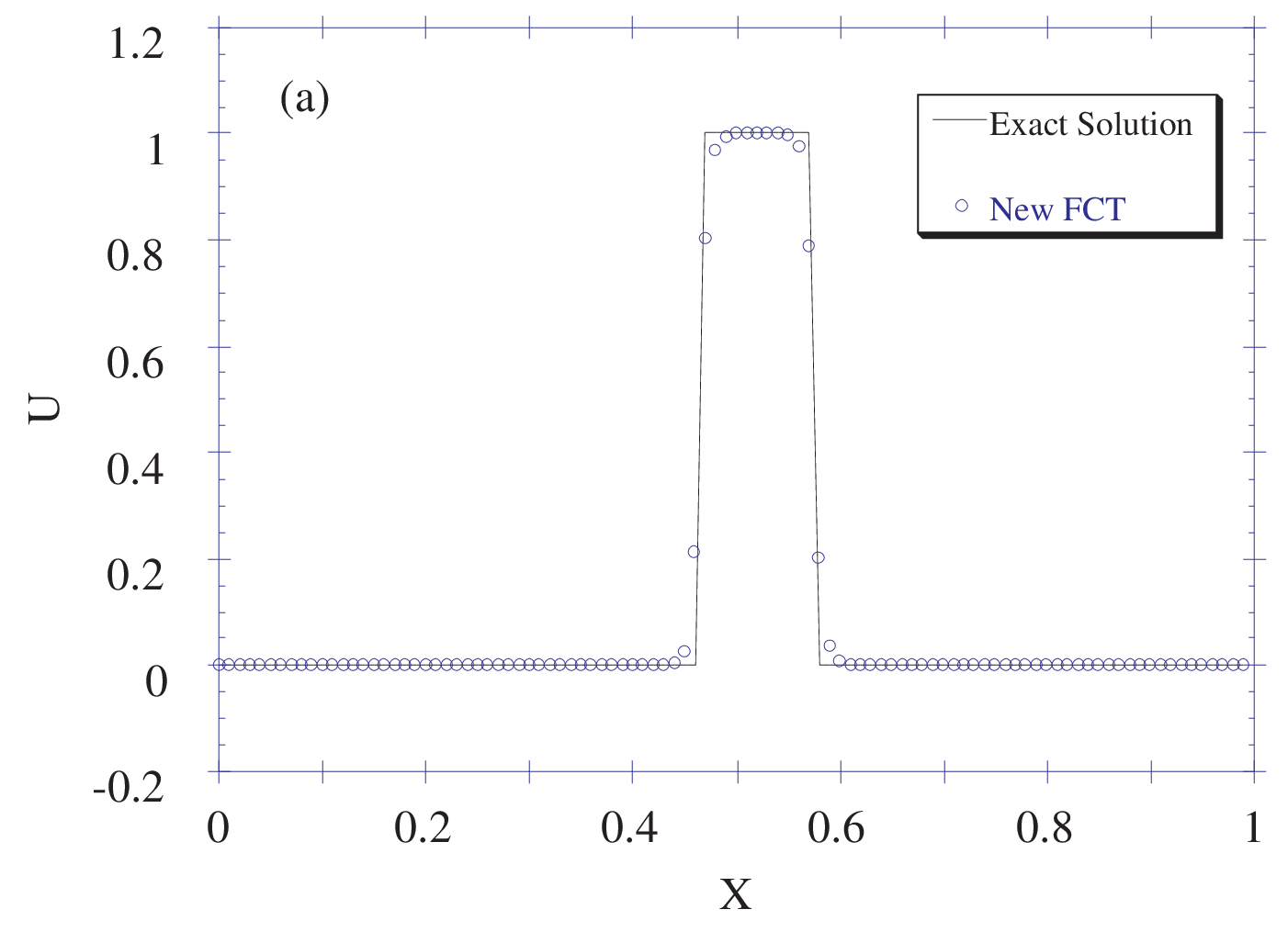}
\includegraphics[width=0.5\textwidth]{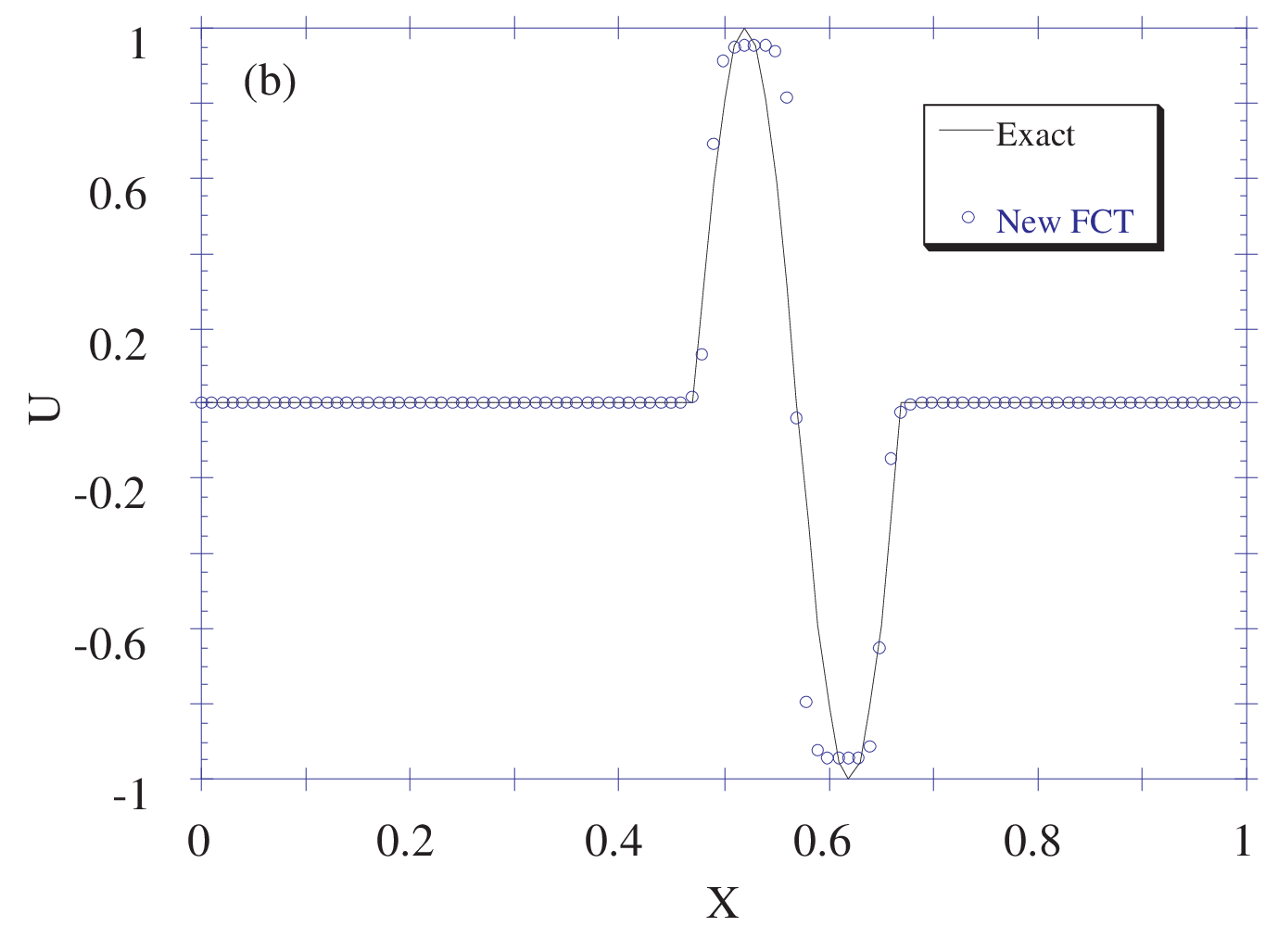}
\label{fig: new fct scalar}
\end{figure}

As shown in Fig.~\ref{fig: zalesak fct scalar}~(a), the square wave is 
captured quite well by the difference scheme, however, there is a 
distinct lack of symmetry in the solution.  This lack of symmetry is 
evident in this version of the FCT despite the choice of the CFL number 
(which should lead to symmetric results, ideally because of the phase 
error properties of upwind methods at a CFL number of 
$\half$~\cite{hirsch88}).  The lack of symmetry can be attributed to the 
support of the limiter, which can result in anti-upwind data being used 
in the flux definition~\cite{rider91d}.  This is more evident in 
Fig.~\ref{fig: zalesak fct scalar}~(b), but also evident is the 
overcompressive nature of the scheme.  The sine wave is in the process 
of being compressed into two square waves.  This behavior is clearly 
unacceptable because the character of the waves is largely destroyed by 
this algorithm.  Figure~\ref{fig: new fct scalar} shows that the new FCT 
algorithm is somewhat more diffusive (less compressive) and has the more 
of the expected symmetry in the solution.  Figure~\ref{fig: new fct 
scalar}~(b) still shows that this algorithm remains too compressive 
despite being TVD.  One negative aspect of this calculation is the 
clipping of the extrema with respect to the previous figure, although 
overall this solution is superior in most respects to Zalesak's FCT.

\begin{figure}
Solution of the scalar advection equation with Zalesak's FCT with 
the high-order flux defined by Lax-Wendroff differencing.
\caption{
}
\includegraphics[width=0.5\textwidth]{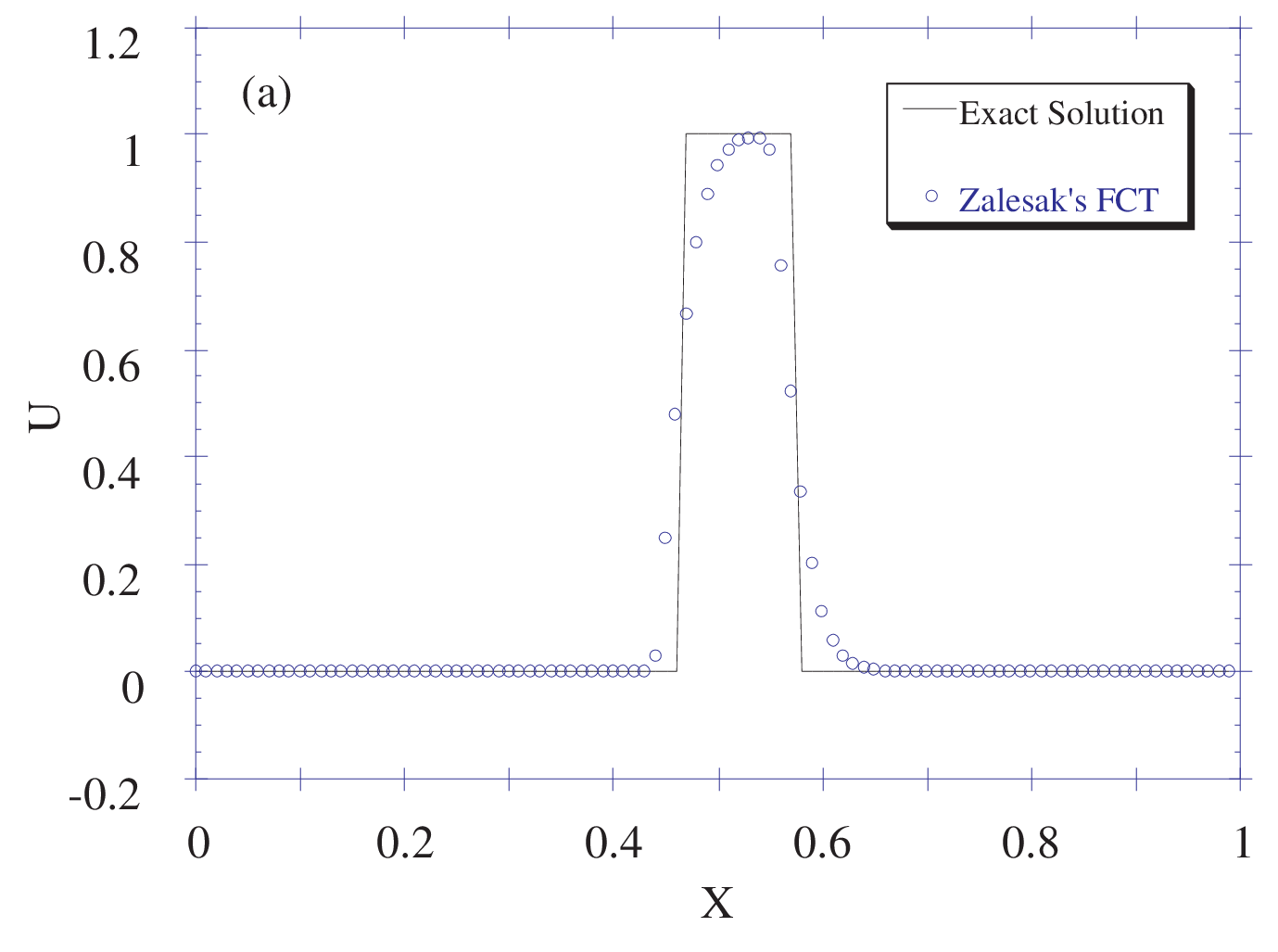}
\includegraphics[width=0.5\textwidth]{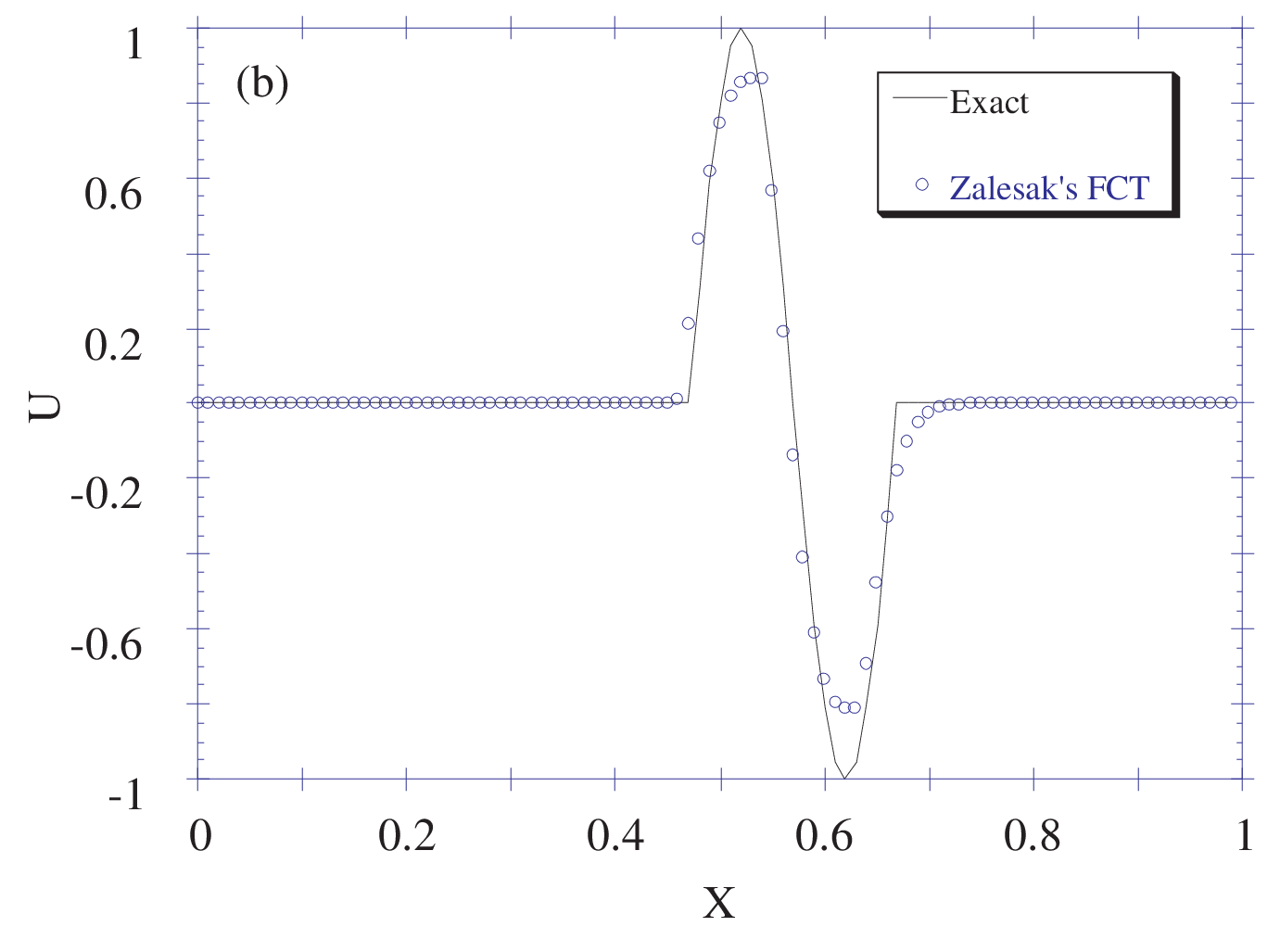}
\label{fig: zalesak fct scalar lw}
\end{figure}

\begin{figure}
\caption{
Solution of the scalar advection equation with the new FCT with 
the high-order flux defined by Lax-Wendroff differencing.
}
\includegraphics[width=0.5\textwidth]{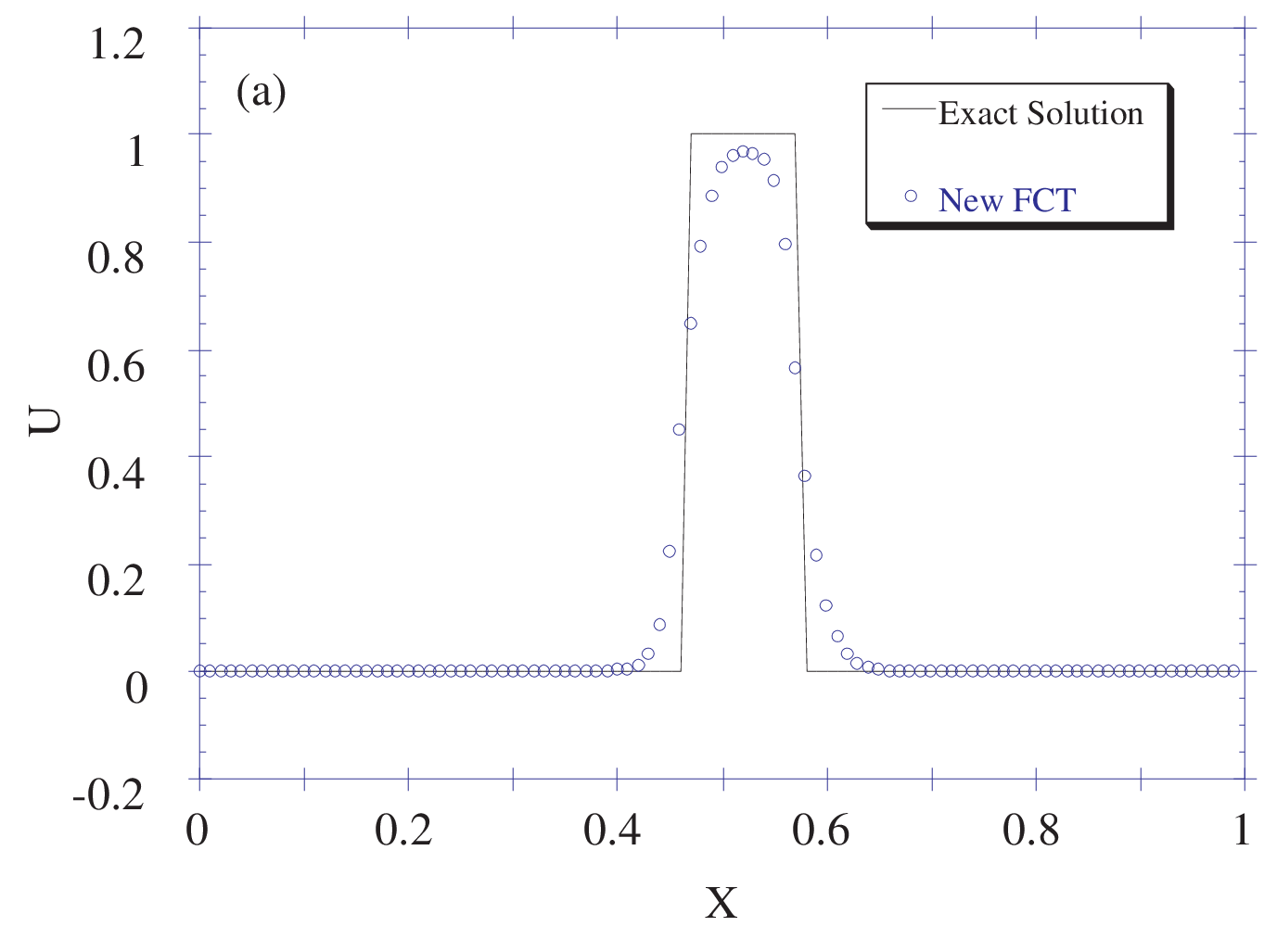}
\includegraphics[width=0.5\textwidth]{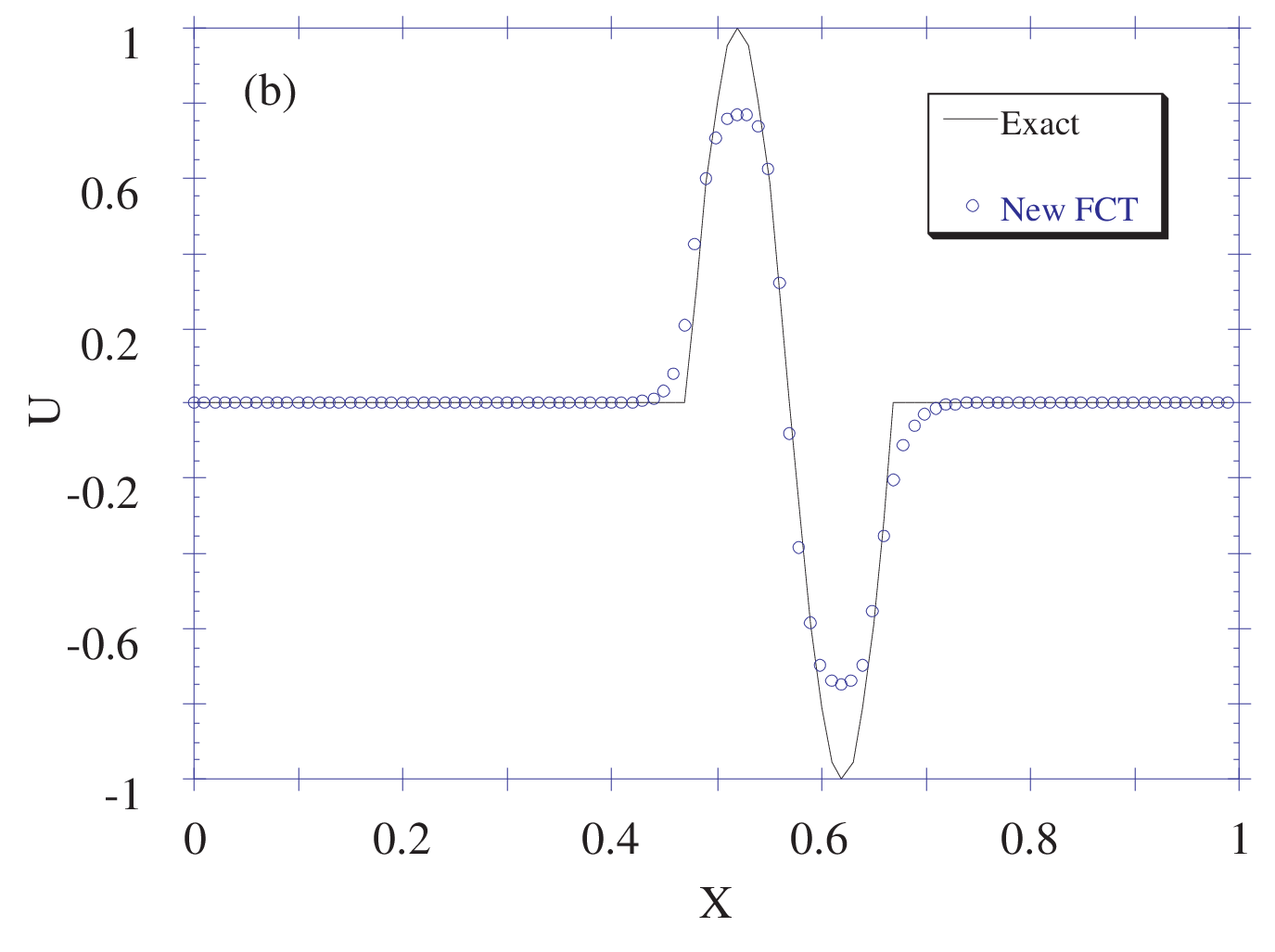}
\label{fig: new fct scalar lw}
\end{figure}

By using the Lax-Wendroff fluxes as the base for the antidiffusive 
fluxes, the problem of overcompression is eliminated from both 
algorithms.  This is at the cost of some clipping of the solution's 
extrema.  The clipping in Fig.~\ref{fig: zalesak fct scalar lw} is less 
than that in Fig.~\ref{fig: new fct scalar lw}, but at the cost of the 
symmetry of the solution.  The lack of symmetry is also present in these 
results.     

\begin{figure}
\caption{
Solution of the scalar advection equation with the modified-flux 
FCT ($n=1$ limiter).
}
\includegraphics[width=0.5\textwidth]{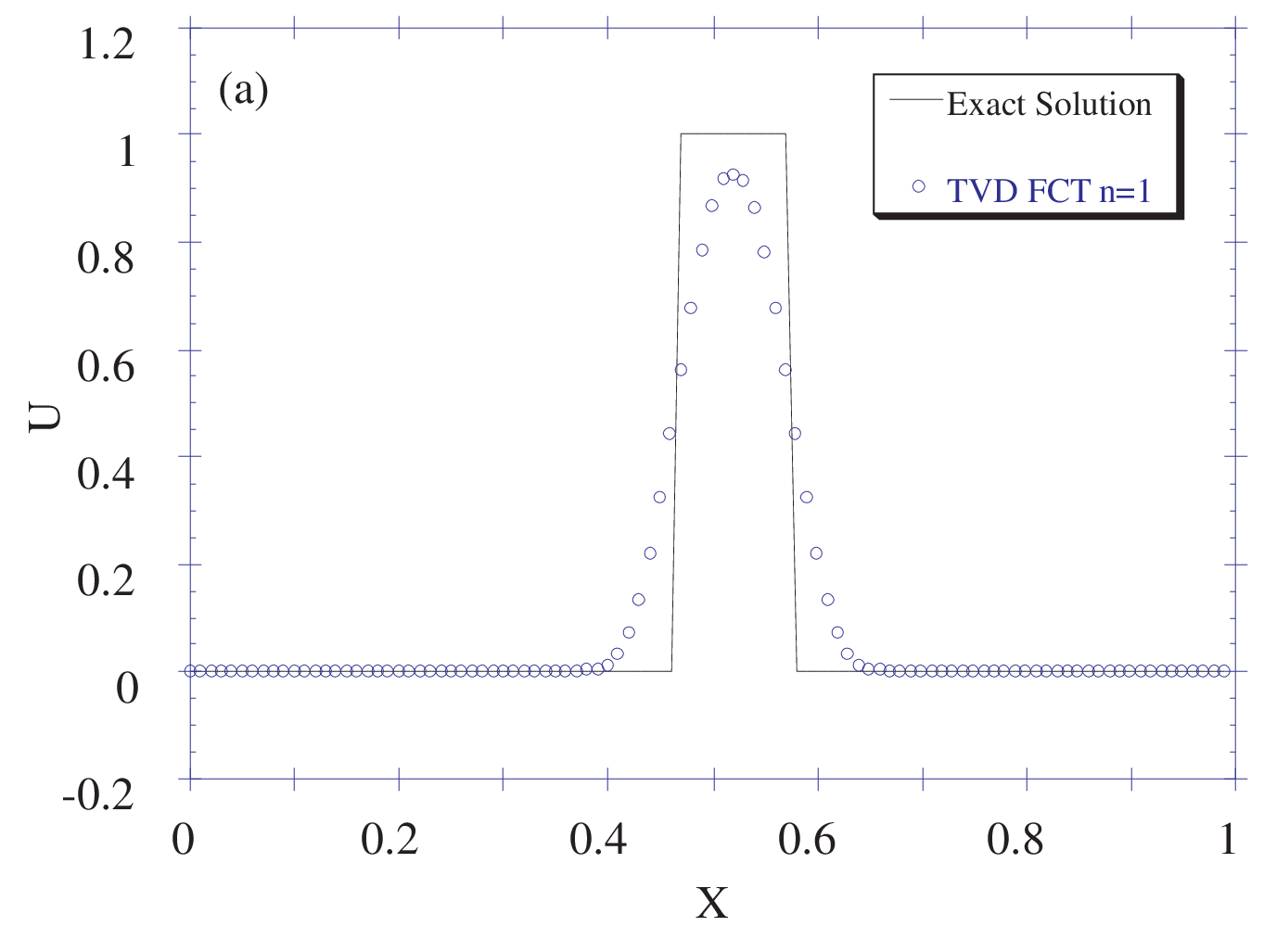}
\includegraphics[width=0.5\textwidth]{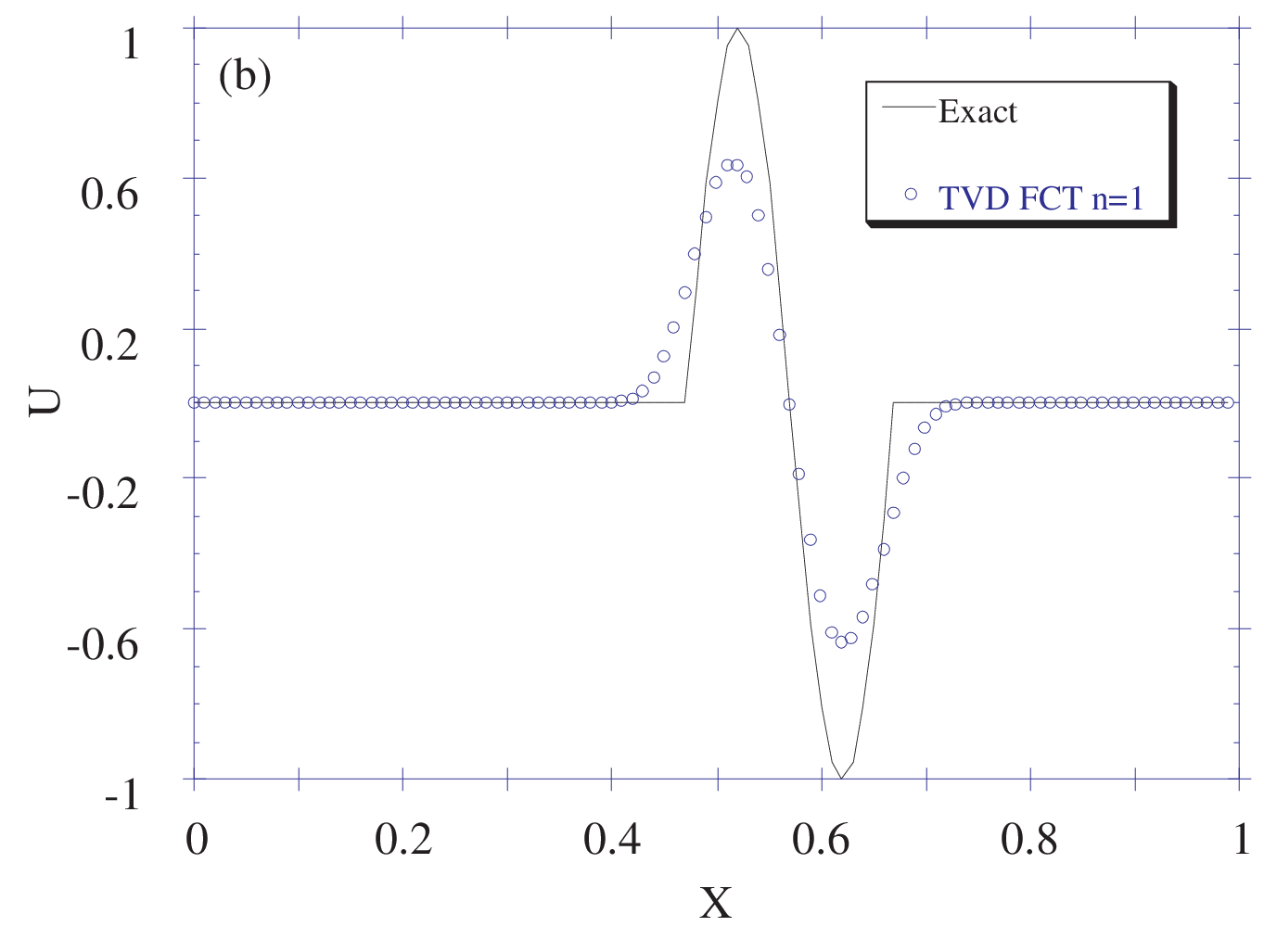}
\label{fig: mod flux fct scalar n=1}
\end{figure}

Figures~\ref{fig: mod flux fct scalar n=1} and~\ref{fig: mod flux fct 
scalar n=2} show the impact of the choice of $n$ in the modified-flux 
FCT formulation (and for that matter other implementations of limiters).  
Here the Lax-Wendroff based high-order fluxes are used.  The lower value 
of $n$ results in solutions that exhibit a great deal of dissipation and 
clipping of extrema.  For the $n=2$, solution is of high quality with 
the clipping of extrema quite controlled.  This solution nearly equals 
that of the other FCT formulations for the square wave.  For the sine 
wave, despite some clipping, the overcompression has disappeared with 
the character of the original profile well preserved.

\begin{figure}
\caption{
Solution of the scalar advection equation with a symmetric TVD 
scheme. 
}
\includegraphics[width=0.5\textwidth]{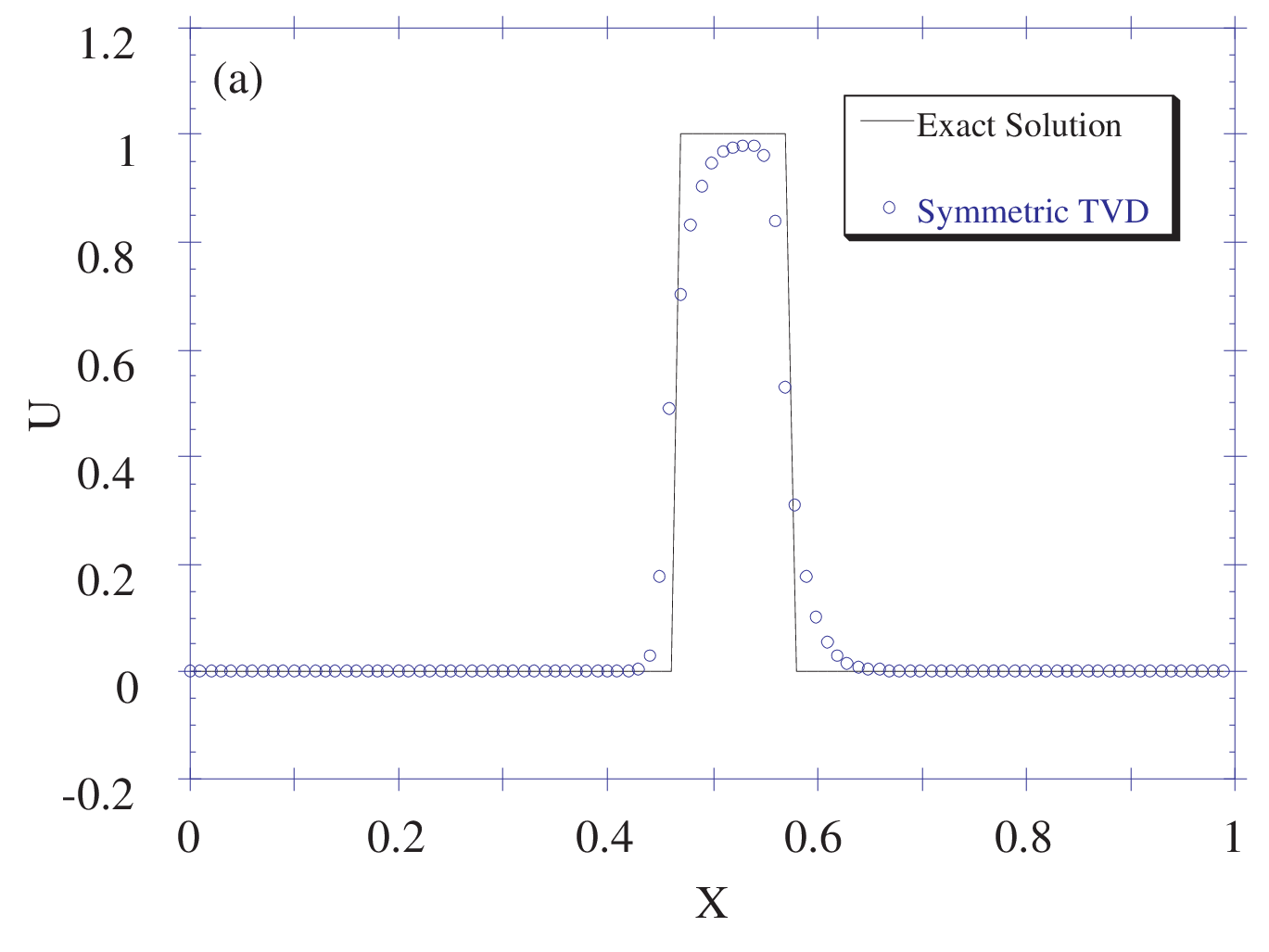}
\includegraphics[width=0.5\textwidth]{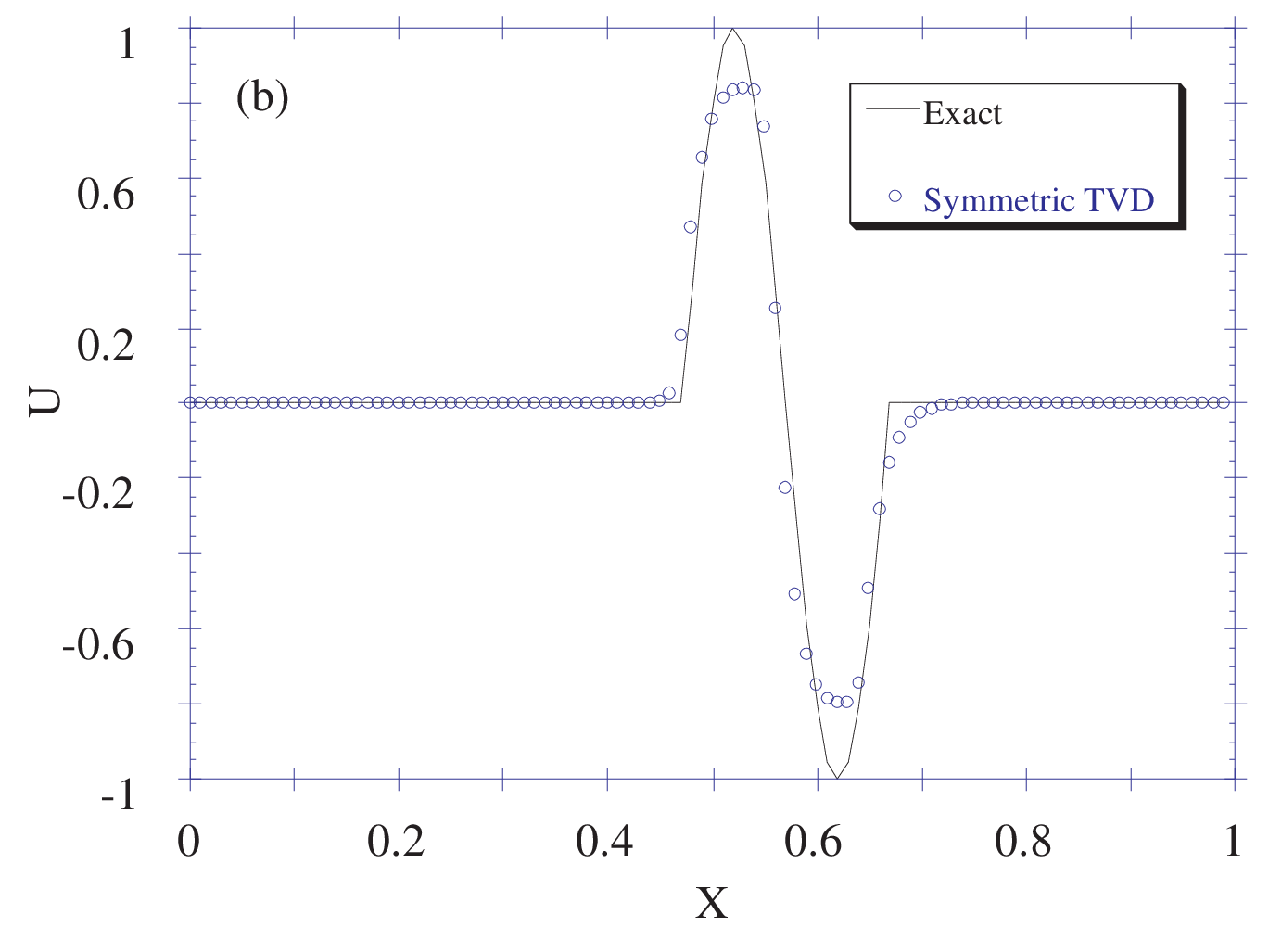}
\label{fig: symm tvd scalar}
\end{figure}

The symmetric TVD algorithm (second-order in both time and space) 
produces results similar to the new FCT, but with a lack of symmetry.  
This can be cured with a predictive first step as with the FCT.  As 
Fig.~\ref{fig: symm tvd scalar} shows, both exhibit a fair amount of 
extrema clipping and lack of symmetry.  These are similar to the results 
obtained in Fig.~\ref{fig: zalesak fct scalar lw} with Zalesak's FCT, 
but are more diffused.

\subsection{Burger's Equation}\label{sect: burg}
In all cases, the solutions obtained by using the high-resolution 
algorithms on Burger's equation are quite good in terms of quality.  
Little would be gained by simply viewing their profiles (they are 
similar to the results in~\cite{yee85} for a TVD algorithm).  By nature 
these high-resolution methods produce results that are first-order 
accurate in the $L_\infty$ norm and approach second-order accuracy in 
the $L_1$ norm.  In the next four figures discussed, figure (a) will be 
for time equal to 0.2 when the solution remains smooth, and (b) will 
show the error norms ($L_1$, $L_2$ and $L_\infty$) at time equal 1.0 
after a shock has formed.  For the methods used, each is second-order in 
time and space with the exception of the fourth-order FCT method, which 
is fourth-order in space.  Second-order temporal accuracy is obtained by 
using a Lax-Wendroff type formulation.  These calculations are all done 
with $\lambda$ held constant.  The norms shown in the figures are 
defined as follows:
\bdm
L_1 = \sum_j^N \Frac{\abs{e_j}}{N}\;,
\edm
\bdm
L_2 = \fn{\sum_j^N \Frac{e_j^2}{N}}^\half\;,
\edm
\bdm
L_\infty=\sup\fn{\abs{e_j}}\;,
\edm
where
\bdm
e_j = U_j^{exact}-U_j^{approx.}\;.
\edm

\begin{figure}
\caption{
Convergence of error norms for Burger's equation for Zalesak's FCT 
with the high-order flux defined by Lax-Wendroff differencing.
}
\includegraphics[width=0.5\textwidth]{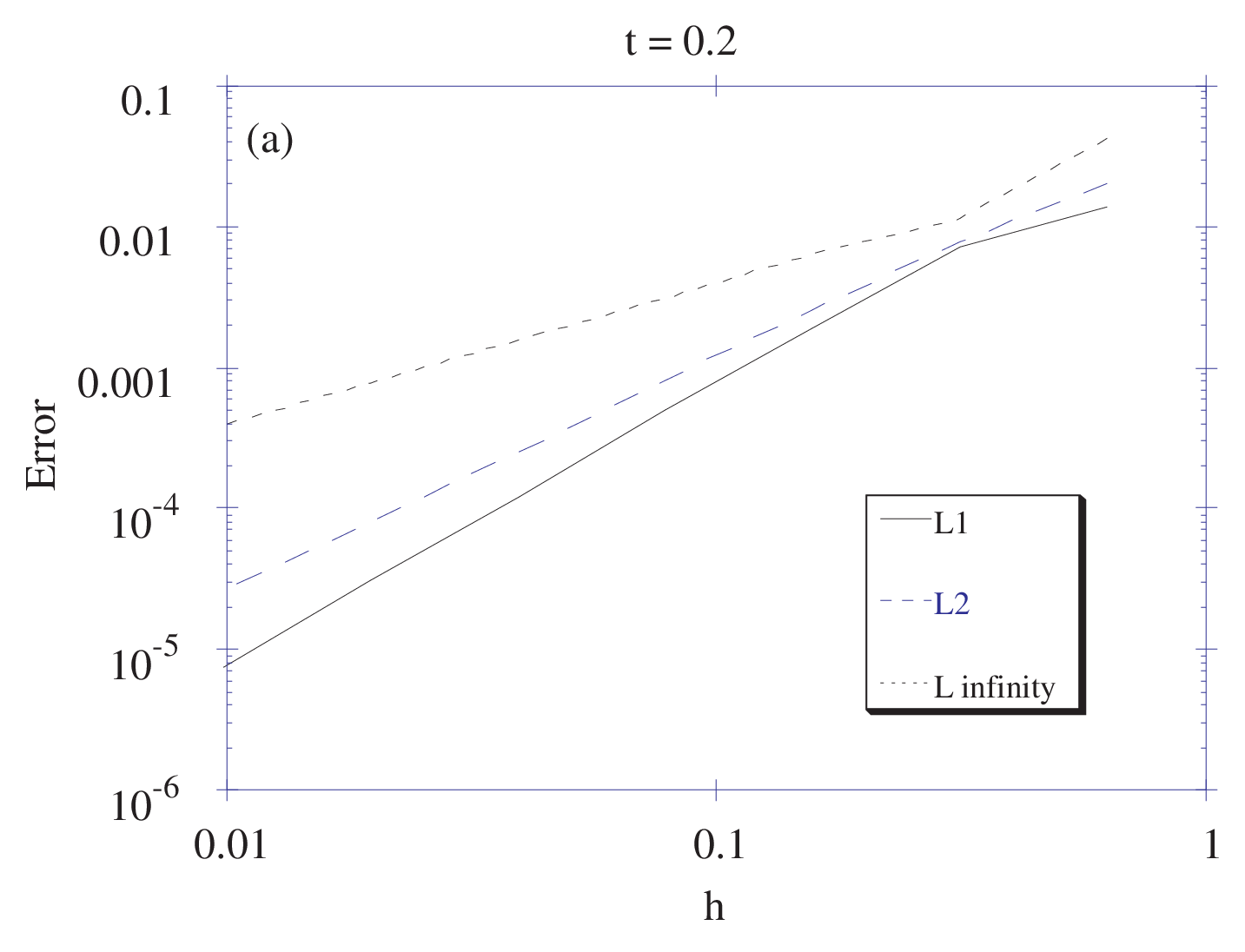}
\includegraphics[width=0.5\textwidth]{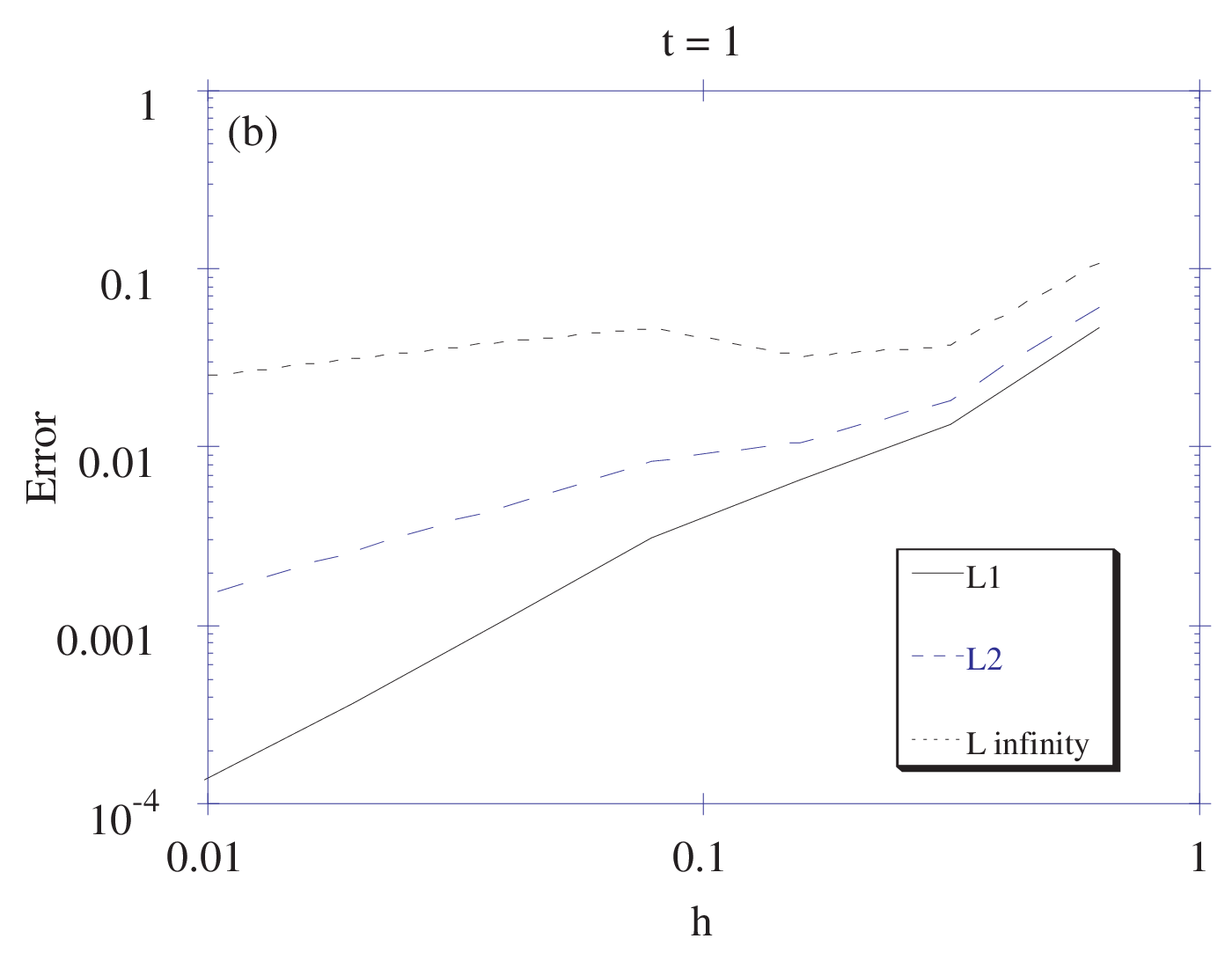}
\label{fig: old fct norms}
\end{figure}

\begin{figure}
\caption{
Convergence of error norms for Burger's equation for Zalesak's FCT 
with the high-order flux defined by fourth-order central differencing.
}
\includegraphics[width=0.5\textwidth]{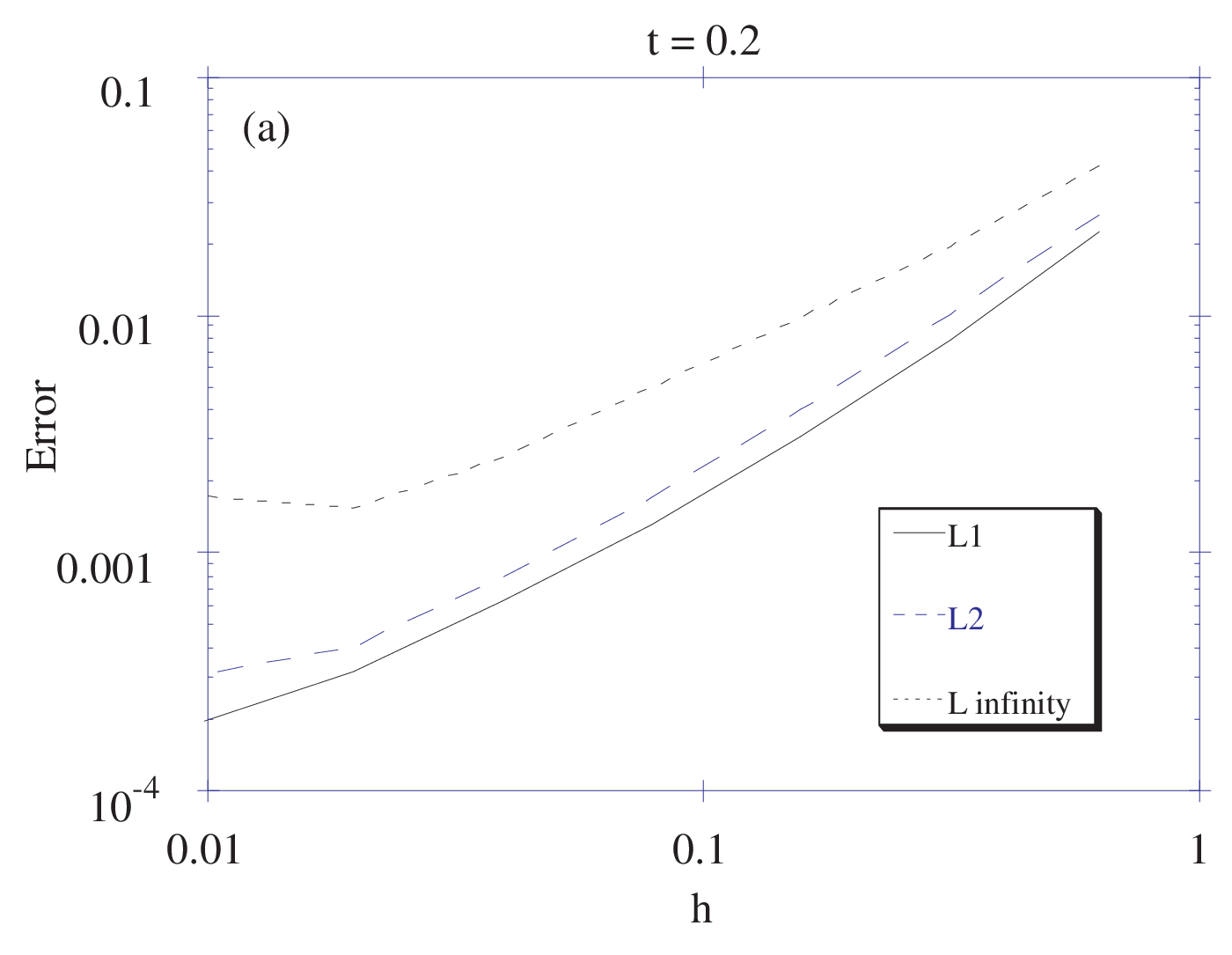}
\includegraphics[width=0.5\textwidth]{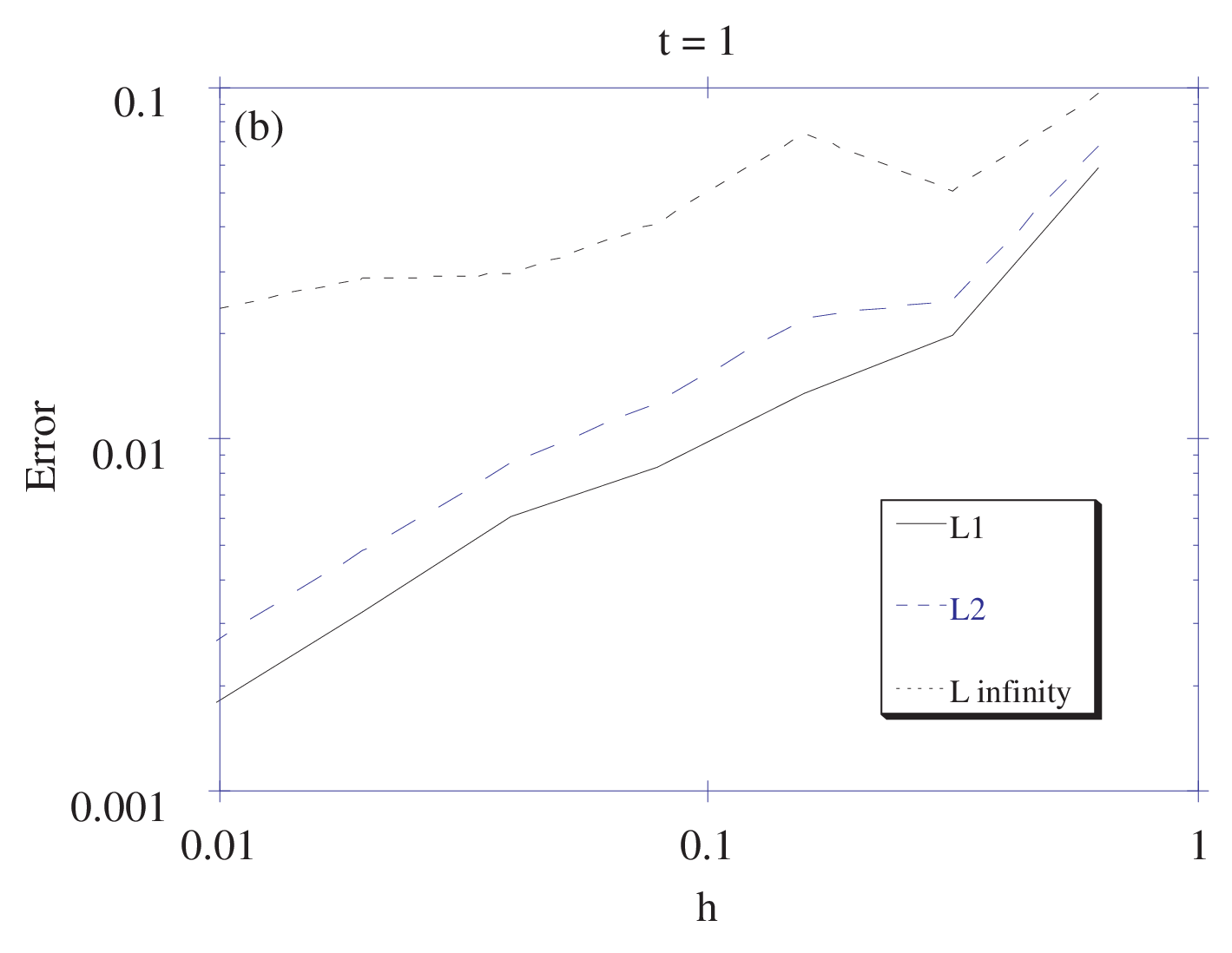}
\label{fig: old fct norms 4th}
\end{figure}

\begin{figure}
\caption{
Convergence of error norms for Burger's equation for the new FCT 
with the high-order flux defined by Lax-Wendroff differencing.
}
\includegraphics[width=0.5\textwidth]{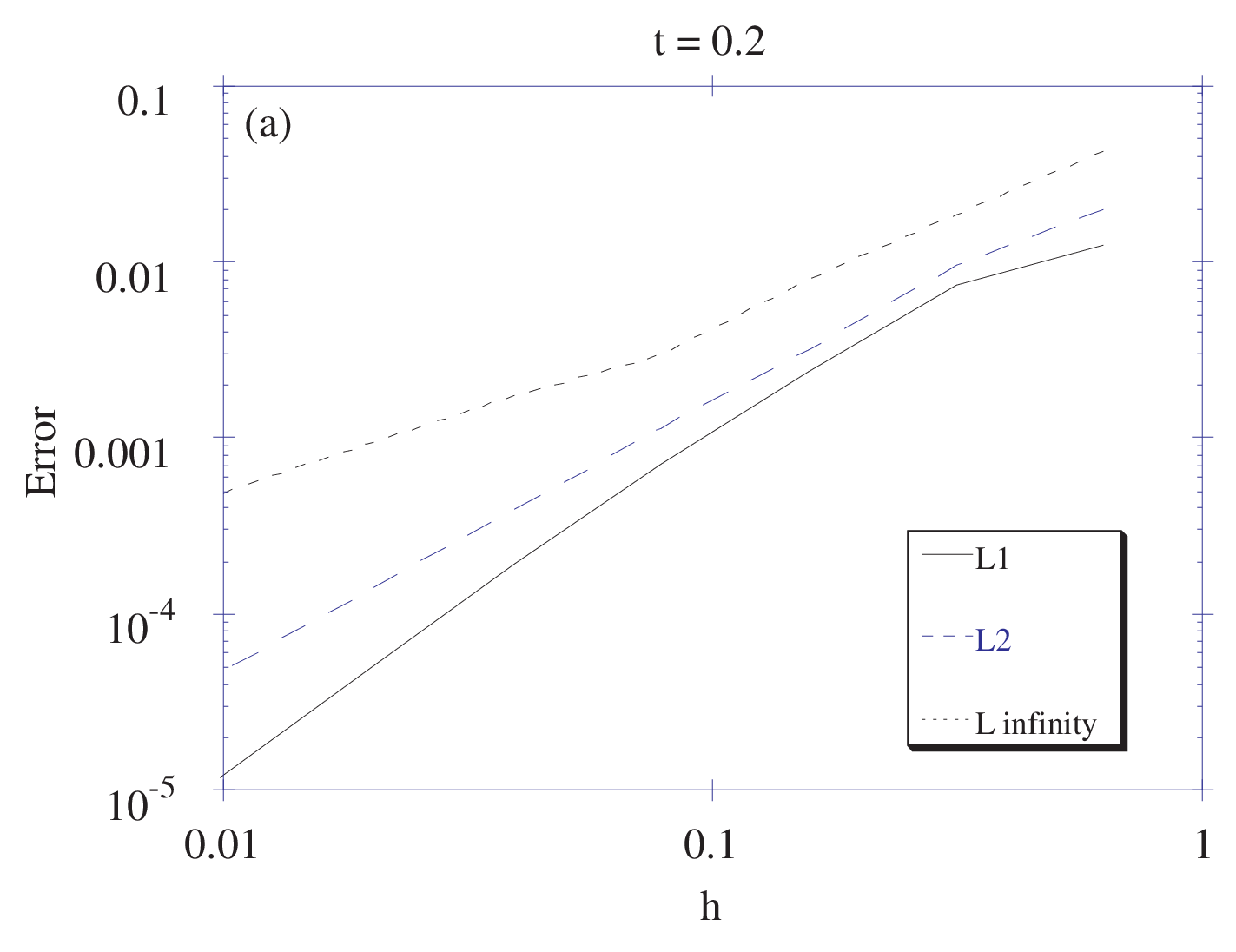}
\includegraphics[width=0.5\textwidth]{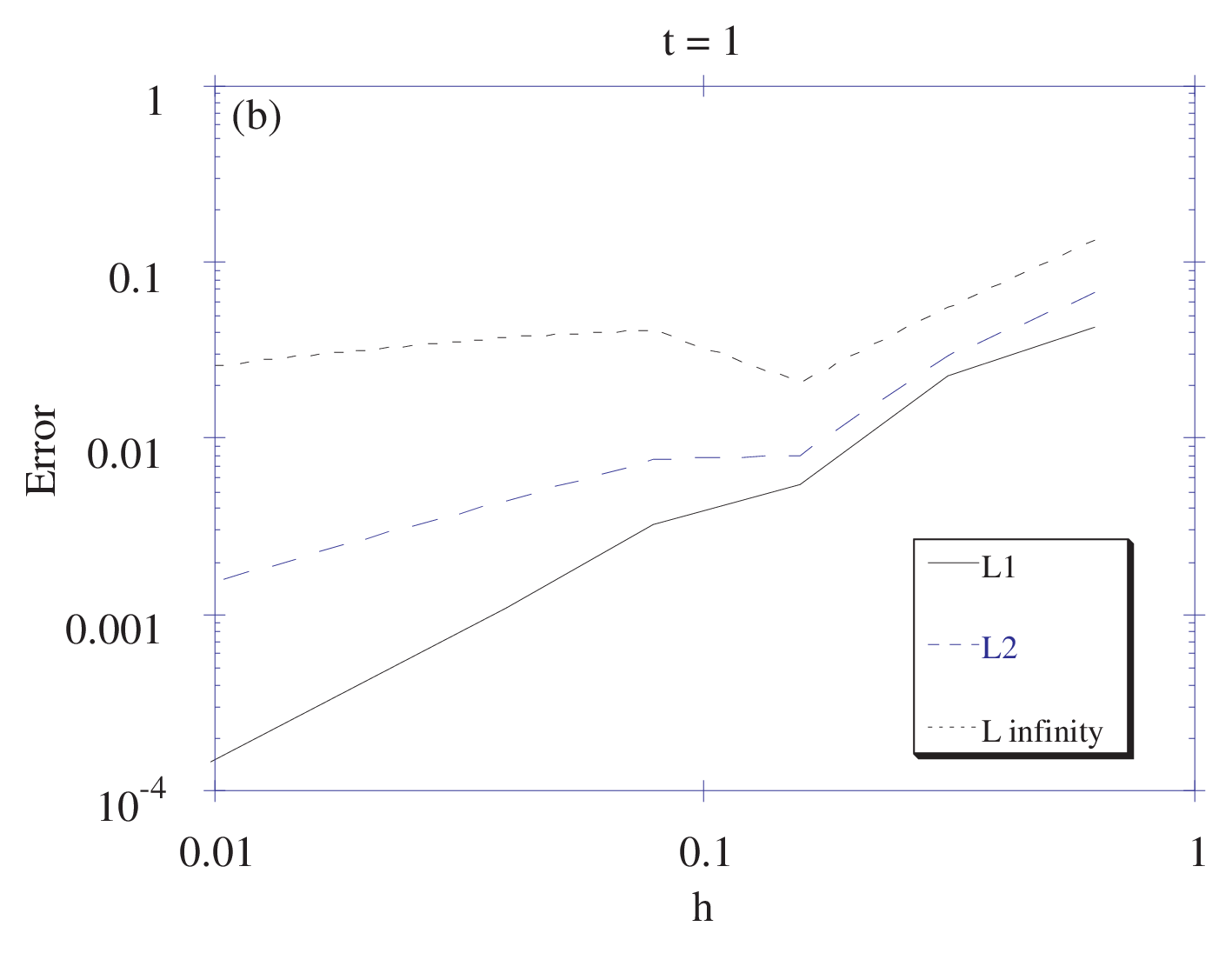}
\label{fig: new fct norms}
\end{figure}

\begin{figure}
\caption{
Convergence of error norms for Burger's equation for a symmetric 
TVD algorithm.
}
\includegraphics[width=0.5\textwidth]{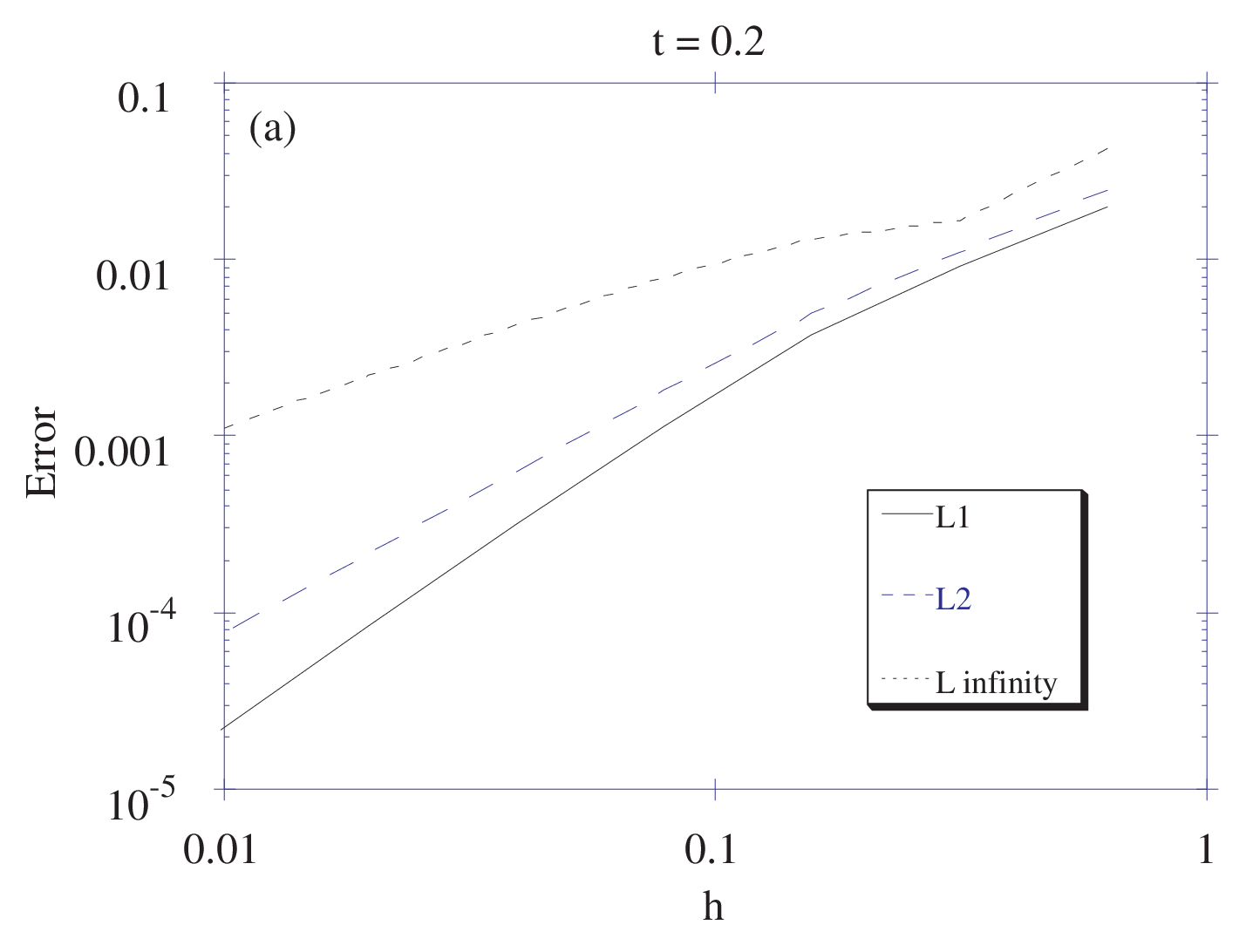}
\includegraphics[width=0.5\textwidth]{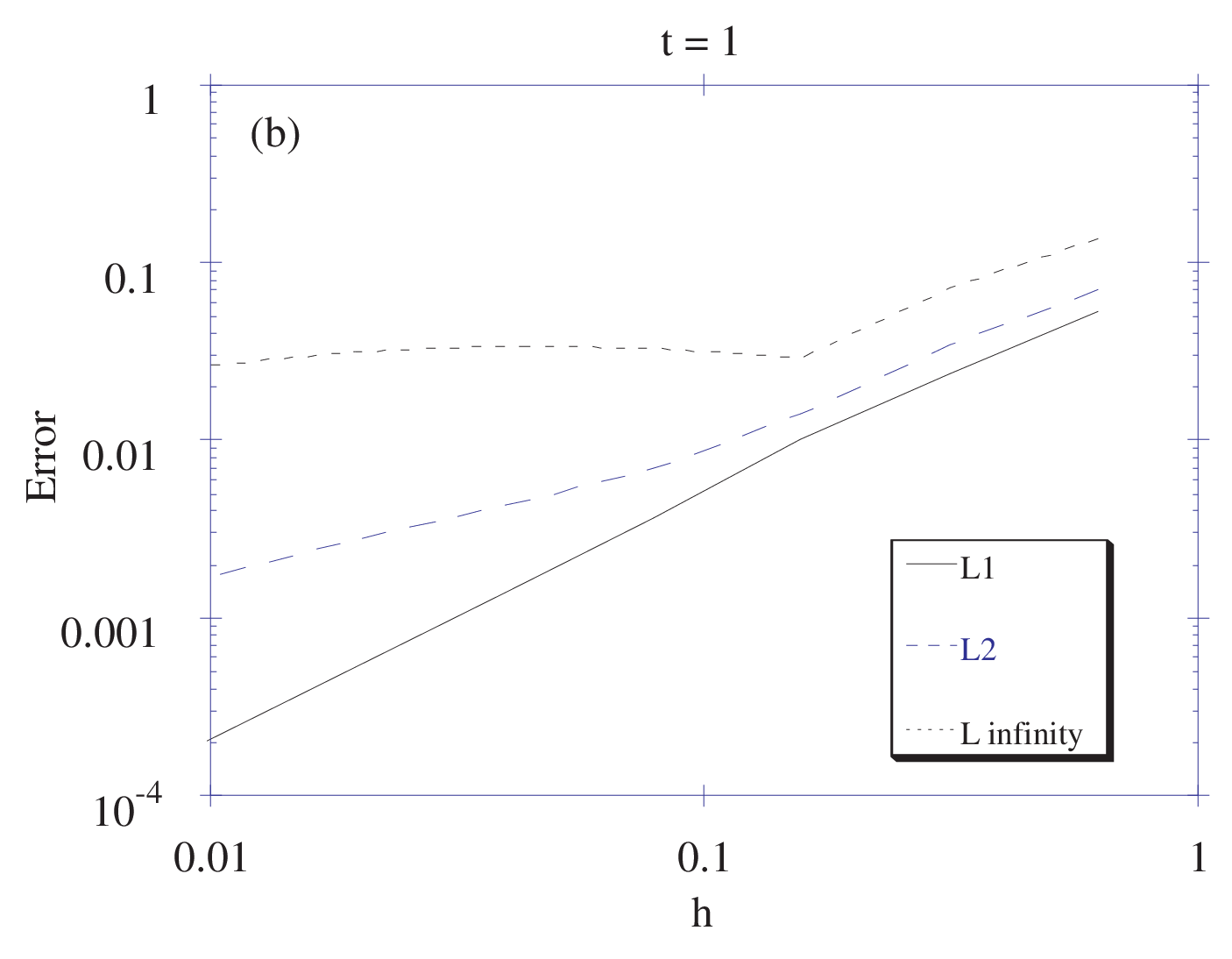}
\label{fig: sym tvd norms}
\end{figure}

In Fig.~\ref{fig: old fct norms} the solution for $t=0.2$ converges in 
the expected fashion, but at $t=1$ problems are present with the 
convergence in the $L_\infty$ norm.  As the grid is refined, the 
$L_\infty$ norm error increases rather than decreases as expected.  As 
the grid size is further decreased convergence resumes, but is quite 
slow (about order $1/4$).  Figure~\ref{fig: old fct norms 4th} shows 
that the convergence properties of the fourth-order antidiffusive flux 
do not converge at a fourth-order rate and are in fact worse than those 
shown in the previous figure.  The nonconvergence in the $L_\infty$ norm 
for intermediate grid sizes for the $t=1$ case is comparable.  The new 
FCT algorithm shows slight improvements over both of these cases, but 
still has the same difficulties after a shock has formed in the 
solution.  As shown by Fig.~\ref{fig: new fct norms}, the solutions 
converge faster than Zalesak's FCT, but are still plagued by some of the 
same problems.  This behavior is also shared by the symmetric TVD's 
results in Fig.~\ref{fig: sym tvd norms}.  The symmetric TVD does not 
converge as well as the new FCT method, but the nonconvergence problem 
is not as pronounced although it is clearly present.

The similarity of the solutions for the two FCT methods and the 
symmetric TVD algorithm, and the lack of such a problem in the 
modified-flux FCT (or TVD) method points to the form of the limiter as 
being the problem.  The FCT and symmetric TVD use cell-edged limiters 
rather than cell-centered limiters.  This difference requires that each 
limiter has a wider spatial stencil than the cell-centered limiter, and 
as a result the resulting algorithm is not as sensitive to the presence 
of a discontinuity.  This lack of sensitivity results in a poorer 
handling of shocks and discontinuities.  The FCT is less diffusive than 
the symmetric TVD method, and this lack of diffusion increases the 
problem.  The results for the fourth-order spatial limiter point out two 
problems: because the fourth-order spatial difference is more 
compressive than the second-order difference scheme, the convergence 
difficulty in the $L_\infty$ norm at a shock is increased slightly.  
Experiments with a second-order Runge-Kutta time integration scheme show 
improvements in the $L_1$ convergence of the FCT.

\subsection{Sod's Shock Tube Problem}
The third problem involves the solution of Sod's test problem which 
tests the mettle of each algorithm against a difficult physical problem.  
For the FCT methods [in the modified-flux $\sigma = 1/2\fn{\abs{a} - 
\lambda a^2}$], the Lax-Wendroff flux is used to define the 
antidiffusive flux.  All results were produced for $\Delta t = 0.4 
\Delta x$ and shown for $t=0.24$.  The initial conditions are the same 
as Sod uses, but are listed here for completeness, for $X<0.5$,
\bdm
\brac{\begin{array}{c}\rho^l\\u^l\\p^l\end{array}} = 
\brac{\begin{array}{c}1.0\\0.0\\1.0\end{array}}\;,
\edm
and for $X\geq 0.5$
\bdm
\brac{\begin{array}{c}\rho^r\\u^r\\p^r\end{array}} = 
\brac{\begin{array}{c}0.125\\0.0\\0.1\end{array}}\;,
\edm
with $\gamma = 1.4$.  

\begin{figure}
\caption{
Solution of Sod's shock tube problem with Zalesak's FCT.
}
\includegraphics[width=0.5\textwidth]{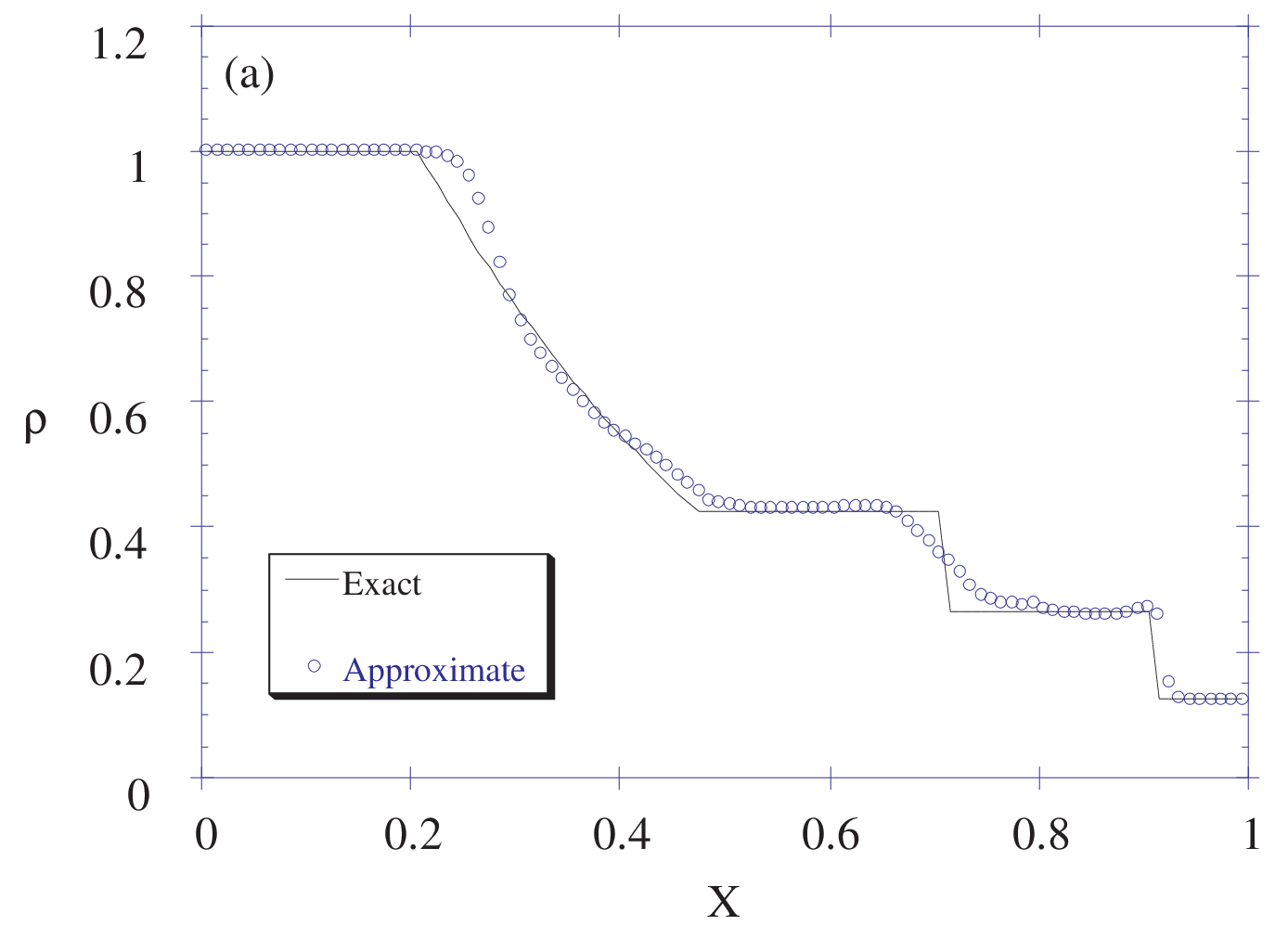}
\includegraphics[width=0.5\textwidth]{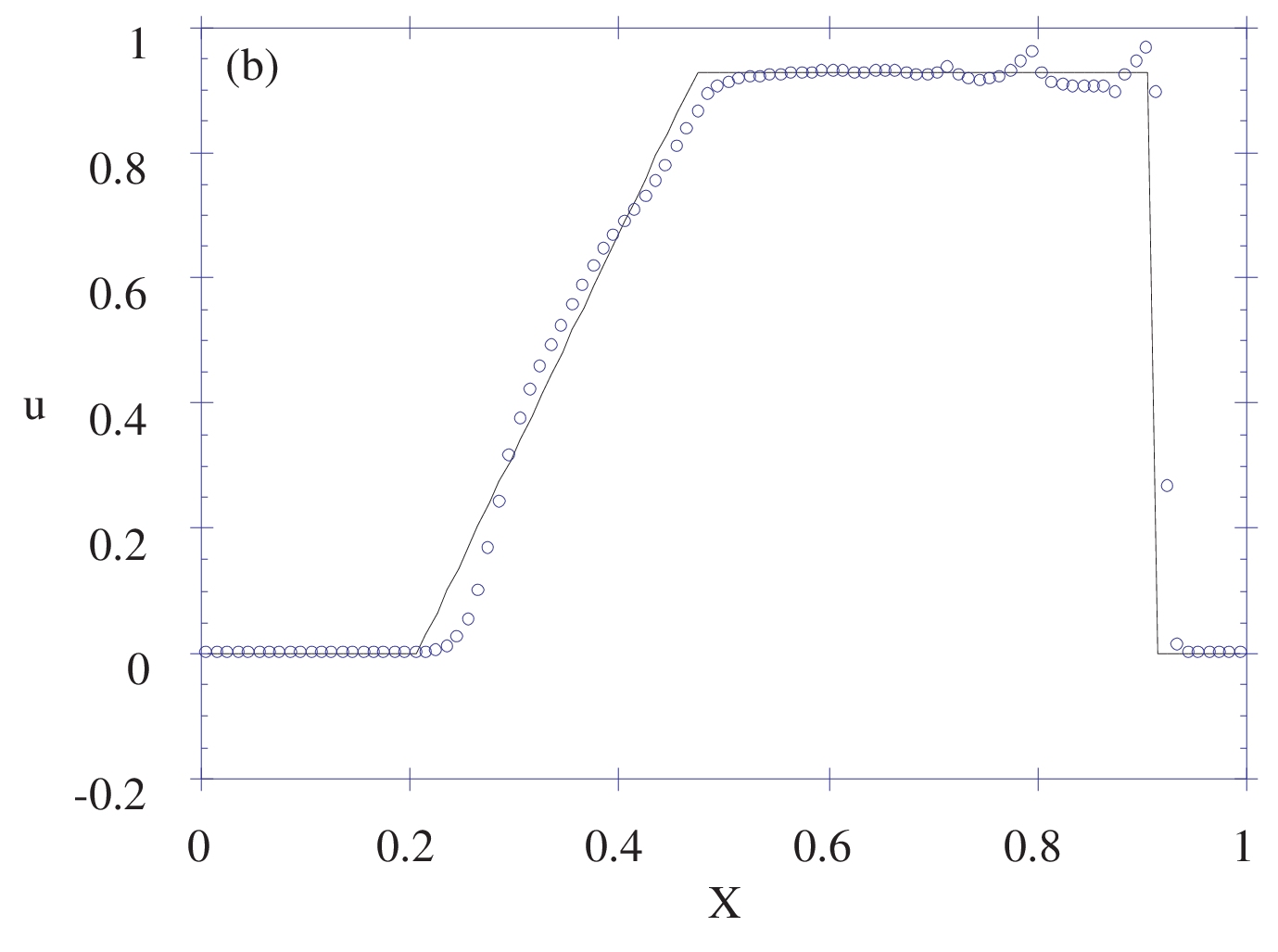}
\includegraphics[width=0.5\textwidth]{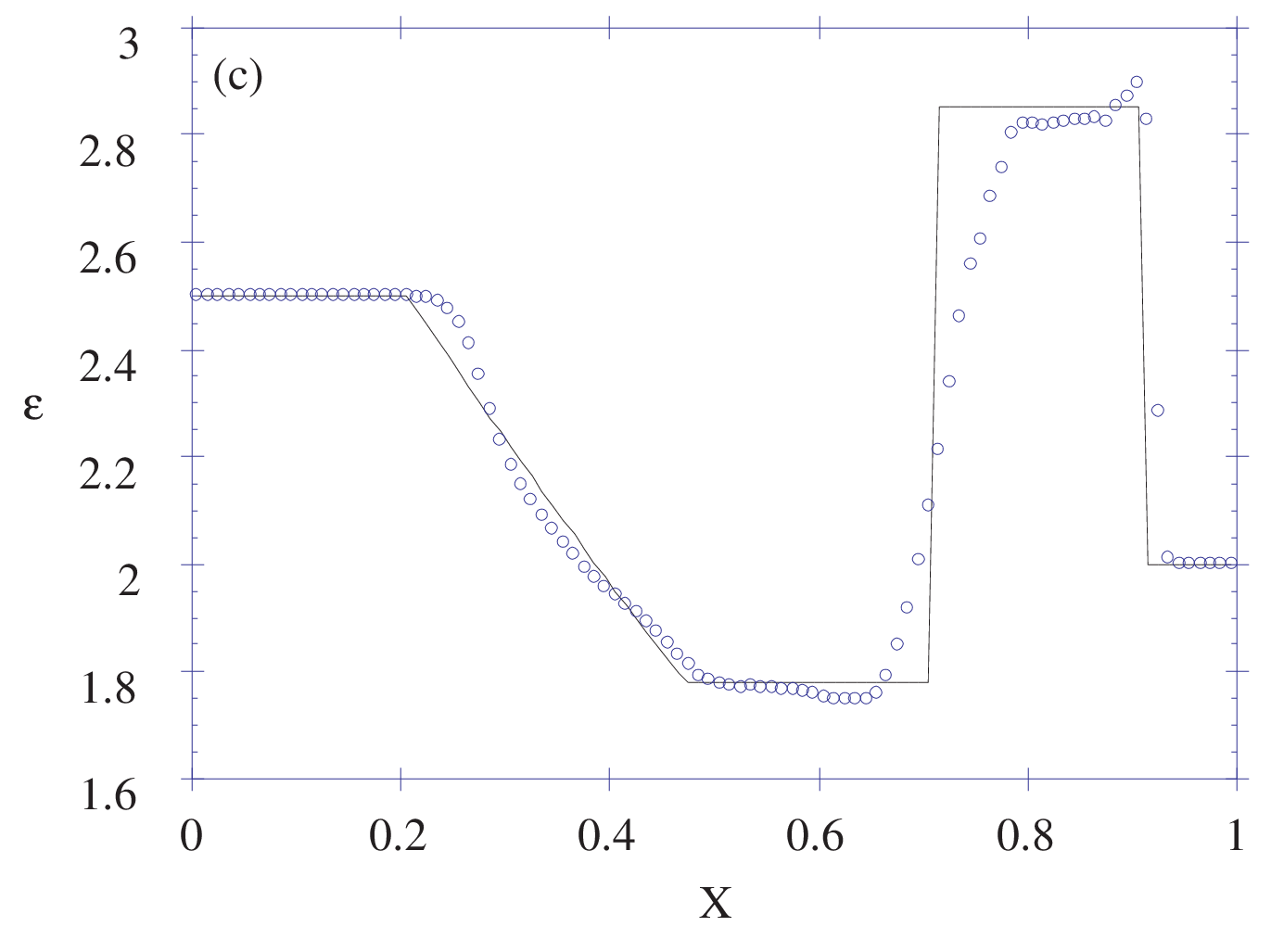}
\includegraphics[width=0.5\textwidth]{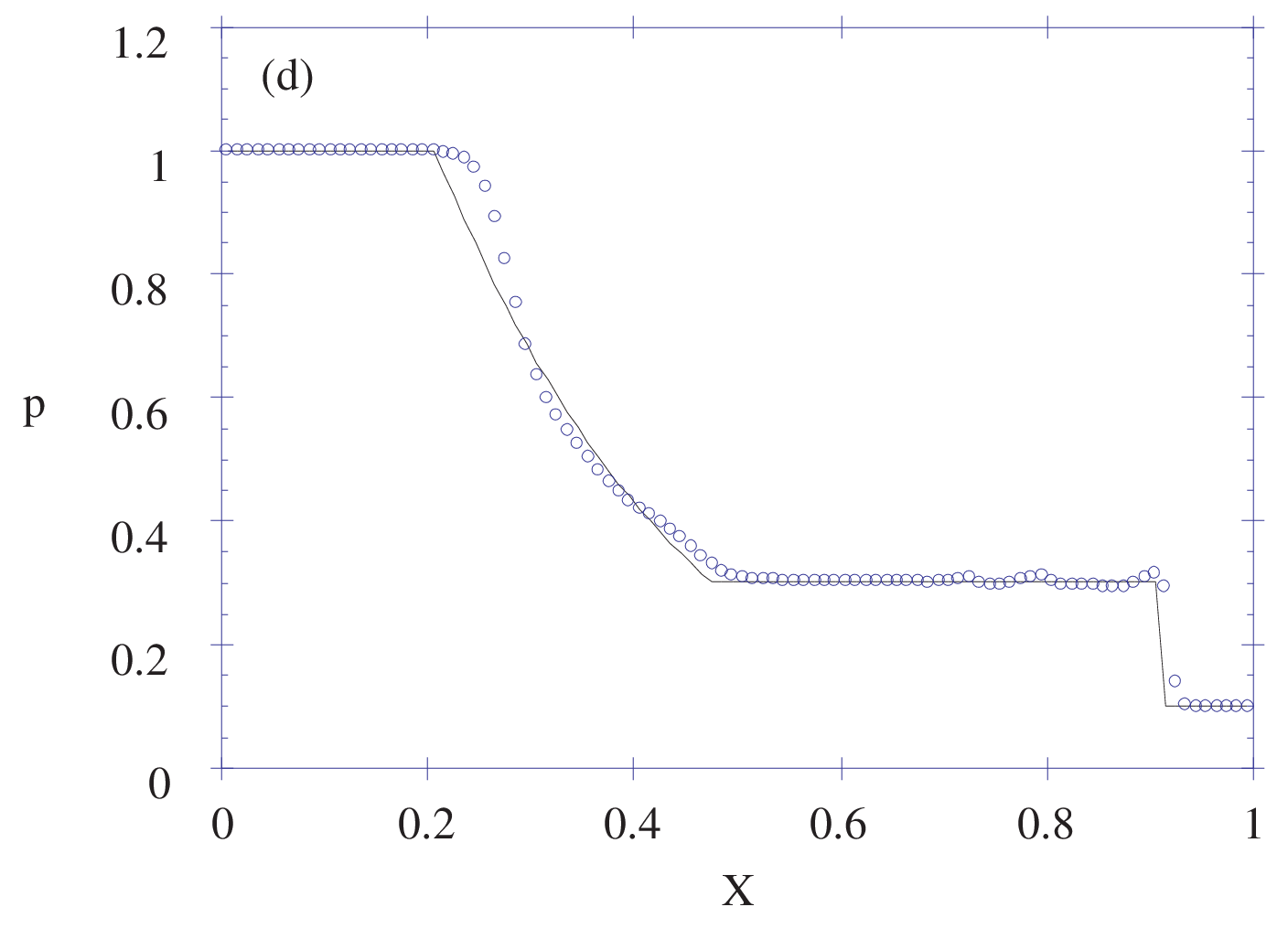}
\label{fig: riemann zalesak fct}
\end{figure}

\begin{figure}
\caption{
Solution of Sod's shock tube problem with the new FCT.
}
\includegraphics[width=0.5\textwidth]{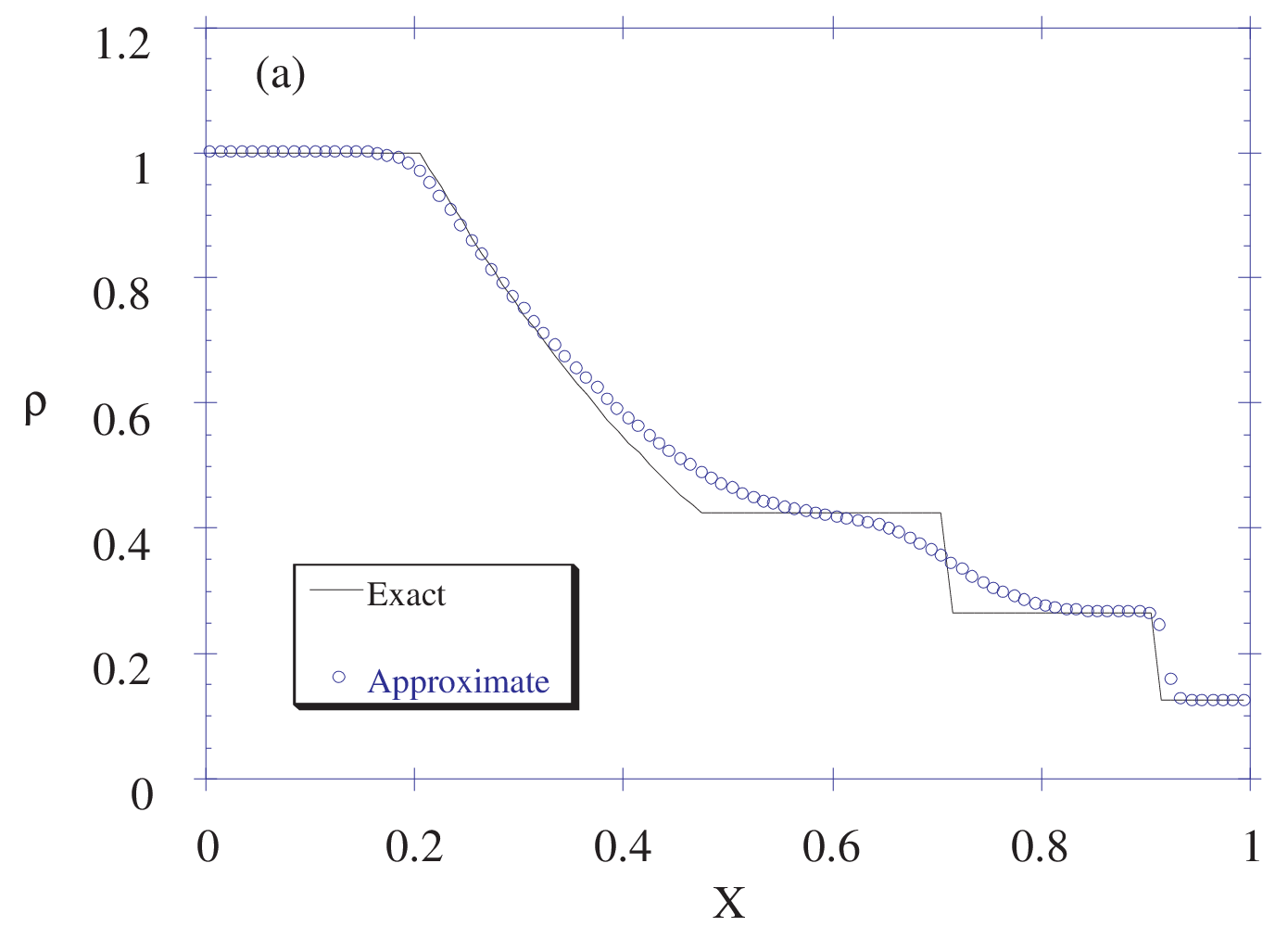}
\includegraphics[width=0.5\textwidth]{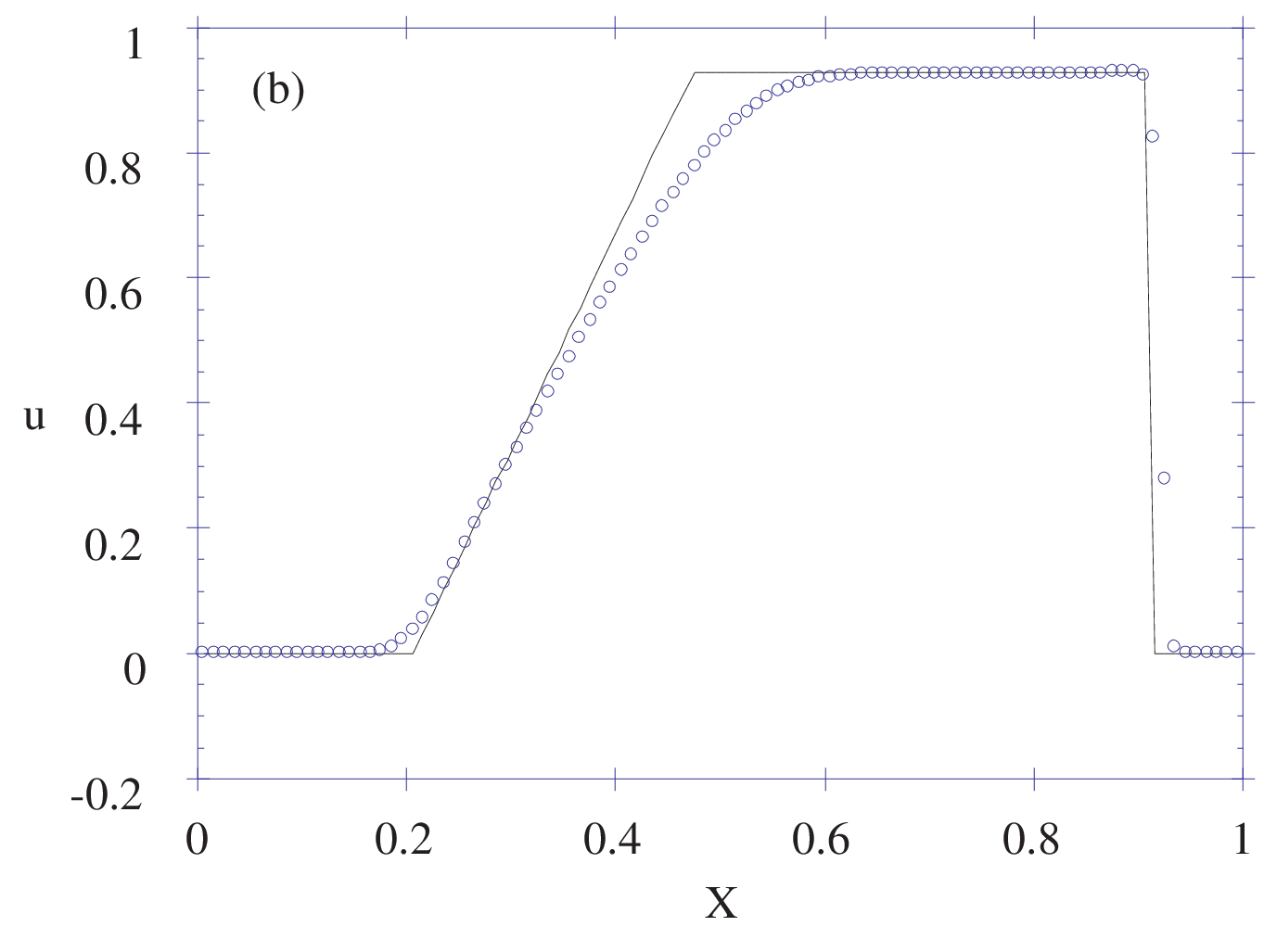}
\includegraphics[width=0.5\textwidth]{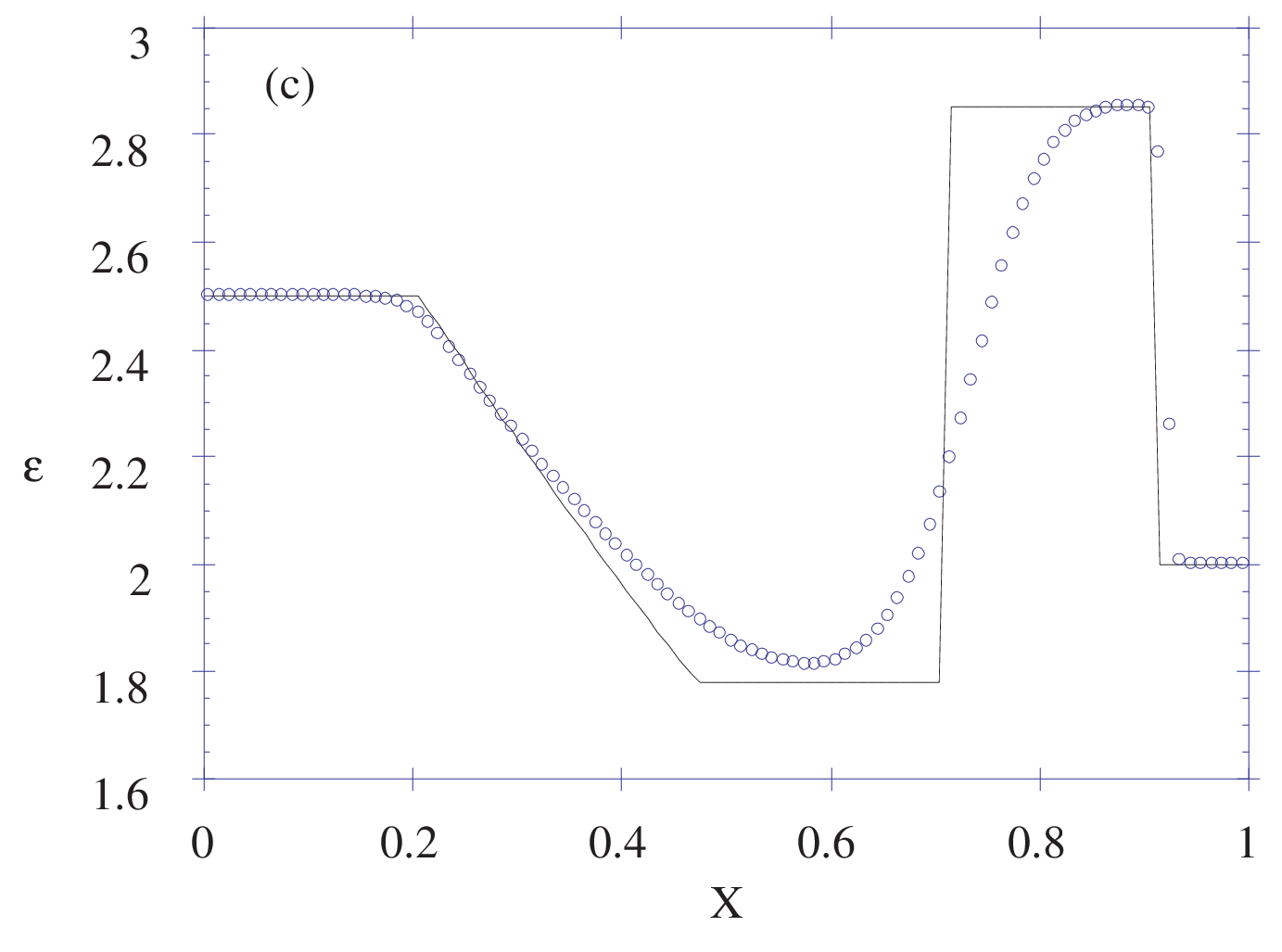}
\includegraphics[width=0.5\textwidth]{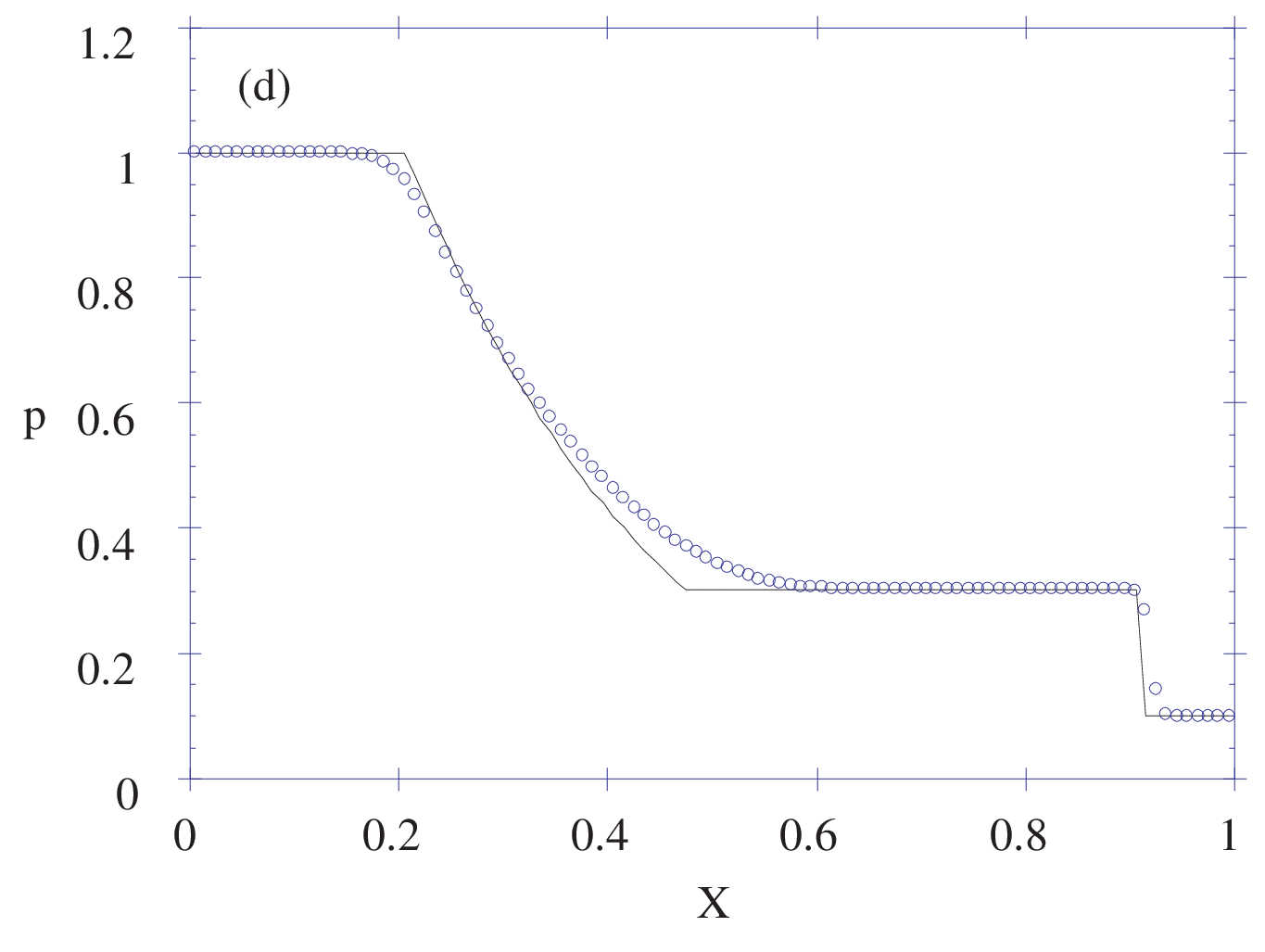}
\label{fig: riemann new fct}
\end{figure}

Figure~\ref{fig: riemann zalesak fct} shows that the results using 
Zalesak's FCT are reasonable, but are polluted with a fair number of 
nonlinear instabilities.  These instabilities are significantly worse if 
the limiter is based on a second-order central differences with numerous 
small expansion shocks present in the rarefaction fan.  Even with the 
extra diffusion produced by the Lax-Wendroff flux, an expansion shock is 
present in the rarefaction wave and oscillations are present in the 
preshock region of the flow.  The overall quality of this solution is 
quite poor.  The new FCT formulation produces qualitatively better 
results that appear to be due to greater dissipation in the scheme.  The 
expansion shock is no longer present.   The overall quality of this 
solution is not high because of the considerable smearing of the 
features of the flow. In Fig.\ \ref{fig: riemann new fct}, the results 
show that a great deal of smearing is present except at the shock wave 
where the solution is very sharp.  In both of these figures the 
pressure-related terms in the momentum and energy equations are 
incorporated as source terms rather than as convective fluxes, and are 
central differenced.

By computing the first step of the new FCT with Roe's first-order 
scheme, and using an approximate Riemann solver to compute the flux 
correction, the results are extremely good.  As Fig.\ \ref{fig: riemann 
new/w fct} shows, the smearing of a standard FCT implementation of the 
new FCT is gone, with the shock being computed with the same crispness.  
The rarefaction fan is smooth and in good agreement with the exact 
solution.  The resolution of the contact discontinuity is somewhat 
smeared but is acceptable.  

\begin{figure}
\caption{
Solution of Sod's shock tube problem with the modified-flux FCT 
and $n=1.5$ limiters on all fields.
}
\includegraphics[width=0.5\textwidth]{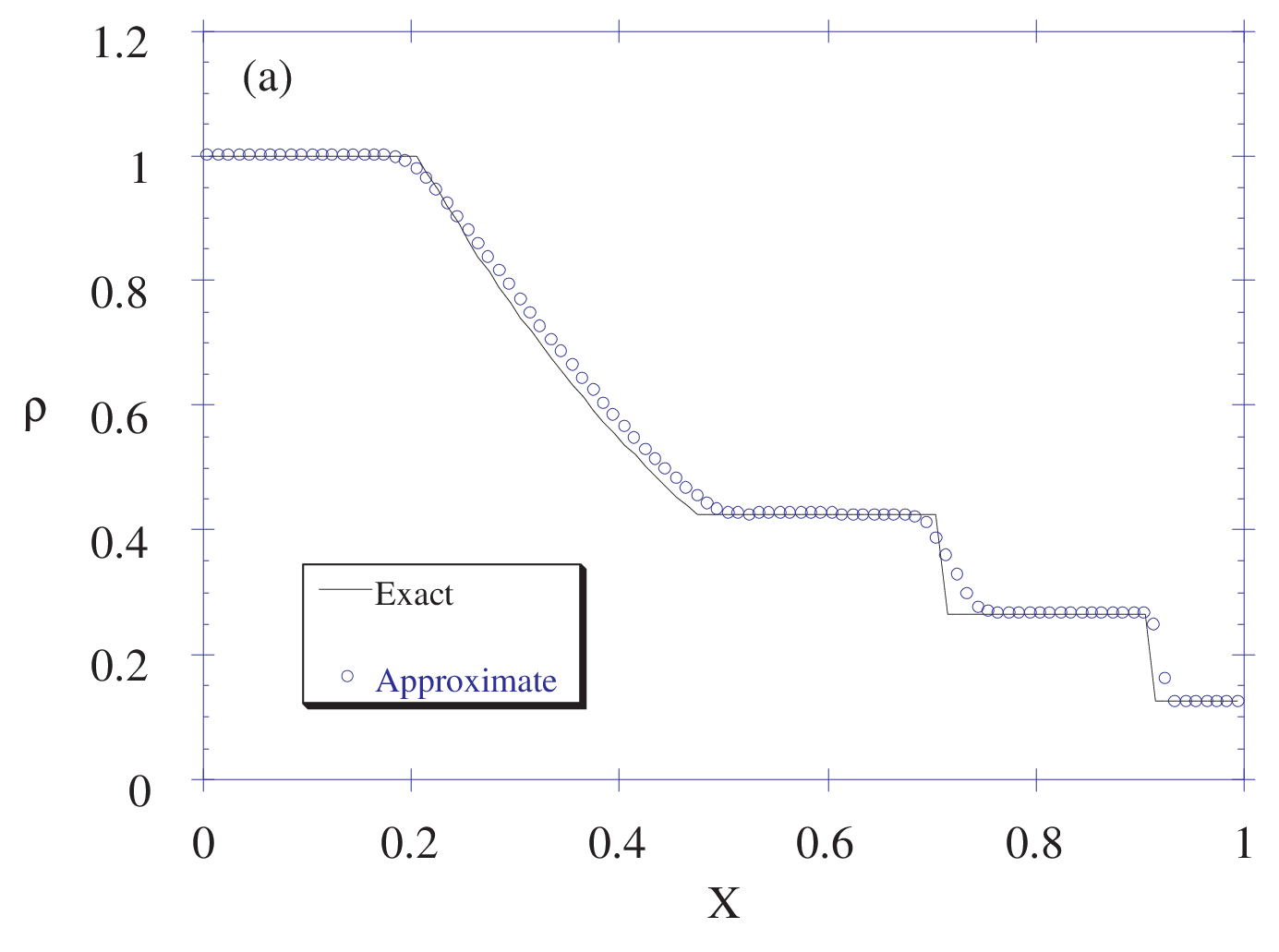}
\includegraphics[width=0.5\textwidth]{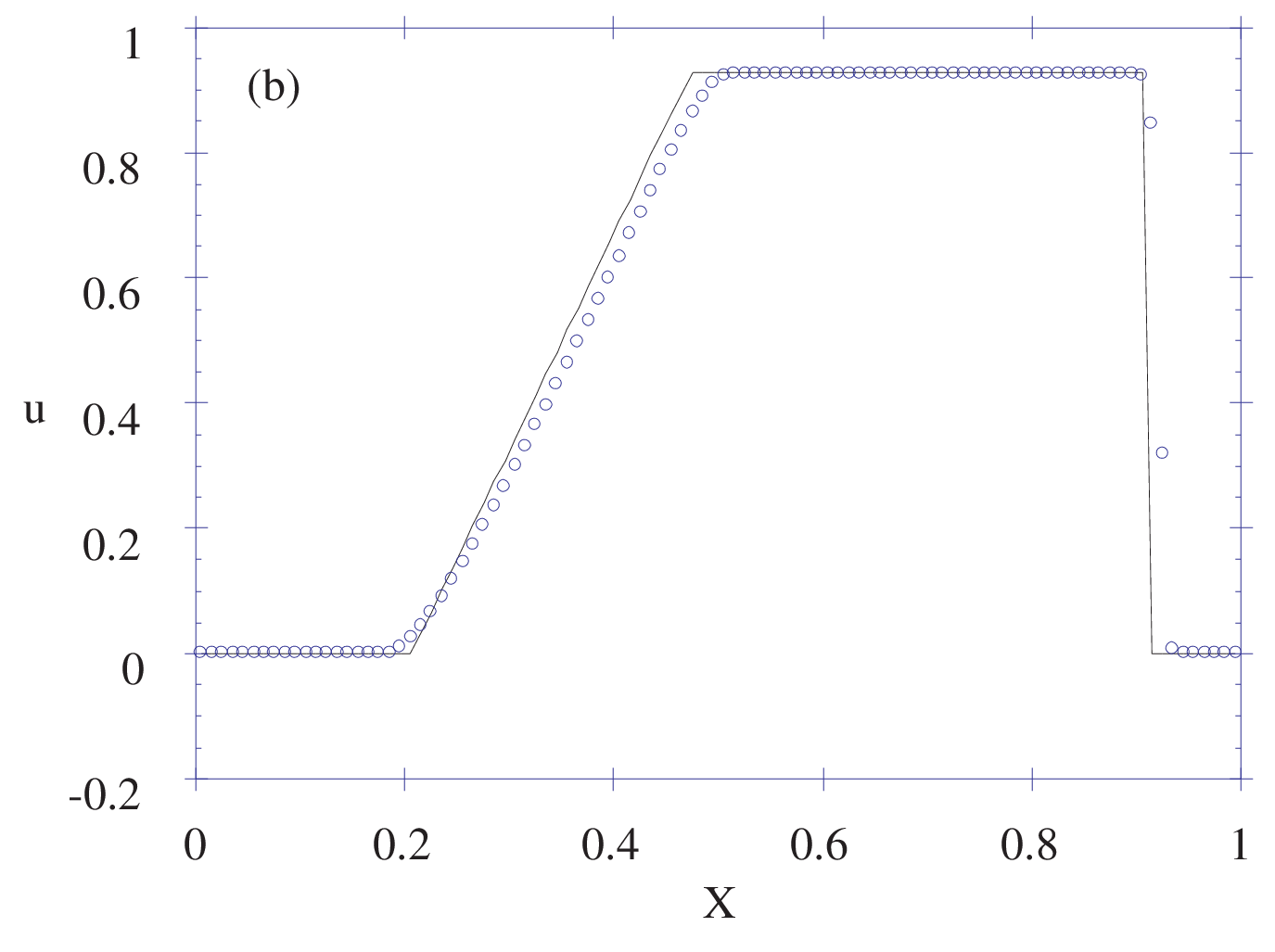}
\includegraphics[width=0.5\textwidth]{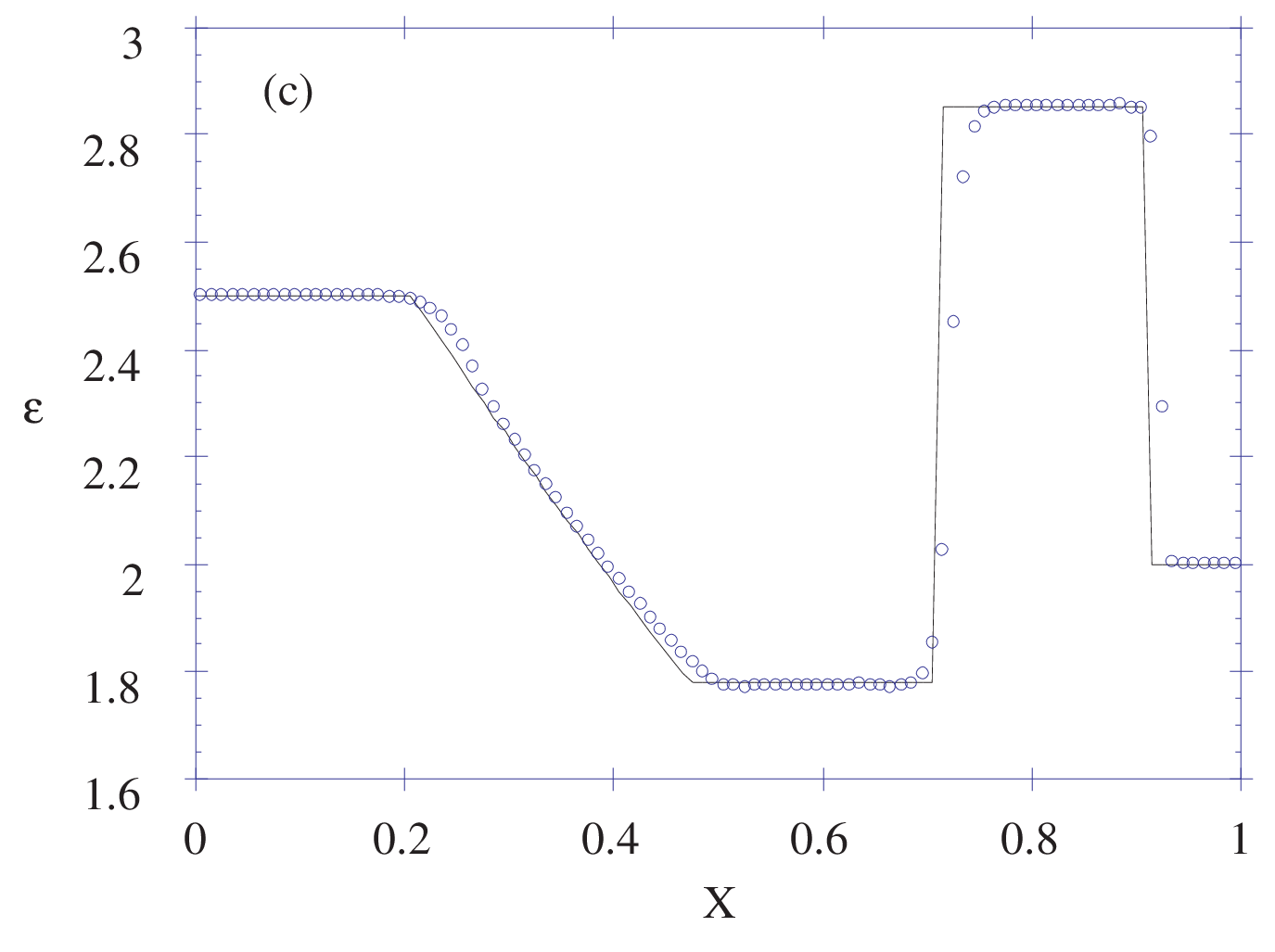}
\includegraphics[width=0.5\textwidth]{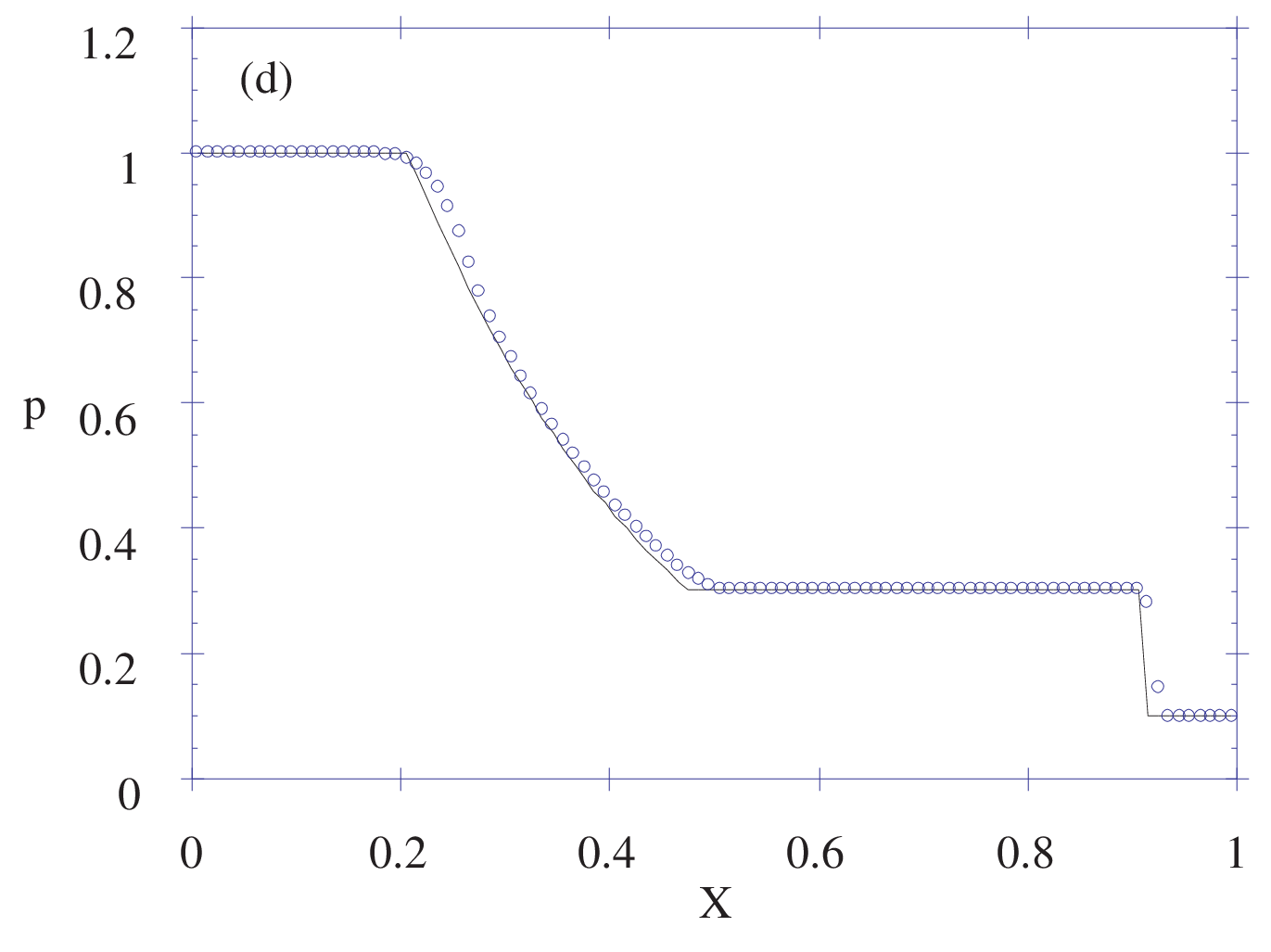}
\label{fig: riemann mod flux fct}
\end{figure}

\begin{figure}
\caption{
Solution of Sod's shock tube problem with a symmetric TVD 
algorithm.
}
\includegraphics[width=0.5\textwidth]{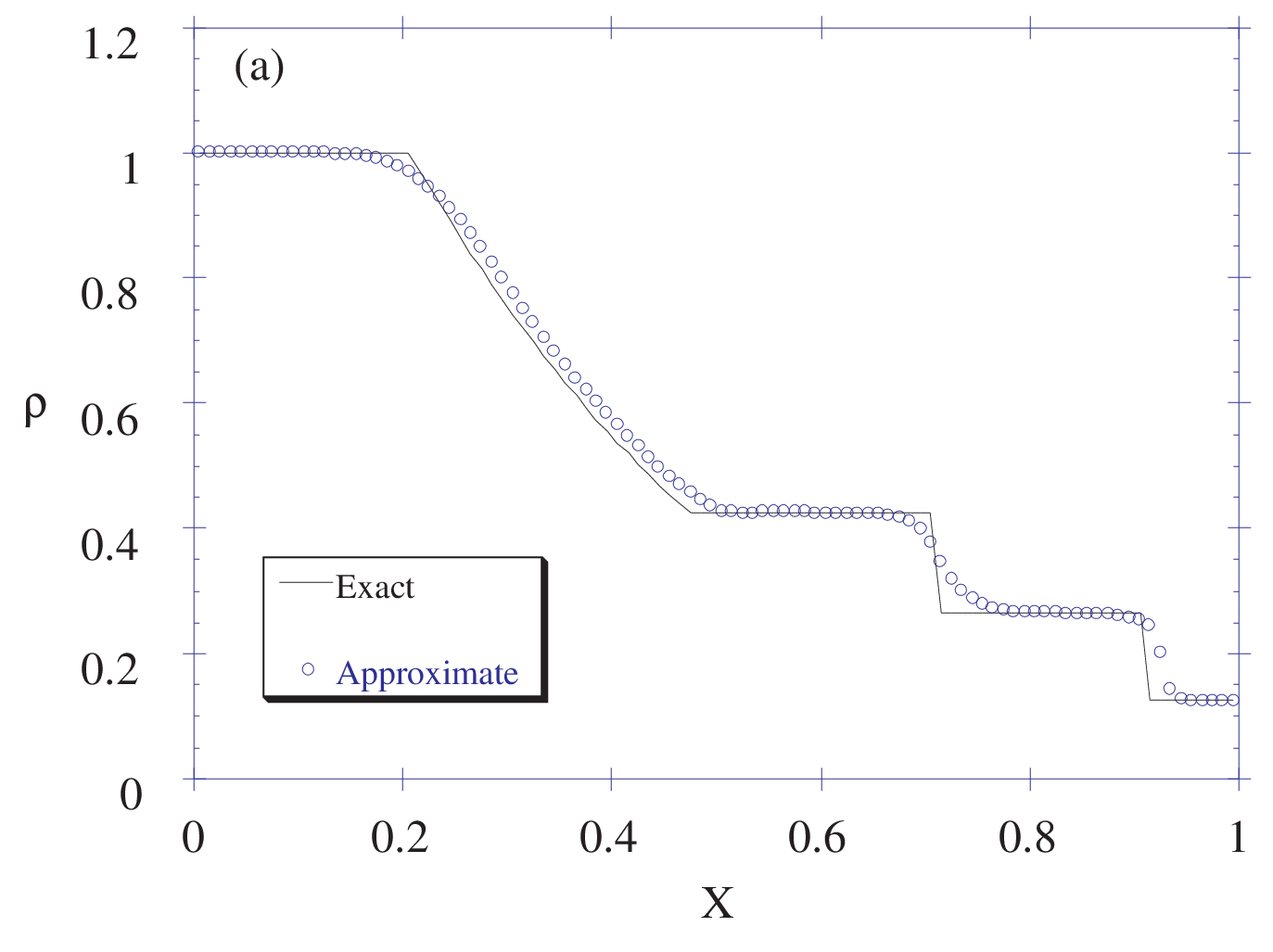}
\includegraphics[width=0.5\textwidth]{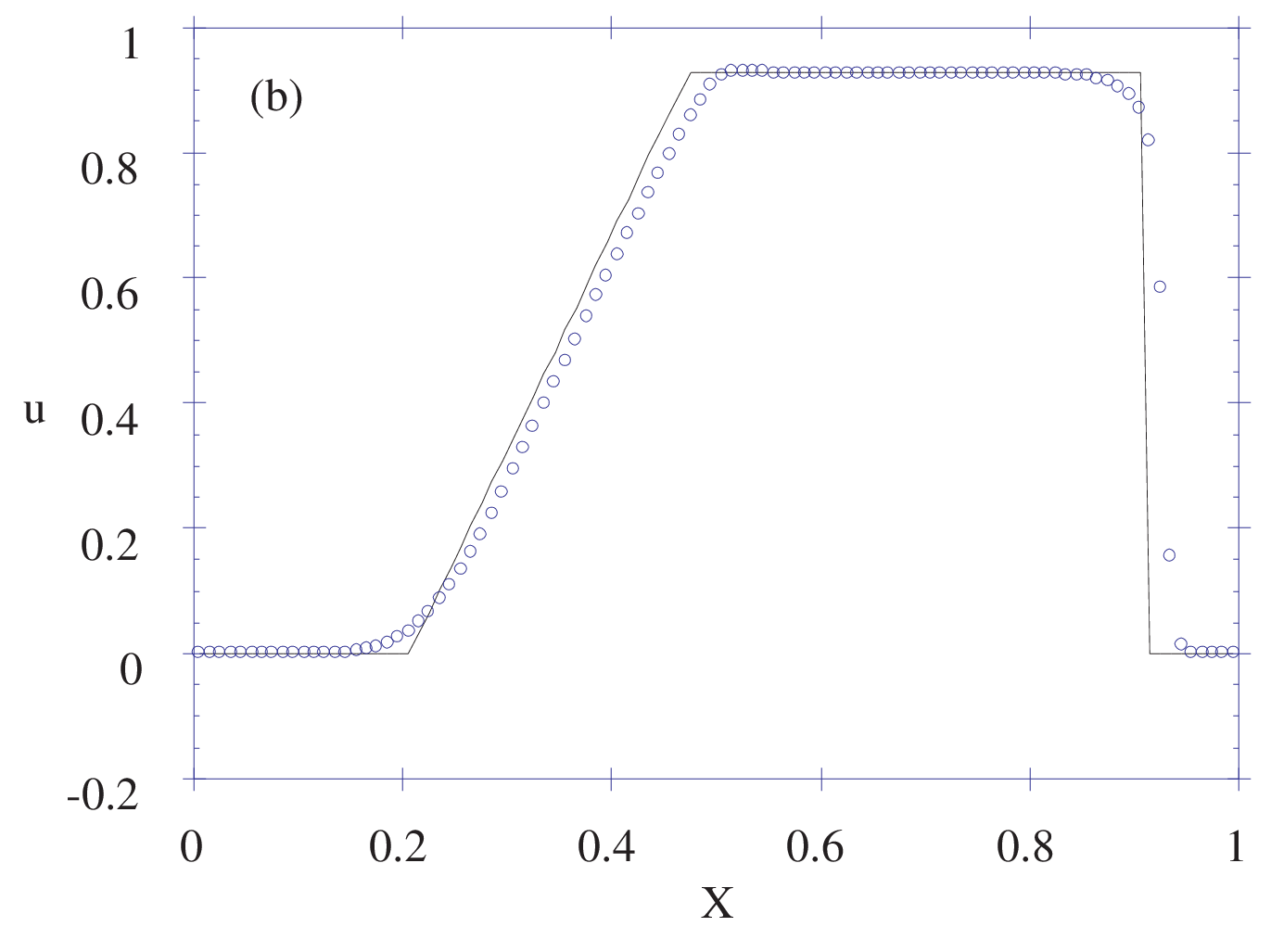}
\includegraphics[width=0.5\textwidth]{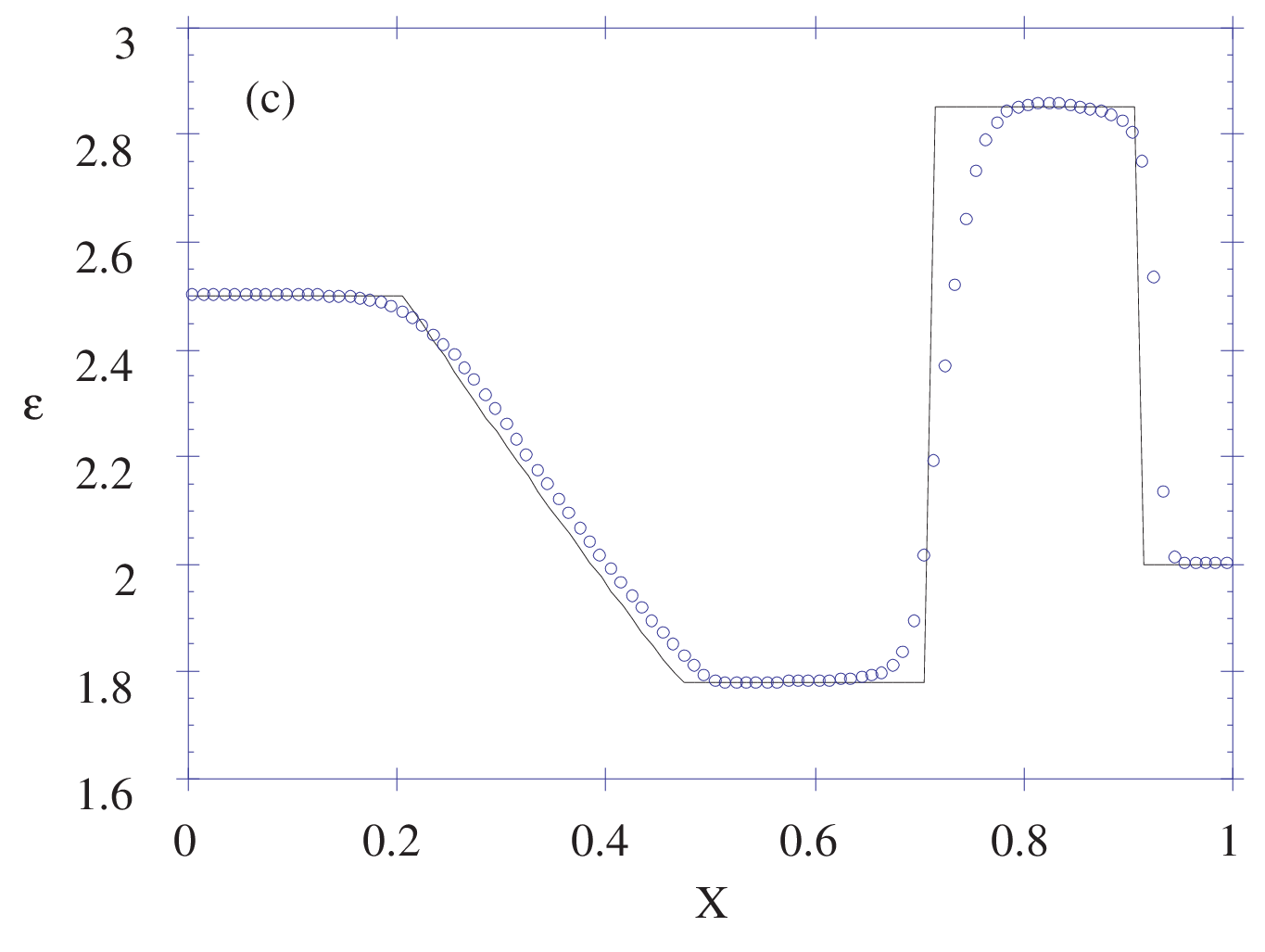}
\includegraphics[width=0.5\textwidth]{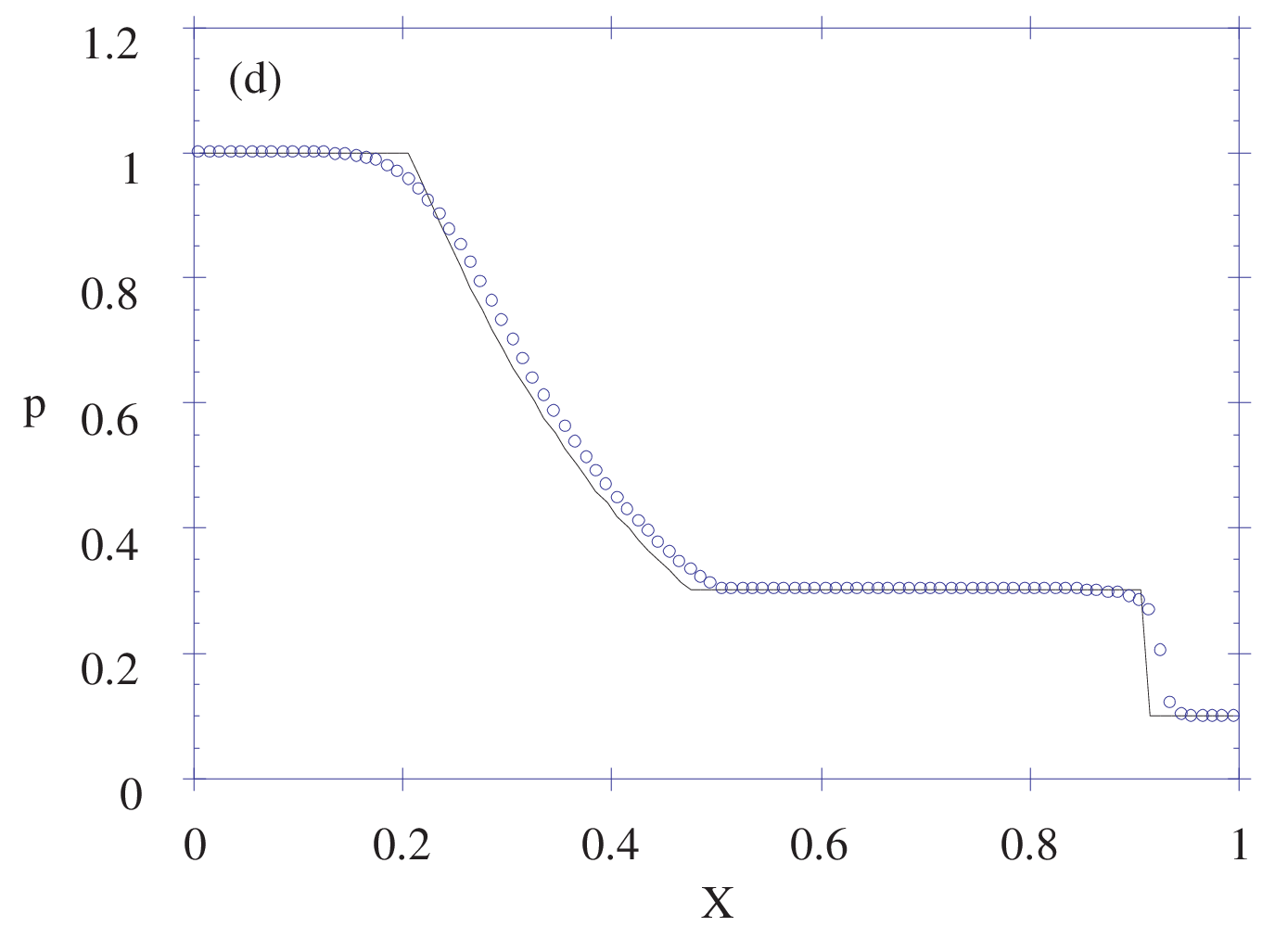}
\label{fig: riemann sym tvd}
\end{figure}

\begin{figure}
\caption{
Solution of Sod's shock tube problem with a UNO limiter and a 
modified-flux TVD algorithm.
}
\includegraphics[width=0.5\textwidth]{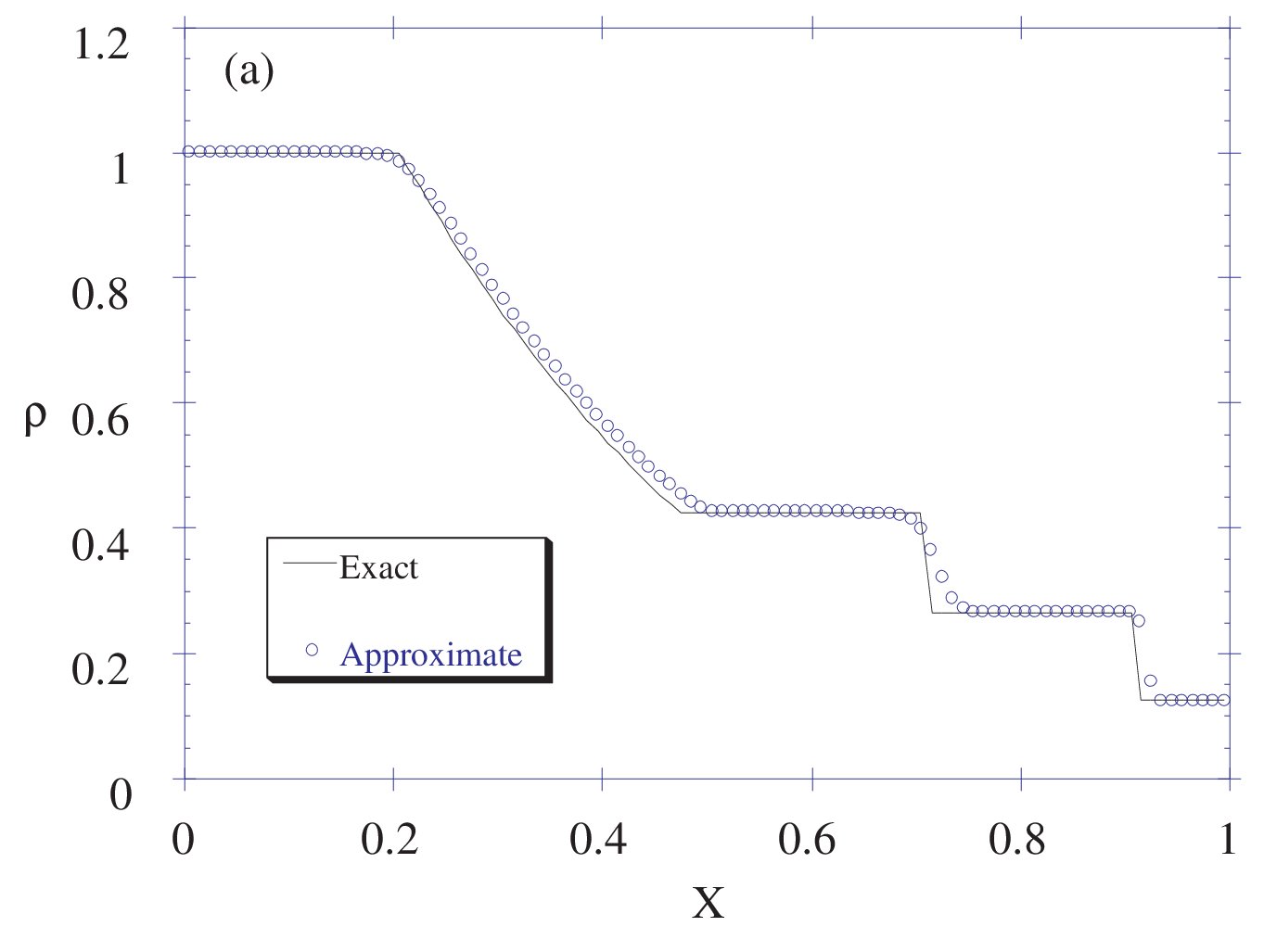}
\includegraphics[width=0.5\textwidth]{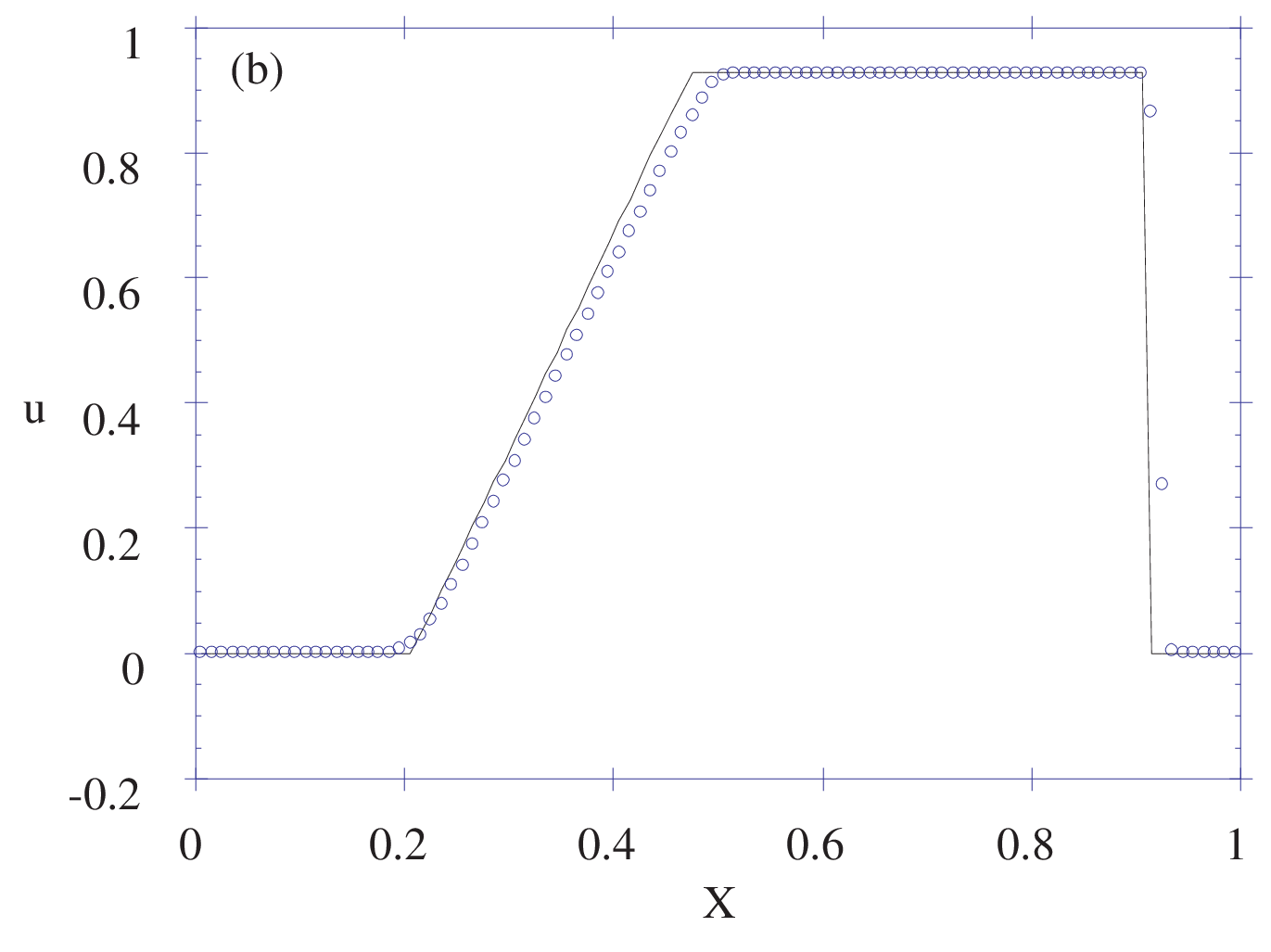}
\includegraphics[width=0.5\textwidth]{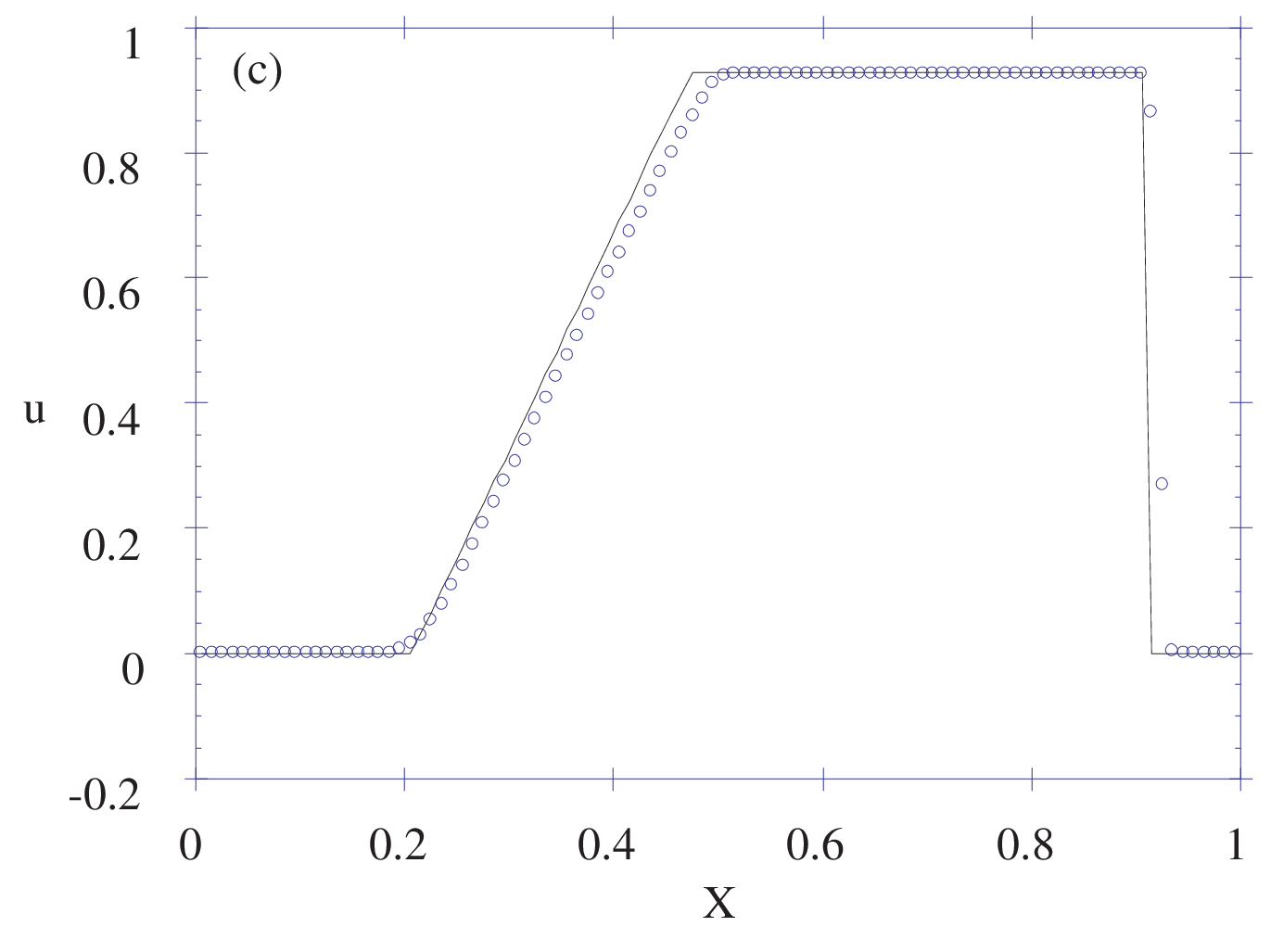}
\includegraphics[width=0.5\textwidth]{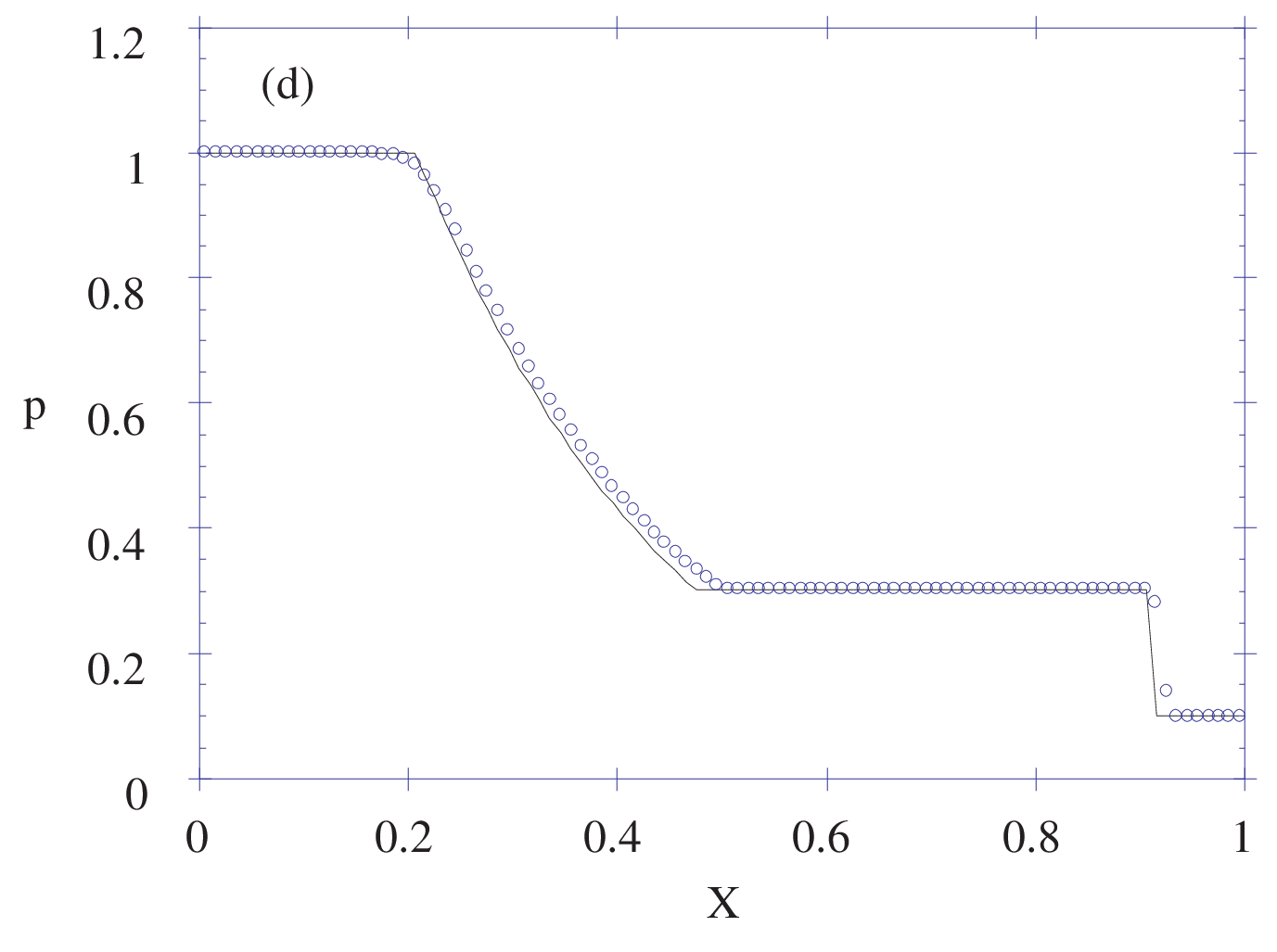}
\label{fig: riemann uno}
\end{figure}

The modified-flux FCT (Fig.\ \ref{fig: riemann mod flux fct}) has 
slightly poorer resolution of the contact discontinuity, but computes 
the shock in a sharper fashion.  The overall quality of the solution is 
nearly identical to the previous case.  In this case the value of 
$n=1.5$ was used on all three fields.  Better resolution of the contact 
discontinuity could be obtained with the $n=2$ limiter.  The use of the 
$n=2$ limiter with the nonlinear fields is generally 
ill-advised~\cite{sweby84}.  The final two figures are shown for 
comparison with the previous figures.  The symmetric TVD method 
(Fig.~\ref{fig: riemann sym tvd}), gives adequate solution although the 
amount of smearing exceeds that of the other methods incorporating Roe's 
approximate Riemann solver.  The UNO method (implemented with a method 
similar to the modified-flux TVD algorithm~\cite{harten87a,munz88}) was 
used to compute the solution shown in Fig.~\ref{fig: riemann uno}.  This 
solution is of a quality similar to that found in Fig.~\ref{fig: riemann 
mod flux fct} with slightly better resolution of each of the features of 
the flow.  Issues related to limiter construction in FCT may need 
additional work to improve the results found using new FCT algorithms.  
It is not clear that simply using higher order antidiffusive fluxes will 
yield better results near extrema and discontinuities in the solution 
without improvements in the other terms in the limiter.

\section{Concluding Remarks}
A generalized FCT algorithm is shown to be TVD under certain conditions.  
The method does not need the standard diffusive first step in older FCT 
algorithms.  The new algorithm has improved properties from the 
standpoint of both use and analysis.  Results show that the new FCT 
algorithms (both ``new FCT'' and modified-flux FCT) perform better than 
the older FCT algorithms, and are comparable with other modern methods.  
This is shown to be especially important for systems of equations where 
the improvement is greater than with the scalar wave equation.  The new 
formulation allows Riemann solvers to be used effectively with FCT 
methods.  Additionally, this paper has more clearly defined the link 
between FCT methods and TVD methods.  This will allow advances made with 
one method to more easily fertilize improvements in the other type of 
method.

The initial motivation of this work was to tie together in a more 
coherent fashion the various modern high-resolution methods for 
numerically solving hyperbolic conservation laws.   This work should be 
considered a start, with the advances mentioned above, as progress 
toward this goal.

Future work will include the modification of the FCT to include 
MUSCL-type schemes as well as the appropriate generalization of 
Zalesak's multidimensional limiter to these types of methods.  As 
mentioned earlier, these methods, once cast in the appropriate form, can 
be used for implicit time integration where the necessary form is 
similar to that found in TVD implicit formulations. Tests on simple test 
problems indicate that these methods are unconditionally stable.

In the next paper of this series we will extend the current analysis to 
a more geometric approach.

\begin{center}
{\bf Acknowledgments}
\end{center}

This work was performed under the auspices of the U. S. Department of 
Energy.

\begin{figure}
\caption{
The characteristics of the FCT limiters for the modified-flux 
formulation.
}
\includegraphics[width=0.5\textwidth]{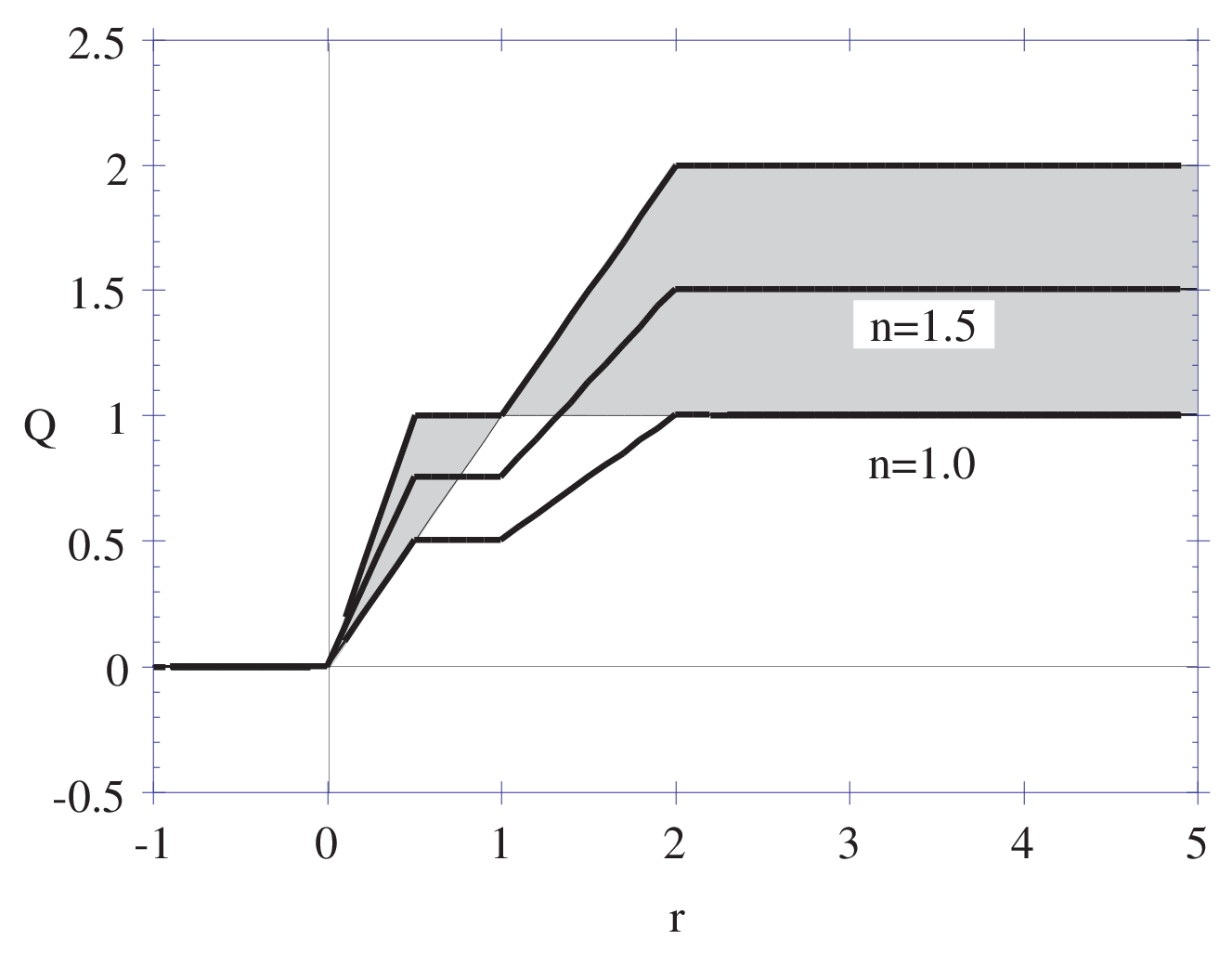}
\label{fig: limiters}
\end{figure}

\begin{figure}
\caption{
Solution of the scalar advection equation with the modified-flux 
FCT ($n=2$ limiter).
}
\includegraphics[width=0.5\textwidth]{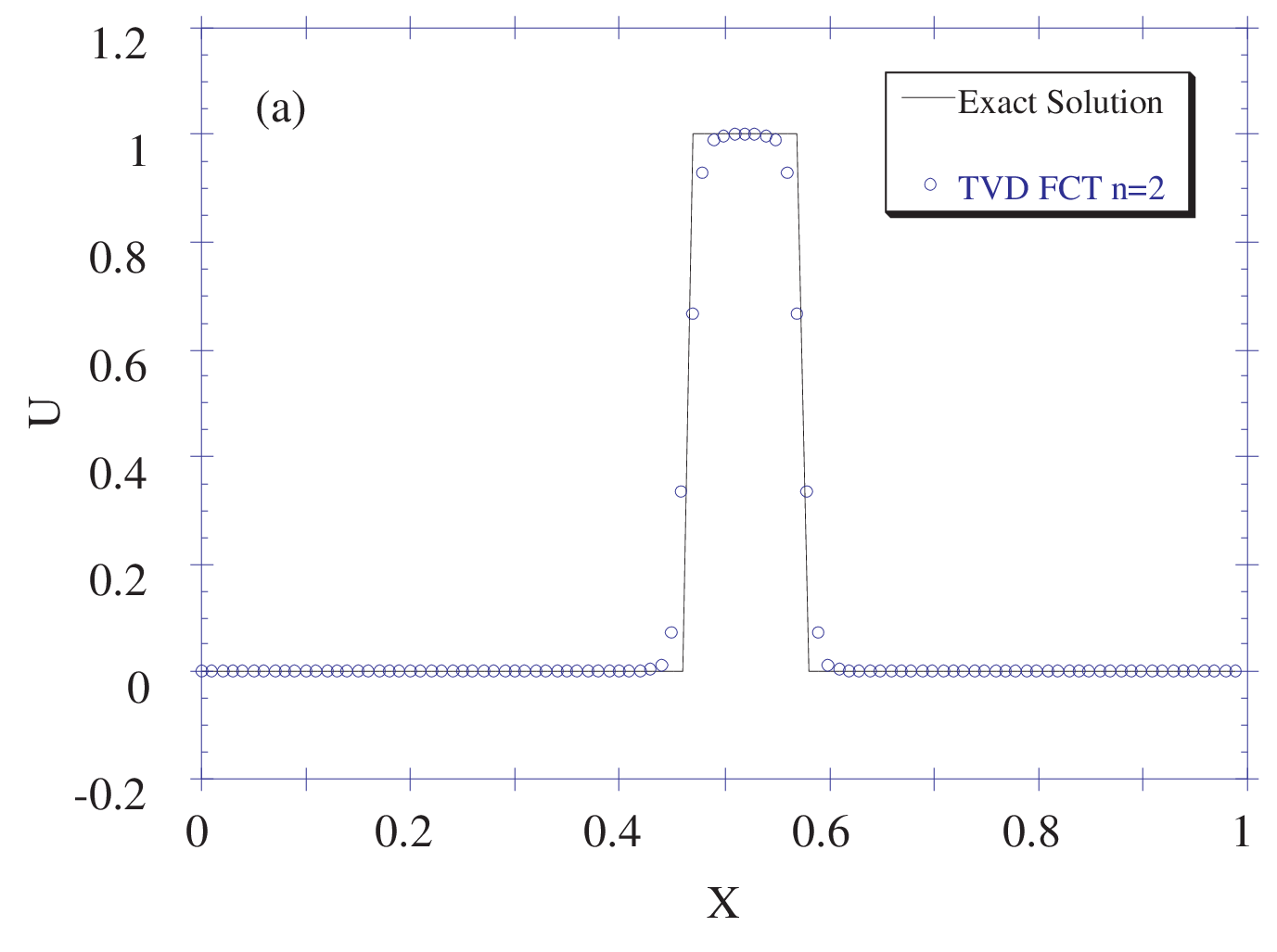}
\includegraphics[width=0.5\textwidth]{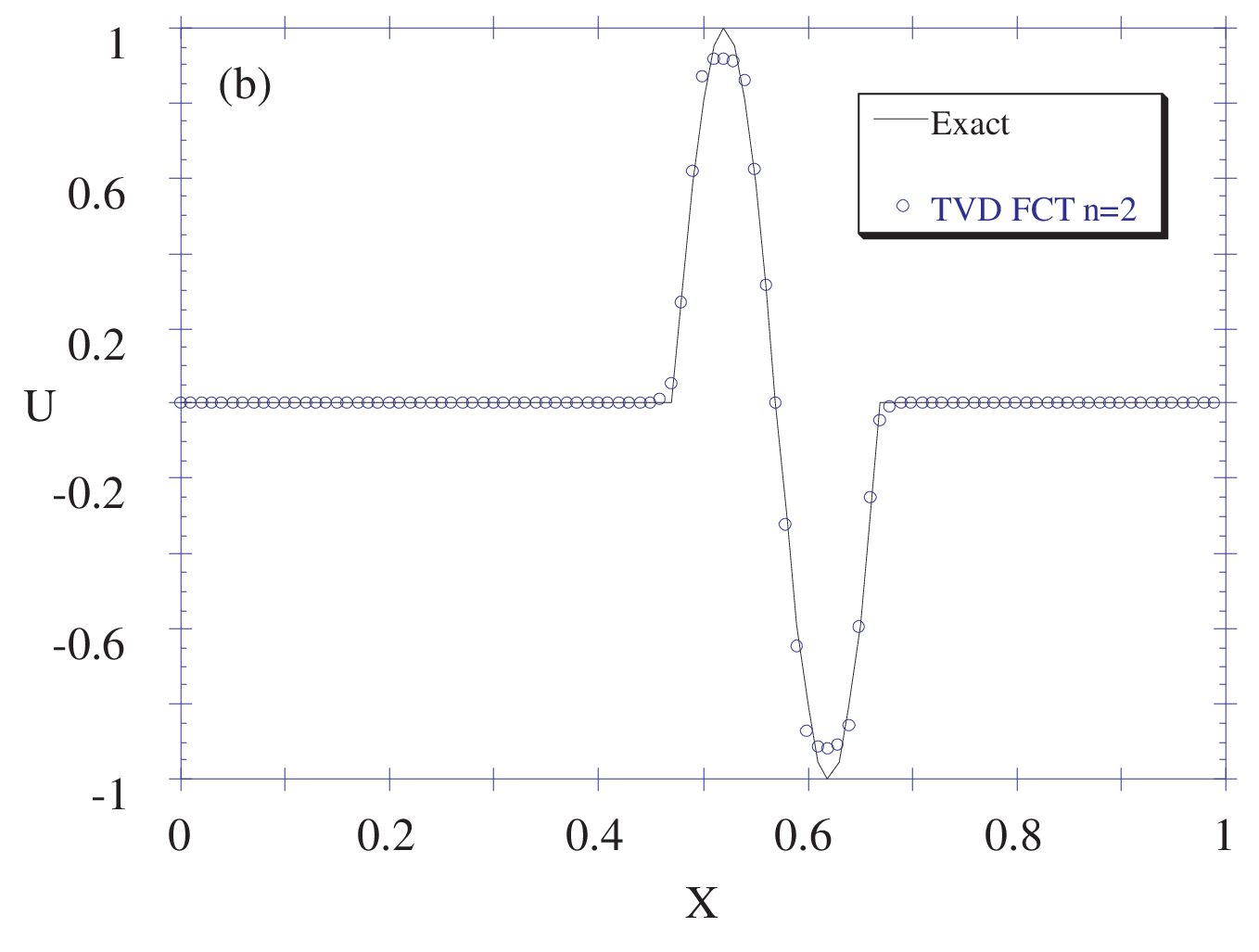}
\label{fig: mod flux fct scalar n=2}
\end{figure}

\clearpage

\clearpage

\begin{figure}
\caption{
Solution of Sod's shock tube problem with new FCT with Roe's 
approximate Riemann solver used to define both low- and high-order 
fluxes.
}
\includegraphics[width=0.5\textwidth]{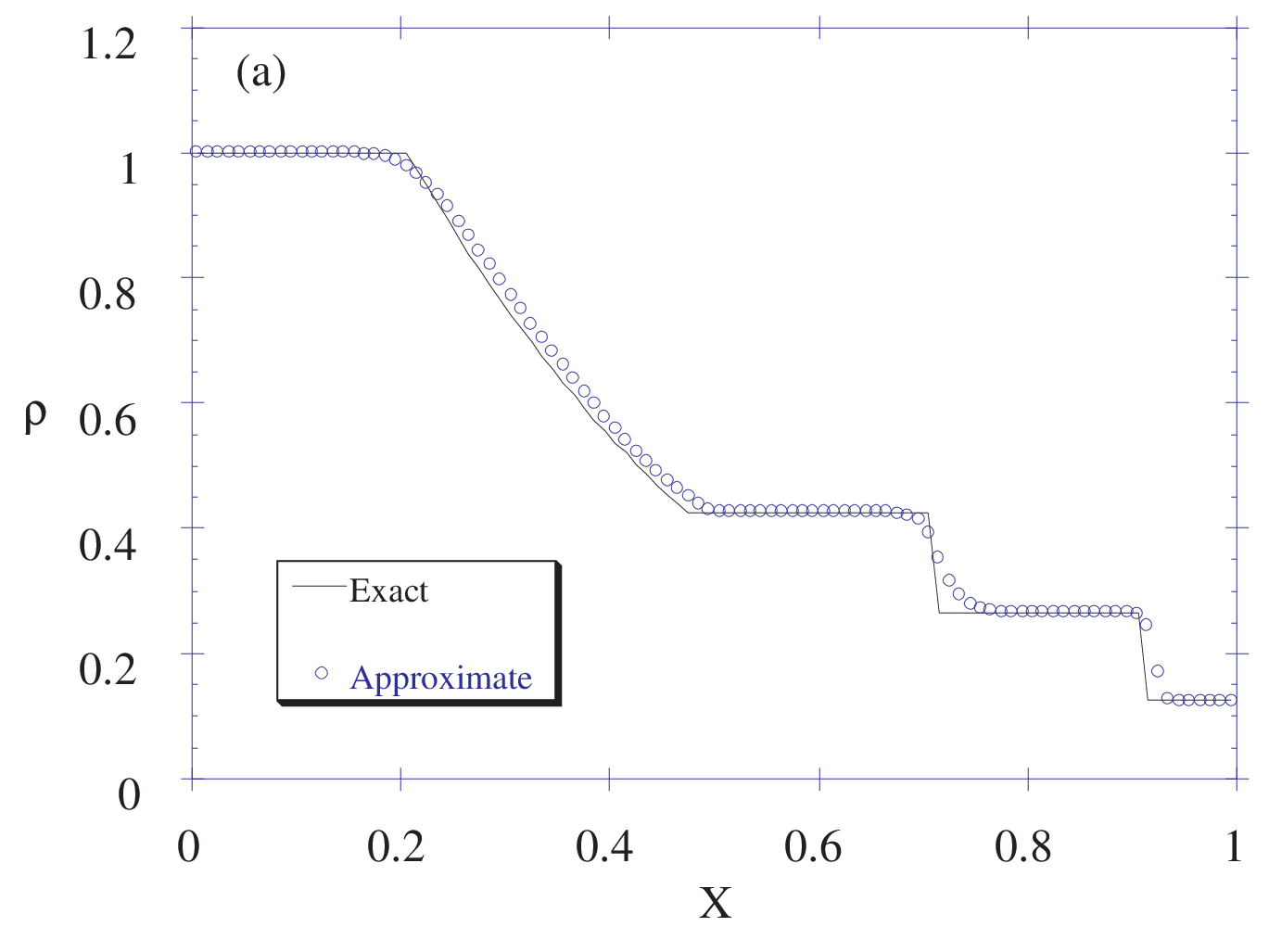}
\includegraphics[width=0.5\textwidth]{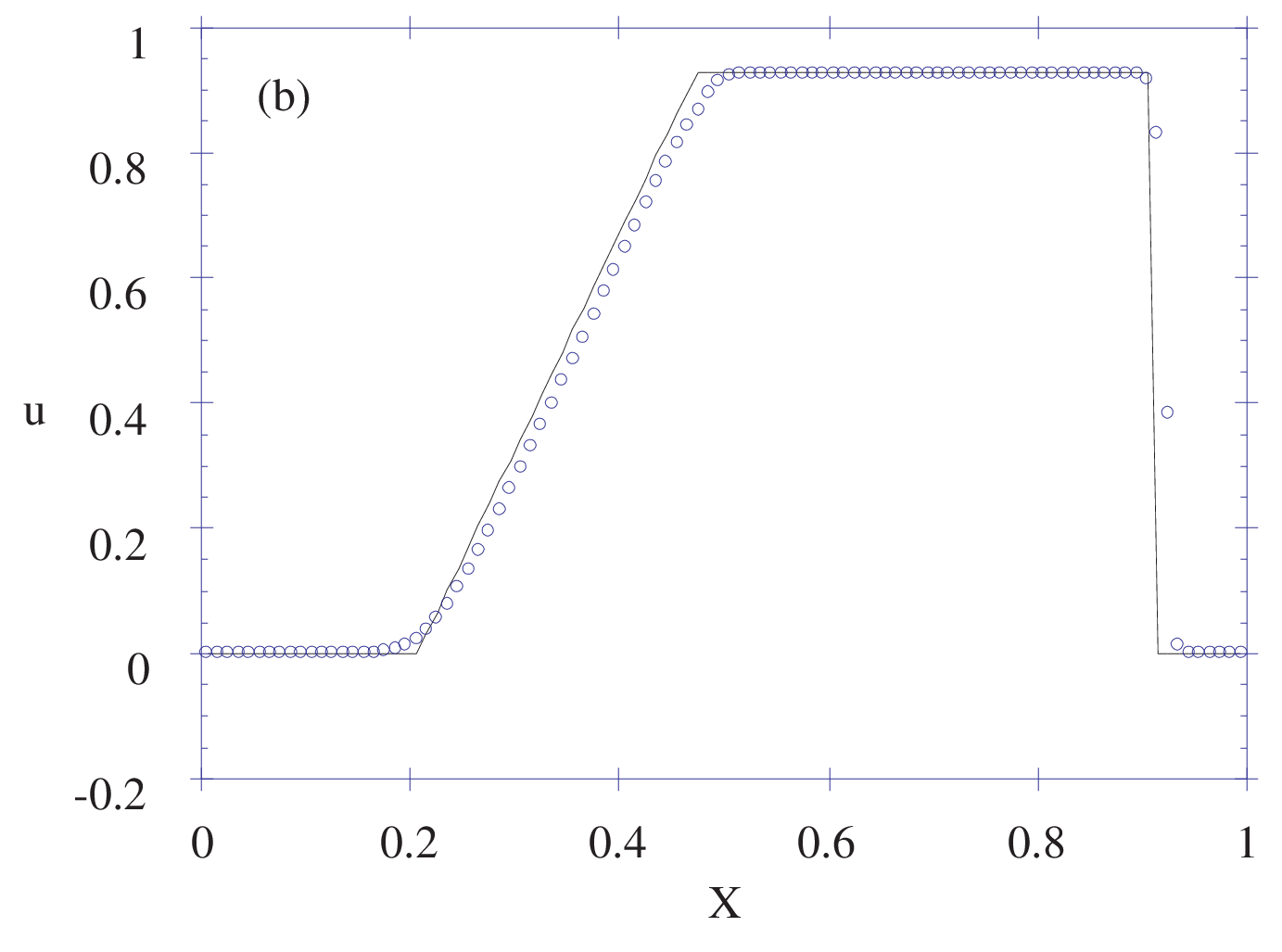}
\includegraphics[width=0.5\textwidth]{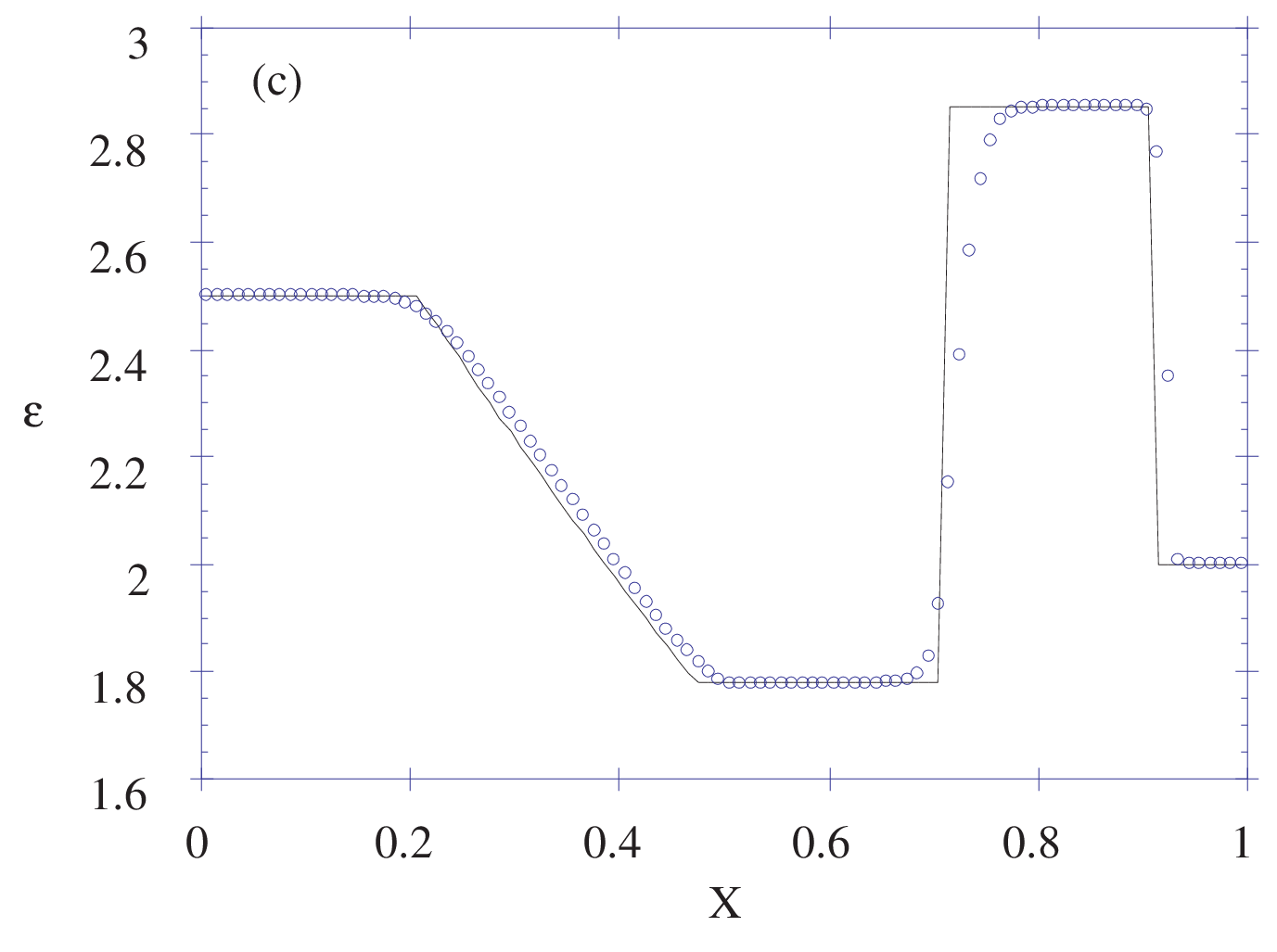}
\includegraphics[width=0.5\textwidth]{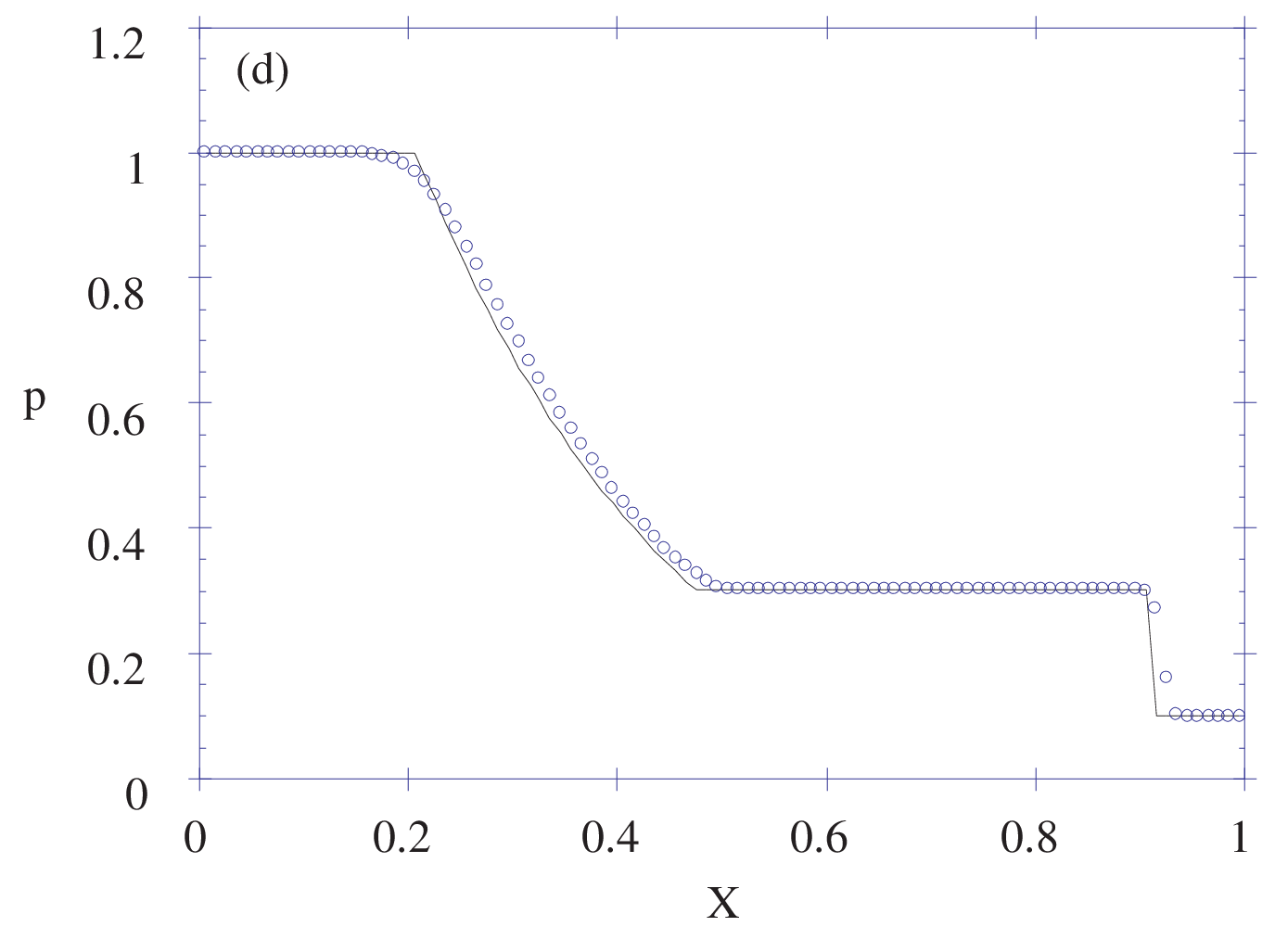}
\label{fig: riemann new/w fct}
\end{figure}

\end{document}